\documentclass{aa}  
\bibpunct{(}{)}{;}{a}{}{,} 

\usepackage{graphicx}
\usepackage{subfig}
\usepackage[varg]{txfonts}
\begin{document}

   \title{Pan-STARRS1 variability of XMM-COSMOS AGN }

   \subtitle{II. Physical correlations and power spectrum analysis}

   \author{T. Simm\inst{1}
          \and M. Salvato\inst{1}
          \and R. Saglia\inst{1,2}
          \and G. Ponti\inst{1}
          \and G. Lanzuisi\inst{3,4}
          \and K. Nandra\inst{1}
          \and R. Bender\inst{1,2}
          }

   \institute{Max Planck Institute for Extraterrestrial Physics, Giessenbachstrasse, Postfach 1312, 85741 Garching, Germany\\
   \email{tsimm@mpe.mpg.de} 
         \and University Observatory Munich, Ludwig-Maximilians Universitaet, Scheinerstrasse 1, 81679 Munich, Germany
         \and INAF – Osservatorio Astronomico di Bologna, Via Ranzani 1, 40127, Bologna, Italy
         \and Dipartimento di Fisica e AstronomiaUniversità di Bologna, viale Berti Pichat 6/2, 40127, Bologna, Italy	      
             }


 
  \abstract
   {}
   {The goal of this work is to better understand correlations between the rest-frame UV/optical variability amplitude of QSOs and physical quantities like redshift, luminosity, black hole mass and Eddington ratio. Previous analysis of the same type found evidence for correlations between the variability amplitude and these AGN parameters. However, most of the found relations exhibit considerable scatter and the trends obtained by various authors are often contradictory. Moreover the shape of the optical power spectral density (PSD) is currently available for only a handful of objects.}
   {We search for scaling relations between the fundamental AGN parameters and rest-frame UV/optical variability properties for a sample of $\sim$90 X-ray selected AGNs covering a wide redshift range from the XMM-COSMOS survey, with optical light curves in four bands ($g_{\mathrm{P1}}$, $r_{\mathrm{P1}}$, $i_{\mathrm{P1}}$, $z_{\mathrm{P1}}$) provided by the Pan-STARRS1 (PS1) Medium Deep Field 04 survey. To estimate the variability amplitude we utilize the normalized excess variance ($\sigma_{\mathrm{rms}}^{2}$) and probe variability on rest-frame timescales of several months and years by calculating $\sigma_{\mathrm{rms}}^{2}$ from different parts of our light curves. In addition, we derive the rest-frame optical PSD for our sources using continuous-time autoregressive moving average (CARMA) models.}
   {We observe that the excess variance and the PSD amplitude are strongly anti-correlated with wavelength, bolometric luminosity and Eddington ratio. There is no evidence for a dependency of the variability amplitude on black hole mass and redshift. These results suggest that the accretion rate is the fundamental physical quantity determining the rest-frame UV/optical variability amplitude of quasars on timescales of months and years. The optical PSD of all of our sources is consistent with a broken power law showing a characteristic bend at rest-frame timescales ranging between $\sim$100 and $\sim$300 days. The break timescale exhibits no significant correlation with any of the fundamental AGN parameters. The low frequency slope of the PSD is consistent with a value of $-1$ for most of our objects, whereas the high frequency slope is characterized by a broad distribution of values between $\sim-2$ and $\sim-4$. These findings unveil significant deviations from the simple "damped random walk" model, frequently used in previous optical variability studies. We find a weak tendency for AGNs with higher black hole mass having steeper high frequency PSD slopes.}
   {}

   \keywords{	accretion, accretion disks --
   				methods: data analysis -- 
   				black hole physics --
   				galaxies: active --
   				quasars: general --
   				X-rays: galaxies
               }

   \maketitle
%

\section{Introduction}

Albeit known for many decades, the physical origin of AGN variability is still an open question. Several mechanisms have been proposed to explain the notorious flux variations, but to date there is no preferred model that is able to predict all the observed features of AGN variability in a self-consistent way \citep{2000ApJ...544..123C,2002MNRAS.329...76H,2006ApJ...642...87P}. Unveiling the source of AGN variability promises better understanding of the physical processes that power these luminous objects. AGN variability is characterized by non-periodic random fluctuations in flux, occurring with different amplitudes on timescales of hours, days, months, years and even decades \citep{2003A&AT...22..661G}. Very strong variability may also be present on much longer timescales of $10^{5}$--$10^{6}$ years \citep{2014ApJ...782....9H,2015MNRAS.451.2517S}. The variability is observed across-wavelength and is particularly strong in the X-ray, UV/optical and radio bands \citep{1997ARA&A..35..445U}. The X-ray band shows very rapid variations, typically with larger amplitude than optical variability on short timescales of days to weeks. However, optical light curves exhibit larger variability amplitudes on longer timescales of months to years on the level of $\sim$10\%--20\% in flux \citep{2003A&AT...22..661G,2014SSRv..183..453U}. Optical variability of AGNs has been studied extensively in the last years, providing a useful tool for quasar selection as well as a probe for physical models describing AGNs  \citep{2009ApJ...698..895K,2011ApJ...730...52K,2013ApJ...779..187K,2010ApJ...708..927K,2011ApJS..194...22K,2012ApJ...746...27K,2013ApJ...775...92K,2010ApJ...721.1014M,2011ApJ...728...26M,2012ApJ...753..106M,2010ApJ...714.1194S,	2012ApJ...744..147S,2011A&A...530A.122P,2011AJ....141...93B,2011ApJ...735...68K,2012ApJ...760...51R,2012ApJ...758..104Z,	2013A&A...554A.137A,2013ApJ...765..106Z,2014ApJ...784...92M,2014MNRAS.439..703G,2015A&A...574A.112D,2015A&A...579A.115F,2015arXiv150708676C}.

Since the optical continuum radiation is believed to be predominantly produced by the accretion disc, it is very likely that optical variability originates from processes intrinsic to the disc. One possible mechanism may be fluctuations of the global mass accretion rate, providing a possible explanation for the observed large variability amplitudes \citep{2006ApJ...642...87P,2008MNRAS.387L..41L,2011ApJ...731...50S,2012ApJ...758..104Z,2013A&A...554A..51G}. However, considering the comparably short timescales of optical variability, a superposition of several smaller, independently fluctuating zones of different temperature at various radii, associated with disc inhomogeneities that are propagating inwards, may be a preferable alternative solution \citep{1997MNRAS.292..679L,2001MNRAS.327..799K,2006MNRAS.367..801A,2011ApJ...727L..24D}. Such localized temperature fluctuations are known to describe several characteristics of AGN optical variability \citep{2013A&A...560A.104M,2014ApJ...783..105R,2014ApJ...792...54S} and may arise from thermal or magnetorotational instabilities in a turbulent accretion flow, as suggested by modern numerical simulations \citep[e.g.][]{2009ApJ...691...16H,2013ApJ...767..148J}. 

The strong temporal correlation of optical and X-ray variability observed in simultaneous light curves on timescales of months to years indicates that inwards moving disc inhomogeneities may drive the long term X-ray variability \citep{2003ApJ...584L..53U,2008MNRAS.387..279A,2009MNRAS.397.2004A,2009MNRAS.394..427B,2010MNRAS.403..605B,2015arXiv150207502C}.	On the other hand, the short time lags of few days between different optical bands \citep{1997ApJS..113...69W,2005ApJ...622..129S} are in favour of a model in which X-ray variability is driving the optical variability approximately on light travel times by irradiating and thereby heating the accretion disc \citep{2007MNRAS.380..669C}. Whichever mechanism actually dominates, it is important to compare the properties of optical and X-ray variability, because understanding their coupling provides a detailed view of the physical system at work that can hardly be obtained by other methods than timing analysis.    	

The power spectral density (PSD) states the variability power per temporal frequency $\nu$. The X-ray PSDs of AGNs are observed to be well described by a broken power law $\mathrm{PSD\left(\nu\right)}\propto\nu^{\gamma}$ with $\gamma=-2$ for frequencies above the break frequency $\nu_{br}$ and $\gamma=-1$ for frequencies below $\nu_{br}$ \citep{1993ApJ...414L..85L,1993MNRAS.265..664G,1997ApJ...476...70N,1999ApJ...514..682E,2002MNRAS.332..231U,2003ApJ...593...96M,2004ApJ...617..939M,2004MNRAS.348..783M,	2012A&A...544A..80G}. Such PSDs are modelled by a stochastic process consisting of a series of independent superimposed events and are termed "red noise" or "flicker noise" PSDs, because low frequencies contribute the most variability power, whereas high frequency variability is increasingly suppressed \citep{1978ComAp...7..103P}. The characteristic frequency $\nu_{br}$ was found to scale inversely with the black hole mass and linearly with the accretion rate \citep{2006Natur.444..730M}. However, the actual dependency on the accretion rate is less clear and was not recovered by \citet{2012A&A...544A..80G}. 

Due to the fact that optical light curves are not continuous and generally suffer from irregular sampling, standard Fourier techniques used in the X-rays can not be applied and therefore the shape of the optical PSD of AGNs is not well known to date. Yet there is evidence that the optical PSD resembles a broken power law as well. For example the high frequency part of the optical PSD has been found to be described reasonably well by a power law of the form $\mathrm{PSD\left(\nu\right)}\propto\nu^{-2}$ \citep{1999MNRAS.306..637G,2001ApJ...555..775C,2003MNRAS.342.1222C,2009ApJ...698..895K,2013ApJ...779..187K,2010ApJ...708..927K,	2010ApJ...721.1014M,2013A&A...554A.137A,2013ApJ...765..106Z}. However, recent PSD analysis performed using high quality Kepler light curves suggest that the high frequency optical PSD may be characterized by steeper slopes between $-2.5$ and $-4$ \citep{2011ApJ...743L..12M,2014ApJ...795....2E,2015MNRAS.451.4328K}. Likewise there is still confusion about the value of the low frequency slope of the optical PSD. Using a sample of $\sim$9000 spectroscopically confirmed quasars in SDSS Stripe 82, \citet{2010ApJ...721.1014M} were unable to differentiate between $\gamma=-1$ and $\gamma=0$ ("white noise") for the low frequency slope. Considering the optical break timescale, typical values between 10--100 days but even up to $\sim$10 years have been reported \citep{2001ApJ...555..775C,2009ApJ...698..895K}. The spread in the characteristic variability timescale is thought to be connected with the fundamental AGN parameters driving the variability. Indeed the optical break timescale was observed to scale positively with black hole mass and luminosity \citep{2001ApJ...555..775C,2009ApJ...698..895K,2010ApJ...721.1014M}.   

Alternatively to performing a PSD analysis, which in general requires well sampled and uninterrupted light curves, it is customary to utilize simpler variability estimators that allow to infer certain properties of the PSD for large samples of objects and sparsely sampled light curves. Convenient variability tools are structure functions \citep[e.g.][]{2010ApJ...714.1194S,2010ApJ...721.1014M,2014ApJ...784...92M} or the excess variance \citep[e.g.][]{1997ApJ...476...70N,2012A&A...542A..83P,2014ApJ...781..105L}. On timescales shorter than the break timescale the X-ray excess variance was found to be anti-correlated with the black hole mass and the X-ray luminosity, whereas there is currently no consensus regarding the correlation with the Eddington ratio \citep{1997ApJ...476...70N,1999ApJ...524..667T,1999ApJS..125..297L,2000ApJ...531...52G,2004MNRAS.348..207P,2005MNRAS.358.1405O,2006MNRAS.370.1534N,2009MNRAS.394..443M,2010ApJ...710...16Z,2011A&A...526A.132G,	2012A&A...537A..87C,2012A&A...542A..83P,2014ApJ...781..105L,2013MNRAS.430L..49M}. Considering the optical variability amplitude an anti-correlation with luminosity and rest-frame wavelength is well established on timescales of $\sim$years \citep{1994MNRAS.268..305H,1999MNRAS.306..637G,2004ApJ...601..692V,2008MNRAS.383.1232W,2009ApJ...696.1241B,2009ApJ...698..895K,	2010ApJ...721.1014M,2012ApJ...758..104Z}.    
Conflicting results have been obtained regarding a dependence of the optical variability amplitude on the black hole mass as some authors found positive correlations, others negative correlations or almost no correlation although probing similar variability timescales \citep{2007MNRAS.375..989W,2008MNRAS.383.1232W,2009ApJ...698..895K,2010ApJ...721.1014M,2012ApJ...758..104Z}.
Finally, an anti-correlation between optical variability and the Eddington ratio has been reported by several authors on timescales of several months \citep{2013ApJ...779..187K} and several years \citep{2008MNRAS.383.1232W,2009ApJ...696.1241B,2010ApJ...716L..31A,2010ApJ...721.1014M,2012ApJ...758..104Z,2013A&A...560A.104M}. However, the observed trends with the AGN parameters show large scatter, with the derived slopes often suggesting a very weak dependence. 

In this work we aim to investigate the correlations between the optical variability amplitude, quantified by the normalized excess variance, and the fundamental AGN physical properties by using a well studied sample of X-ray selected AGNs from the XMM-COSMOS survey with optical light curves in five bands available from the Pan-STARRS1 Medium Deep Field 04 survey. In addition, we perform a PSD analysis of our optical light curves using the CARMA approach introduced by \citet{2014ApJ...788...33K} in order to derive the optical PSD shape for a large sample of objects, including the characteristic break frequency, the PSD normalization and the PSD slopes at high and low frequencies. The paper is organized as follows: in section \ref{sec:catalog} we describe our sample of variable AGNs; the methods used to quantify the variability amplitude and to model the PSD are introduced in section \ref{sec:varmethod}; the correlations between the variability amplitude and the AGN parameters are presented in section \ref{sec:varcorrelations}; the results of the power spectrum analysis are depicted in section \ref{sec:psdanalysis}; we discuss our findings in section \ref{sec:discussion} and section \ref{sec:conclu} summarizes the most important results. Additional information about the sample and the PSD fit results in different wavelength bands are provided in appendix \ref{sec:appendixa} and \ref{sec:appendixb}, respectively.  
 
\section{The sample of variable AGNs}
\label{sec:catalog}

Throughout this work we use the same sample of variable AGNs as defined in \citet{2015arXiv151001739S}, hereafter termed S15. This sample is drawn from the catalog of \citet{2010ApJ...716..348B} presenting the multi-wavelength counterparts to the XMM-COSMOS sources \citep{2007ApJS..172...29H,2009A&A...497..635C}. We have selected the X-ray sources that have a pointlike and isolated counterpart in HST/ACS images and that are detected in single Pan-STARRS1 (PS1) exposures. In addition we focus on the bands for which the observational data are of high quality and available for most of our objects. Thus, the sample comprises 184 ($g_{\mathrm{P1}}$), 181 ($r_{\mathrm{P1}}$), 162 ($i_{\mathrm{P1}}$), 131 ($z_{\mathrm{P1}}$) variable sources detected in the PS1 Medium Deep Field 04 (MDF04) survey. In the following we refer to this sample as the "total sample". We note that this sample contains no upper limit detections of variability and more than 97\% of all sources having MDF04 light curves in a given PS1 band are identified as variable in this band (see table 2 of S15 for detailed numbers in each PS1 band). More than 96\% of our objects are classified as type-1 AGNs\footnote{There are 7 variable type-2 AGNs in our sample which have been classified either spectroscopically (6 objects) or on the basis of the best SED fitting template (1 object).} and 92\% have a specified spectroscopic redshift \citep{2007ApJS..172..383T,2009ApJS..184..218L}. The remaining sources only have photometric redshifts determined in \citet{2011ApJ...742...61S}. However, for the 92\% with known spectroscopic redshifts the accuracy of the photometric redshifts is $\sigma_{\mathrm{NMAD}}=0.009$ with a fraction of outliers of 5.9\%. Therefore we do not differentiate between sources with spectroscopic and photometric redshifts in the following.  

During the whole analysis we only consider the objects classified as type-1 AGN when investigating correlations between the physical AGN parameters and variability. Among the type-1 objects of the total sample 95 ($g_{\mathrm{P1}}$), 97 ($r_{\mathrm{P1}}$), 90 ($i_{\mathrm{P1}}$), 75 ($z_{\mathrm{P1}}$) have known spectroscopic redshifts, SED-fitted bolometric luminosities $L_{\mathrm{bol}}$ \citep{2012MNRAS.425..623L} and black hole masses $M_{\mathrm{BH}}$ \citep{2013A&A...560A..72R}. The black hole masses have all been derived with the same method from the line width of broad emission lines ($\mathrm{H_{\beta}}$ and MgII$\,\lambda\mathrm{2798\AA}$), using virial relationships which were calibrated with reverberation mapping results of local AGNs. For the same sources we therefore also possess the Eddington ratio defined by $\lambda_{\mathrm{Edd}}=L_{\mathrm{bol}}/L_{\mathrm{Edd}}$, where $L_{\mathrm{Edd}}$ is the Eddington luminosity. This sample, hereafter termed "MBH sample", covers a redshift range from 0.3 to 2.5. We stress that this is a large sample of objects with homogeneously measured AGN parameters, spanning a wide redshift range, for which we can study the connection of rest-frame UV/optical variability with fundamental physical properties of AGNs in four wavelength bands. As detailed in appendix \ref{sec:appendixa} our sample does not suffer from strong selection effects, which could significantly bias any detected correlation between variability and the AGN parameters. However, since our sample is drawn from a flux-limited X-ray parent sample there is a tendency for higher redshift sources being more luminous. Yet we found that this effect has negligible impact on the resulting correlations between variability and luminosity. 

\section{Methodology: variability amplitude and power spectrum model}
\label{sec:varmethod}

\subsection{The normalized excess variance}
\label{sec:excessvariance}

In order to quantify the variability amplitude we measure the normalized excess variance \citep{1997ApJ...476...70N} given by 
\begin{flalign}
	\label{eq:nev}	\sigma_{\mathrm{rms}}^{2}=\left(s^{2}-\overline{\sigma_{\mathrm{err}}^{2}}\right)/\left(\bar{f}\right)^{2}=\frac{1}{\left(\bar{f}\right)^{2}}\left(\sum_{i=1}^{N}\frac{\left(f_{i}-\bar{f}\right)^{2}}{\left(N-1\right)}-\sum_{i=1}^{N}\frac{\sigma^{2}_{\mathrm{err},i}}{N}\right)
\end{flalign}    
from the light curve consisting of N measured fluxes $f_{i}$ with individual errors $\sigma_{\mathrm{err},i}$ and arithmetic mean $\bar{f}$. The normalized excess variance, or just excess variance, quotes the residual variance after subtracting the average statistical error $\overline{\sigma_{\mathrm{err}}^{2}}$ from the sample variance $s^{2}$ of the light curve flux. The $\sigma_{\mathrm{rms}}^{2}$ values calculated from the total light curves of our AGNs were already used in S15 and are available in the associated online journal (see appendix C of S15 for details). The error on the excess variance caused by Poisson noise alone \citep{2003MNRAS.345.1271V} is well described by
\begin{flalign}
	\label{eq:errnev}
	err\left(\sigma_{\mathrm{rms}}^{2}\right)=\sqrt{\left(\sqrt{\frac{2}{N}}\cdot\frac{\overline{\sigma_{\mathrm{err}}^{2}}}{\left(\bar{f}\right)^{2}}\right)^{2}+\left(\sqrt{\frac{\overline{\sigma_{\mathrm{err}}^{2}}}{N}}\cdot\frac{2 F_{\mathrm{var}}}{\left(\bar{f}\right)}\right)^{2}},
\end{flalign}  
where $F_{\mathrm{var}}=\sqrt{\sigma_{\mathrm{rms}}^{2}}$ is the fractional variability \citep{1990ApJ...359...86E}. As demonstrated by \citet{2013ApJ...771....9A} there are additional error sources associated with the stochastic nature of AGN variability, red noise leakage, the sampling pattern and the signal to noise of the light curves. In particular these biases depend on the shape of the PSD (see e.g. table 2 in \citet{2013ApJ...771....9A}) and therefore an excess variance measurement can systematically over/underestimate the intrinsic variance of a light curve by a factor of a few (we refer the reader to the discussion in section \ref{sec:rmsvscarma}).

Following the procedure in S15 we consider a source as variable in a given band, if 
\begin{flalign}
	\label{eq:detectnev}
	\sigma_{\mathrm{rms}}^{2}-err\left(\sigma_{\mathrm{rms}}^{2}\right)>0.
\end{flalign} 
We emphasize that this is only a 1$\sigma$ detection of variability. However, in this work we aim to investigate the relation of the amplitude of variability with AGN physical properties down to the lowest achievable level of variability. Using a more stringent variability threshold would dramatically limit the parameter space of $M_{\mathrm{BH}}$, $L_{\mathrm{bol}}$ and $\lambda_{\mathrm{Edd}}$ values we can probe. Finally, the quality of the $\sigma_{\mathrm{rms}}^{2}$ measurements of our sample is generally high, as presented in appendix A of S15.
  
The intrinsic variance of a light curve is defined to measure the integral of the PSD over the frequency range probed by the time series. Since the excess variance is an estimator of the fractional intrinsic variance, it is related to the PSD via     
\begin{flalign}
	\label{eq:psdint}	\sigma_{\mathrm{rms}}^{2}\approx\int_{\nu_{\mathrm{min}}}^{\nu_{\mathrm{max}}}\mathrm{PSD\left(\nu\right)d\nu},
\end{flalign} 
with $\nu_{\mathrm{min}}=1/T$ and the Nyquist frequency $\nu_{\mathrm{max}}=1/\left(2\Delta t\right)$ for a light curve of length $T$ and bin size $\Delta t$, with the PSD normalized to the squared mean of the flux \citep{2003MNRAS.345.1271V,2011A&A...526A.132G,2013ApJ...771....9A}.

\subsection{Measuring $\sigma_{\mathrm{rms}}^{2}$ on different timescales}
\label{sec:rmstimes}

Albeit the excess variance is a variability estimator that is measured from the light curve fluxes and the individual observing times do not appear explicitly in the calculation, the total temporal length and the sampling frequency of the light curve affect the resulting $\sigma_{\mathrm{rms}}^{2}$ value. As described in the previous section the excess variance estimates the integral of the variability power spectrum over the minimal and maximal temporal frequency covered by the light curve. Therefore we can probe different variability timescales by measuring the excess variance from different parts of the light curves. The total sample only contains $\sigma_{\mathrm{rms}}^{2}$ values computed from the nightly-averaged total light curves which typically consist of $\sim$70--80 points and cover a period of $\sim$4 years. The light curves split in several segments with observations performed every $\sim$1--3 days over a period of $\sim$3--4 months, interrupted by gaps of $\sim$7-9 months without observations. Correspondingly the shortest sampled timescale is on the order of few days for the MDF04 survey, depending on weather constraints during the survey, whereas the longest timescale is $\sim4$ years. However, the sampling pattern of the MDF04 light curves additionally allows to measure the excess variance from the well sampled individual segments of the light curves, consisting of typically 10--20 points which span a time interval of $\sim$3--4 months. For each AGN we additionally calculate an excess variance value measured on timescales of months by averaging the $\sigma_{\mathrm{rms}}^{2}$ values of the light curve segments and propagating the $err(\sigma_{\mathrm{rms}}^{2})$ values of each considered segment. In order to be unaffected by sparsely sampled segments, lowering the quality of the variability estimation, we include only the segments with more than 10 observations in the averaging. The sample of variable type-1 AGNs with known physical parameters for this shorter timescale, i.e. the MBH sample on timescales of months fulfilling $\sigma_{\mathrm{rms}}^{2}-err(\sigma_{\mathrm{rms}}^{2})>0$, comprises 76 ($g_{\mathrm{P1}}$), 63 ($r_{\mathrm{P1}}$), 41 ($i_{\mathrm{P1}}$), 43 ($z_{\mathrm{P1}}$) sources, respectively. The considerably smaller sample size follows from the fact that the light curve segments of many AGNs either have less than 10 measurements or are almost flat, leading to very small and even negative $\sigma_{\mathrm{rms}}^{2}$ values. We observe that the variability amplitude on year timescales is on average about an order of magnitude larger than on month timescales. 

\begin{figure}
	\centering
	\includegraphics[width=.4\textwidth]{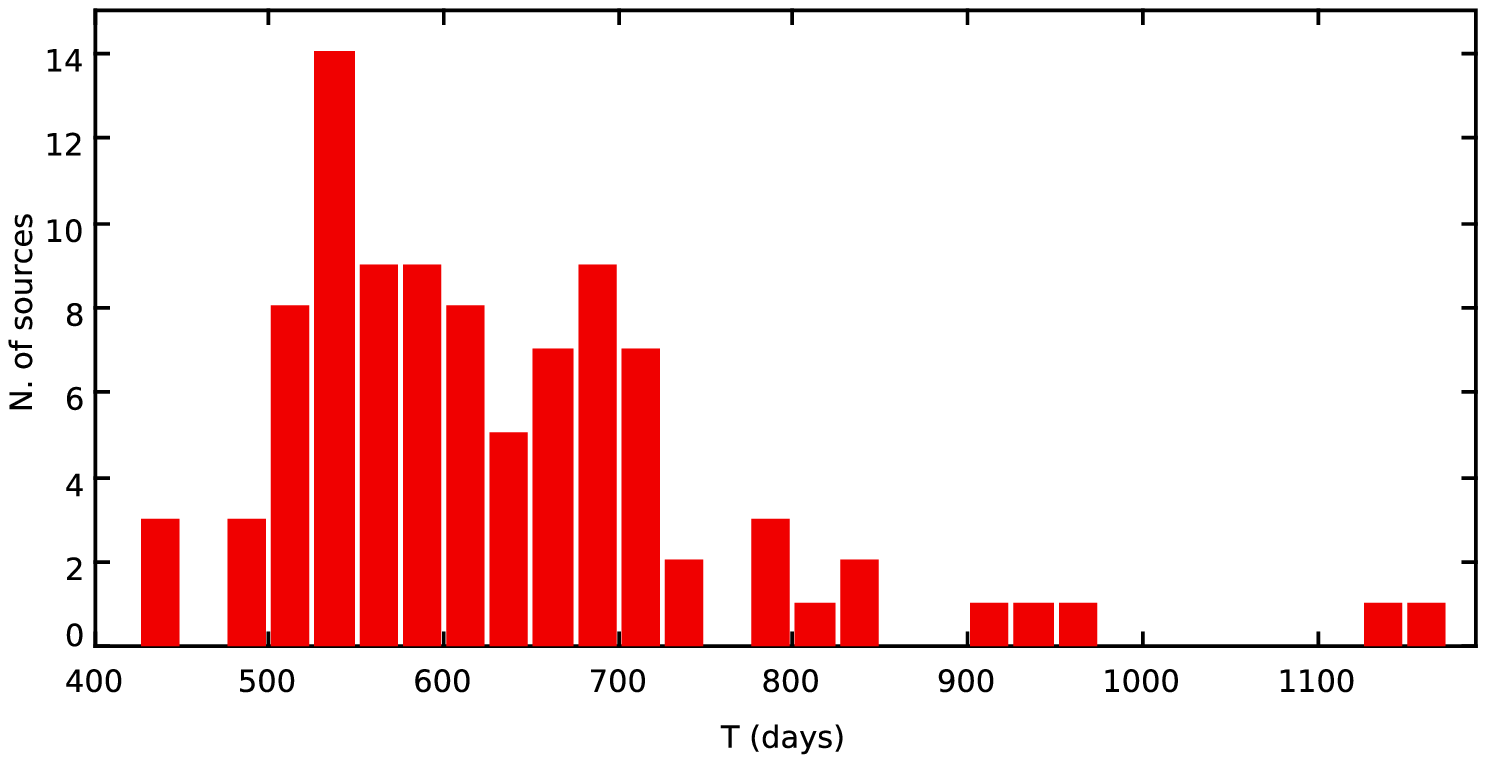}
	\includegraphics[width=.4\textwidth]{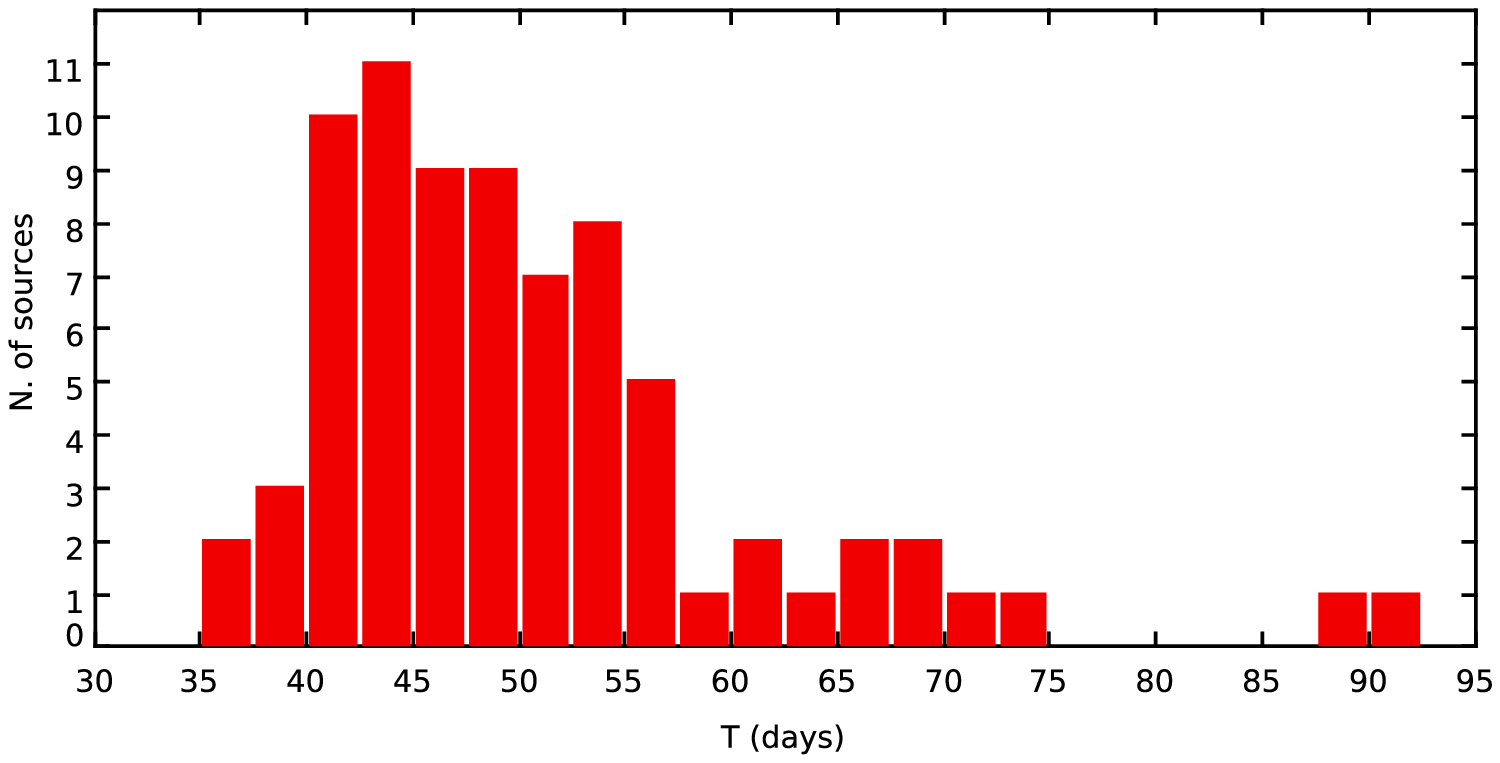}
	\caption{\textit{Top panel}: histogram of the rest-frame observation length of the total light curve for the year timescale MBH sample ($g_{\mathrm{P1}}$ band). \textit{Bottom panel}: histogram of the rest-frame observation length (average value of the light curve segments) for the month timescale MBH sample ($g_{\mathrm{P1}}$ band).}
	\label{fig:histobstime}
\end{figure}

\begin{figure*}
	\centering
	\includegraphics[width=.6\textwidth]{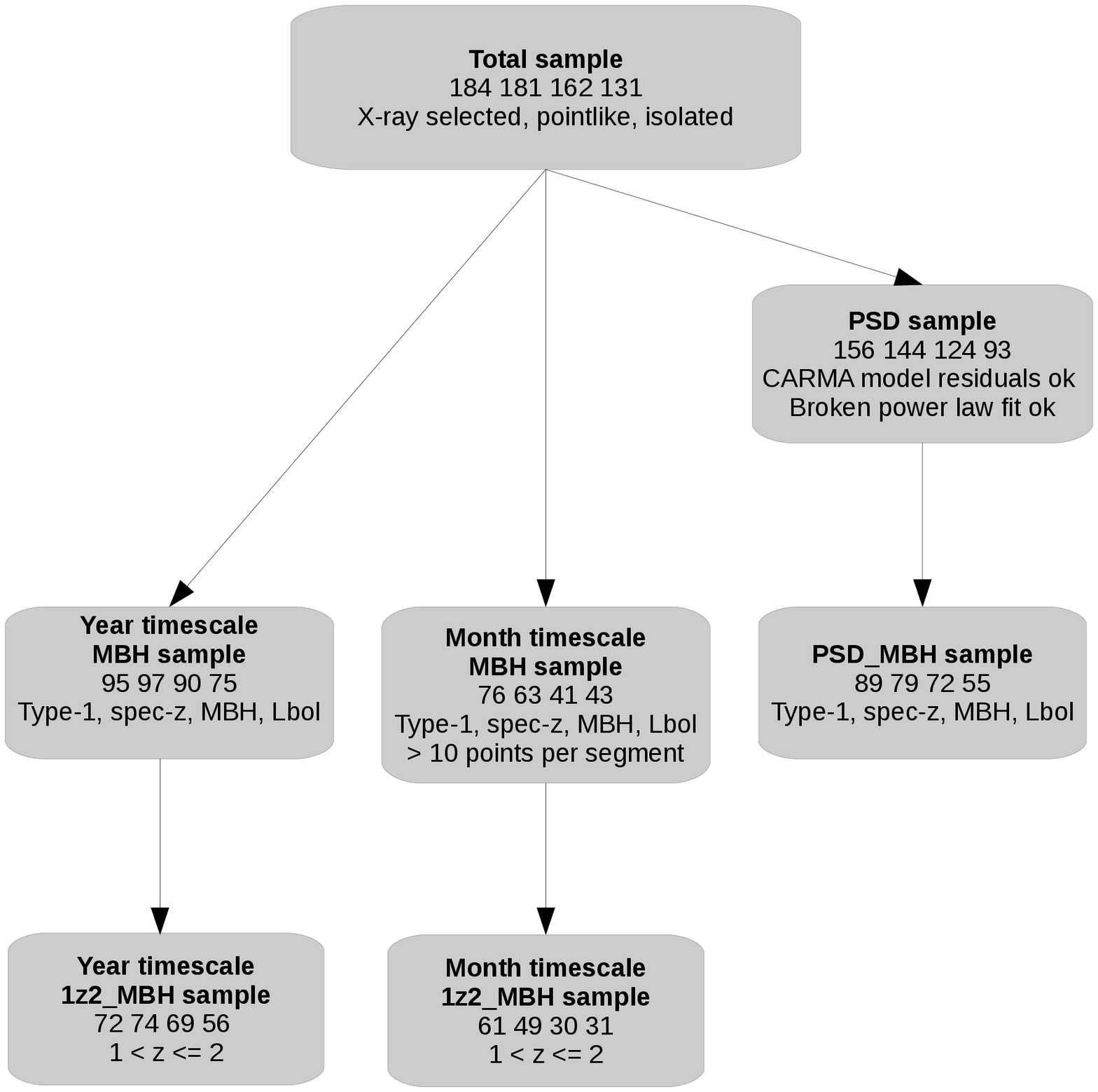}
	\caption{Flow-chart illustrating the selection of all samples considered in this work. Below the sample name (bold face) we list the sample size for each PS1 band in the order $g_{\mathrm{P1}}$, $r_{\mathrm{P1}}$, $i_{\mathrm{P1}}$, $z_{\mathrm{P1}}$. In addition we state the defining properties of each sample, such as objects with known AGN type, spectroscopic redshift (spec-z), black hole mass (MBH), bolometric luminosity (Lbol) or objects within a certain redshift range (see text for details). The two rightmost samples are introduced in section \ref{sec:psdshape}.}
	\label{fig:flowchart}
\end{figure*}

Although the observer-frame timescales covered by the light curves of our sample are very similar for each AGN, the wide redshift range encompassed by our sources leads to a variety of different rest-frame timescales. This is illustrated in Fig. \ref{fig:histobstime}, showing the distribution of the rest-frame observation length $T$ of the total light curve and the average value of the light curve segments, obtained by dividing the observer-frame value by $1+z$ to account for cosmological time dilation. The data of the year and month timescale MBH sample ($g_{\mathrm{P1}}$ band) are displayed. From there we note that the rest-frame length of the total light curve comprises $\sim$1--3 years for our sources, whereas the rest-frame length of the light curve segments corresponds to timescales of $\sim$1--3 months. To reduce possible biases introduced by the spread in redshift, we additionally consider the sources of the MBH sample with redshifts between $1<z\leq 2$ in our investigations, referred to as the "1z2\_MBH sample". On variability timescales of years the 1z2\_MBH sample contains 72 ($g_{\mathrm{P1}}$), 74 ($r_{\mathrm{P1}}$), 69 ($i_{\mathrm{P1}}$), 56 ($z_{\mathrm{P1}}$) AGNs. The corresponding 1z2\_MBH sample on month timescales comprises 61 ($g_{\mathrm{P1}}$), 49 ($r_{\mathrm{P1}}$), 30 ($i_{\mathrm{P1}}$), 31 ($z_{\mathrm{P1}}$) objects. In the following excess variance analysis (section \ref{sec:varcorrelations}) we compare the variability properties of our sources as measured on timescales of years and months whenever applicable. For reference we display the properties of the various samples used throughout this work and how they are selected from the parent sample in Fig. \ref{fig:flowchart}. 
 
\subsection{CARMA modelling of the power spectral density}

Considering equation \ref{eq:psdint}, modelling the PSD of a light curve provides more fundamental variability information than the integrated $\sigma_{\mathrm{rms}}^{2}$ quantity. In fact, the shape of the PSD potentially allows to gain insight into the underlying physical processes connected to variability \citep{1997MNRAS.292..679L,2007ApJ...660..556T}. In order to estimate the PSDs of our light curves we apply the continuous-time autoregressive moving average (CARMA) model presented in \citet{2014ApJ...788...33K}. This stochastic variability model fully accounts for irregular sampling and Gaussian measurement errors. In addition it allows for interpolation and forecasting of light curves by modelling the latter as a continuous-time process. 

A zero-mean CARMA(p,q) process for a time series $y\left(t\right)$ is defined as the solution of the stochastic differential equation
\begin{flalign}
	\label{eq:carmadgl}	
\begin{split}
	\frac{\mathrm{d}^{p}y\left(t\right)}{\mathrm{d}t^{p}}+\alpha_{p-1}\frac{\mathrm{d}^{p-1}y\left(t\right)}{\mathrm{d}t^{p-1}}+...+\alpha_{0}y\left(t\right)=\\
\beta_{q}\frac{\mathrm{d}^{q}\epsilon\left(t\right)}{\mathrm{d}t^{q}}+\beta_{q-1}\frac{\mathrm{d}^{q-1}\epsilon\left(t\right)}{\mathrm{d}t^{q-1}}+...+\epsilon\left(t\right).
\end{split}
\end{flalign}    
It is assumed that the variability is driven by a Gaussian continuous-time white noise process $\epsilon\left(t\right)$ with zero mean and variance $\sigma^{2}$. Apart from $\sigma^{2}$, the free parameters of the model are the autoregressive coefficients $\alpha_{0}$,..., $\alpha_{p-1}$ and the moving average coefficients $\beta_{1}$,..., $\beta_{q}$. In practice the mean of the time series $\mu$ is also a free parameter and the calculation of the likelihood function of the time series sampled from a CARMA process is done on the centered values $\tilde y_{i}=y_{i}-\mu$ for each light curve point $i$. 

The PSD of a CARMA(p,q) process is given by 
\begin{flalign}
	\label{eq:carmapsd}	
\mathrm{PSD\left(\nu\right)}=\sigma^{2}\frac{|\sum_{j=0}^{q}\beta_{j}\left(2\pi i \nu\right)^{j}|^{2}}{|\sum_{k=0}^{p}\alpha_{k}\left(2\pi i \nu\right)^{k}|^{2}}
\end{flalign}   
which forms a Fourier transform pair with the autocovariance function at time lag $\tau$
\begin{flalign}
	\label{eq:carmaacf}	
R\left(\tau\right)=\sigma^{2}\sum_{k=1}^{p}\frac{\left[\sum_{l=0}^{q}\beta_{l}r_{k}^{l}\right]\left[\sum_{l=0}^{q}\beta_{l}\left(-r_{k}\right)^{l}\right]\exp\left(r_{k}\tau\right)}{-2\mathrm{Re}\left(r_{k}\right)\prod_{l=1,l\neq k}^{p}\left(r_{l}-r_{k}\right)\left(r_{l}^{*}+r_{k}\right)}
\end{flalign} 
where $r_{k}^{*}$ is the complex conjugate and Re($r_{k}$) the real part of $r_{k}$, respectively. The values $r_{1}$,..., $r_{p}$ denote the roots of the autoregressive polynomial
\begin{flalign}
	\label{eq:autopoly}	
	 A\left(z\right)=\sum_{k=0}^{p}\alpha_{k}z^{k}.
\end{flalign} 
The CARMA process is stationary if $q<p$ and $\mathrm{Re}\left(r_{k}\right)<0$ for all $k$. The autocovariance function of a CARMA process represents a weighted sum of exponential decays and exponentially-damped sinusoidal functions. Since the autocovariance function is coupled to the PSD via a Fourier transform, the latter can be expressed as a weighted sum of Lorentzian functions, which are known to provide a good description of the PSDs of X-ray binaries and AGNs \citep{2000MNRAS.318..361N,2002ApJ...572..392B,2010LNP...794.....B,2013MNRAS.436.3782D,2015arXiv150900043D}. 

The CARMA model includes the Ornstein-Uhlenbeck process or the "damped random walk", which is depicted in detail in \citet{2009ApJ...698..895K} and was found to accurately describe quasar light curves in many subsequent works,   as the special case of $p=1$ and $q=0$. Considering equations \ref{eq:carmapsd} and \ref{eq:carmaacf} we note that CARMA models provide a flexible parametric form to estimate the PSDs and autocovariance functions of the stochastic light curves of AGNs. For further details on the computational methods, including the calculation of the likelihood function of a CARMA process and the Bayesian method to infer the probability distribution of the PSD given the measured light curve, we refer the reader to \citet{2014ApJ...788...33K} and the references therein.  

\section{Correlations of variability and AGN parameters}
\label{sec:varcorrelations}

\subsection{Wavelength dependence of the excess variance}
\label{sec:rmswave}

\begin{figure*}
\centering
\subfloat{%
	\includegraphics[width=.32\textwidth]{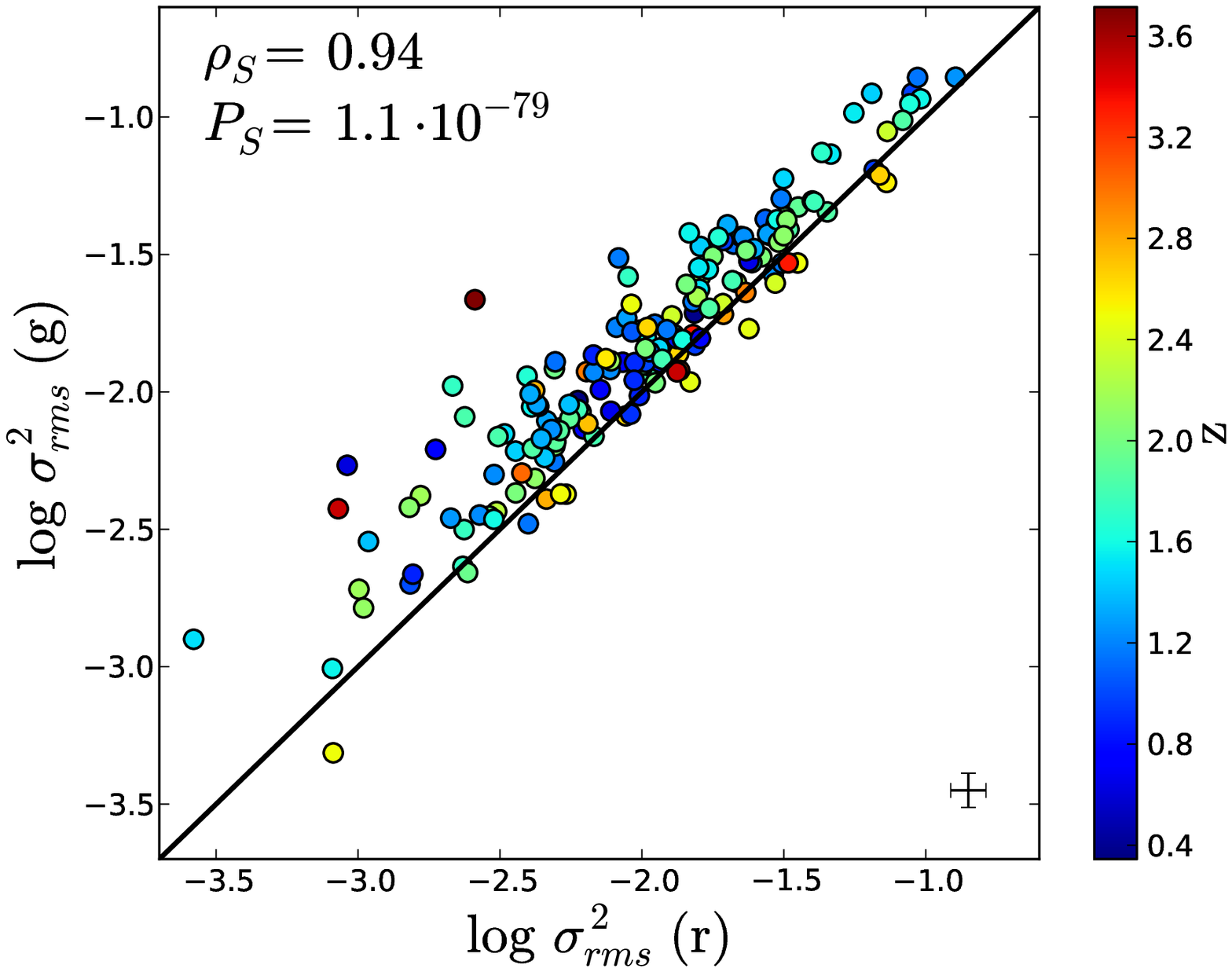}}
\quad
\subfloat{%
	\includegraphics[width=.32\textwidth]{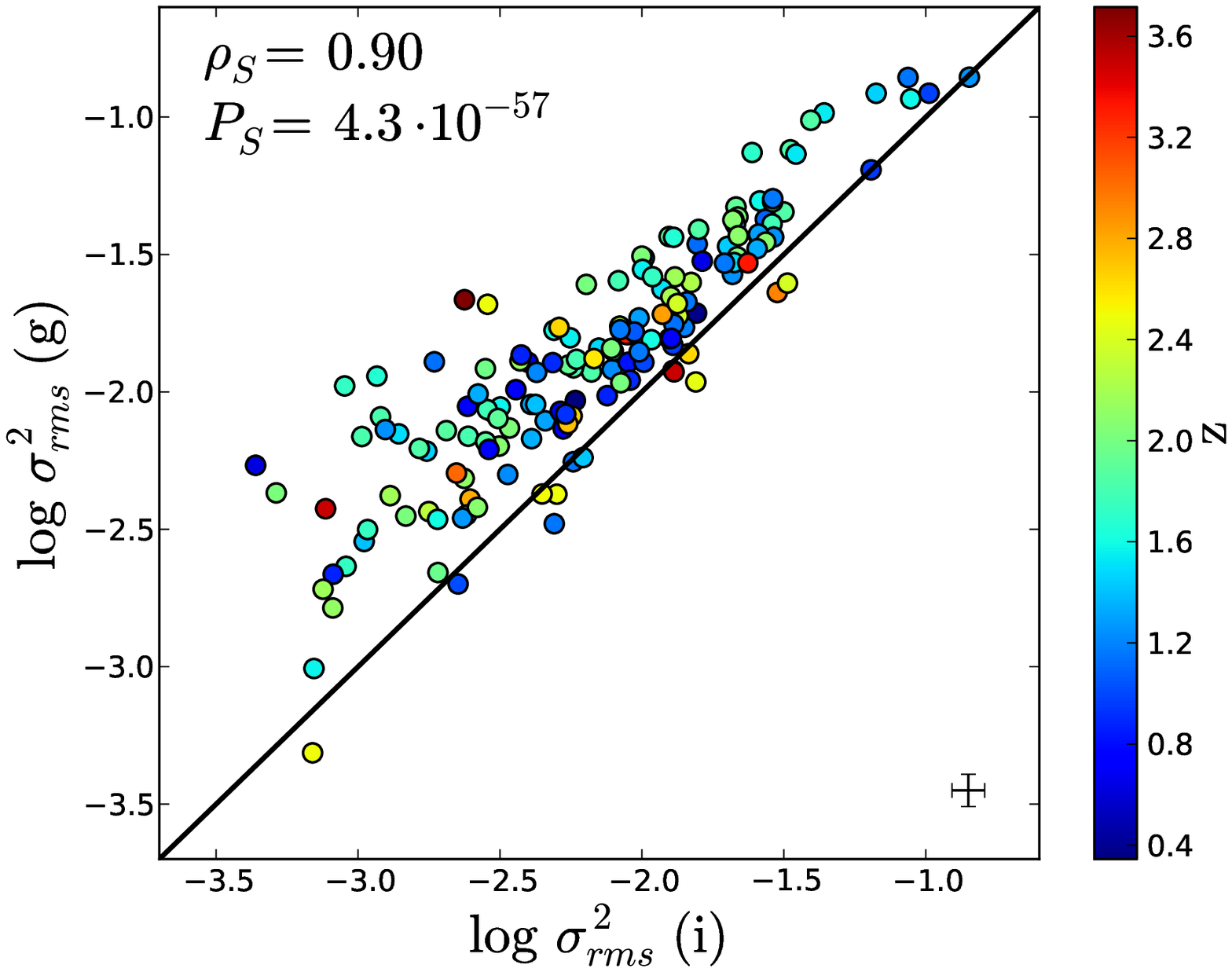}}
\quad
\subfloat{%
	\includegraphics[width=.32\textwidth]{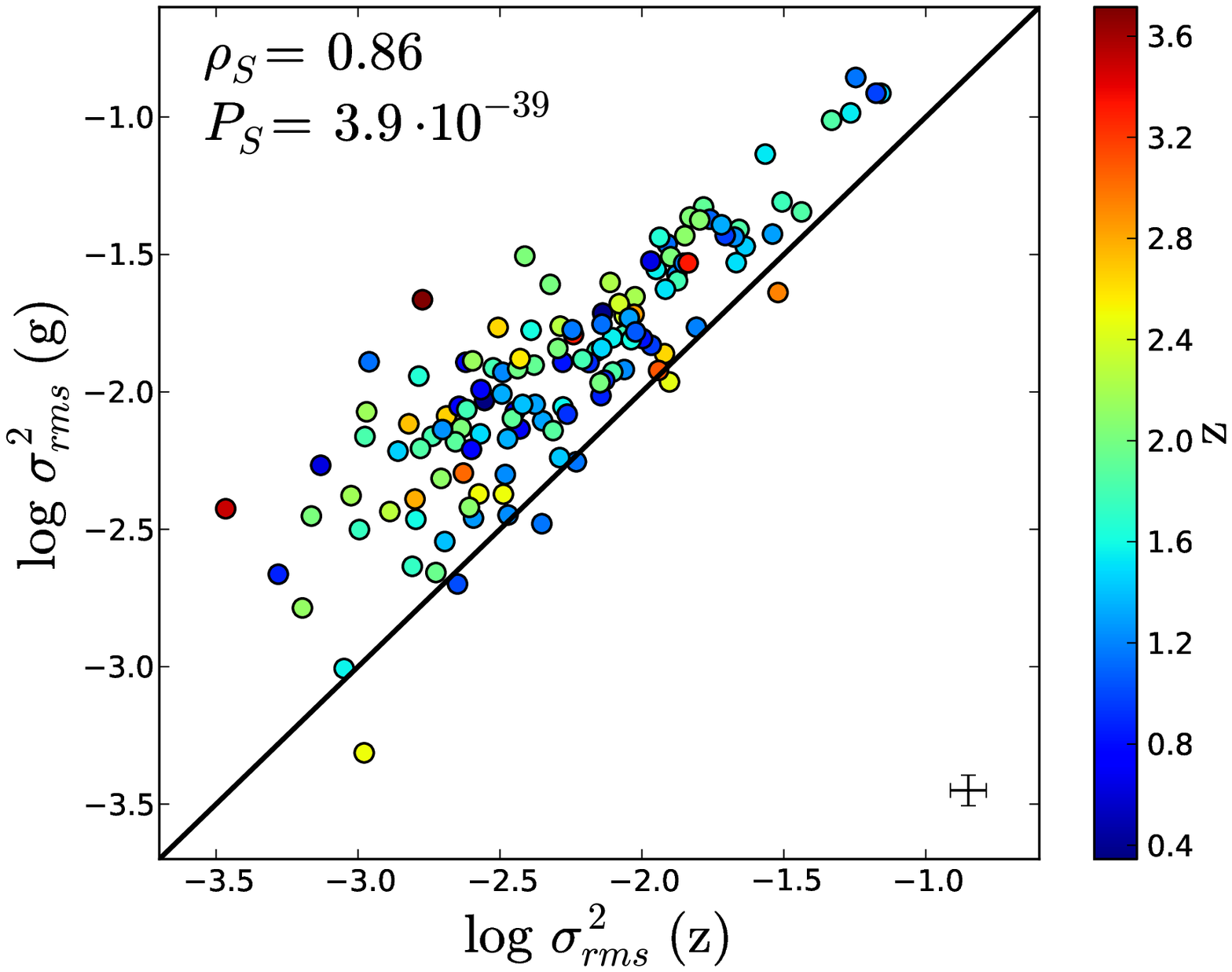}}

\subfloat{%
	\includegraphics[width=.32\textwidth]{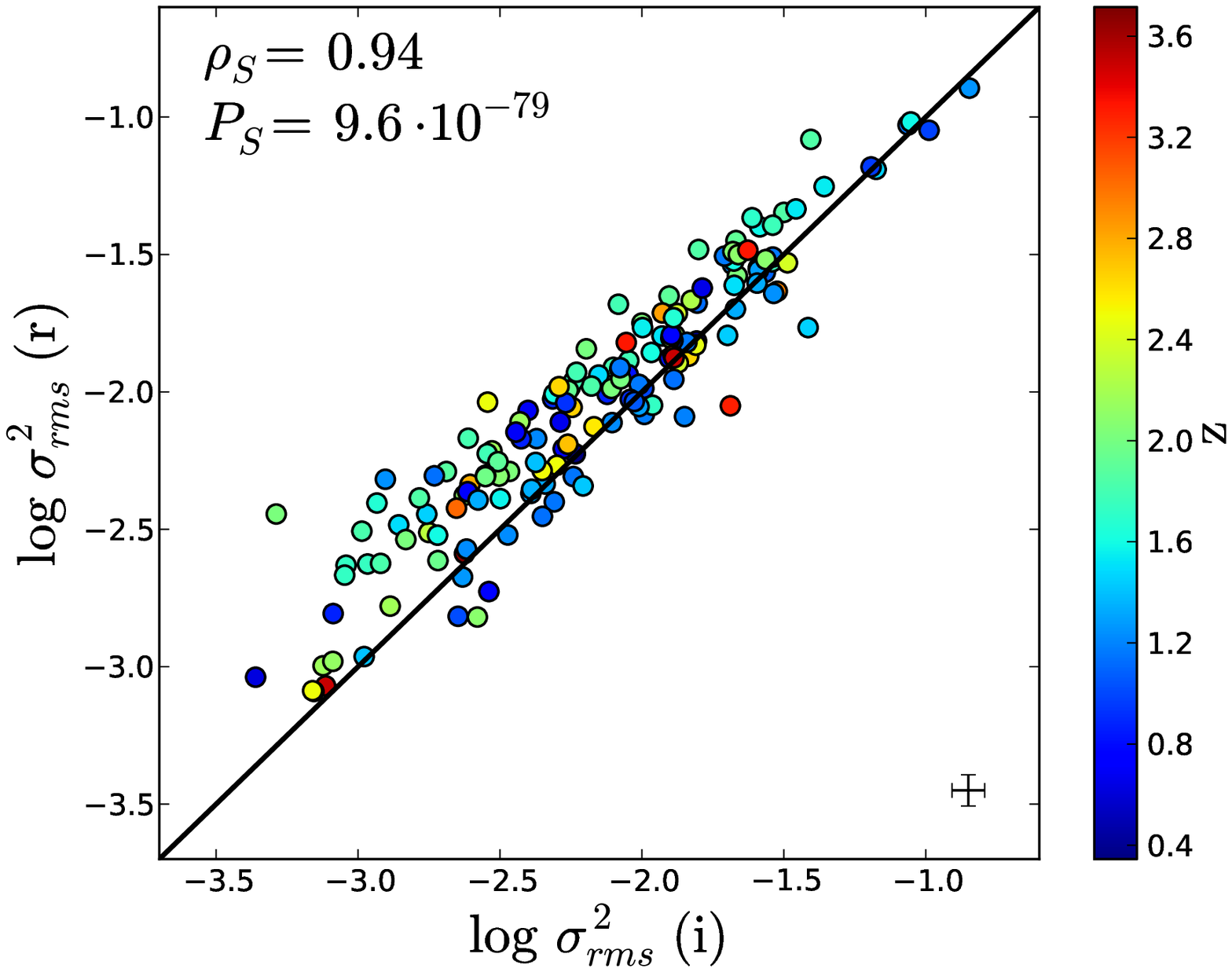}}
\quad	
\subfloat{%
	\includegraphics[width=.32\textwidth]{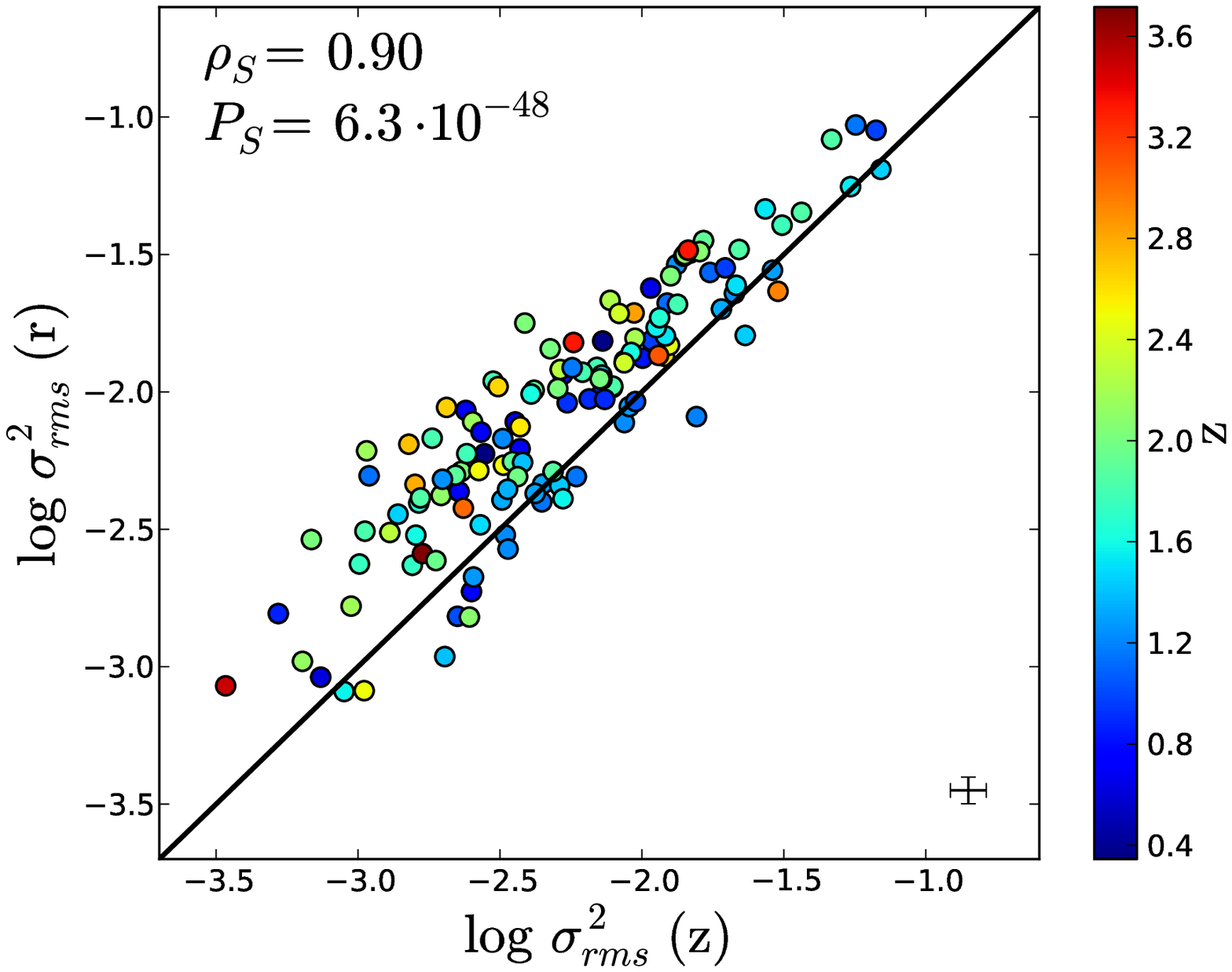}}
\quad
\subfloat{%
	\includegraphics[width=.32\textwidth]{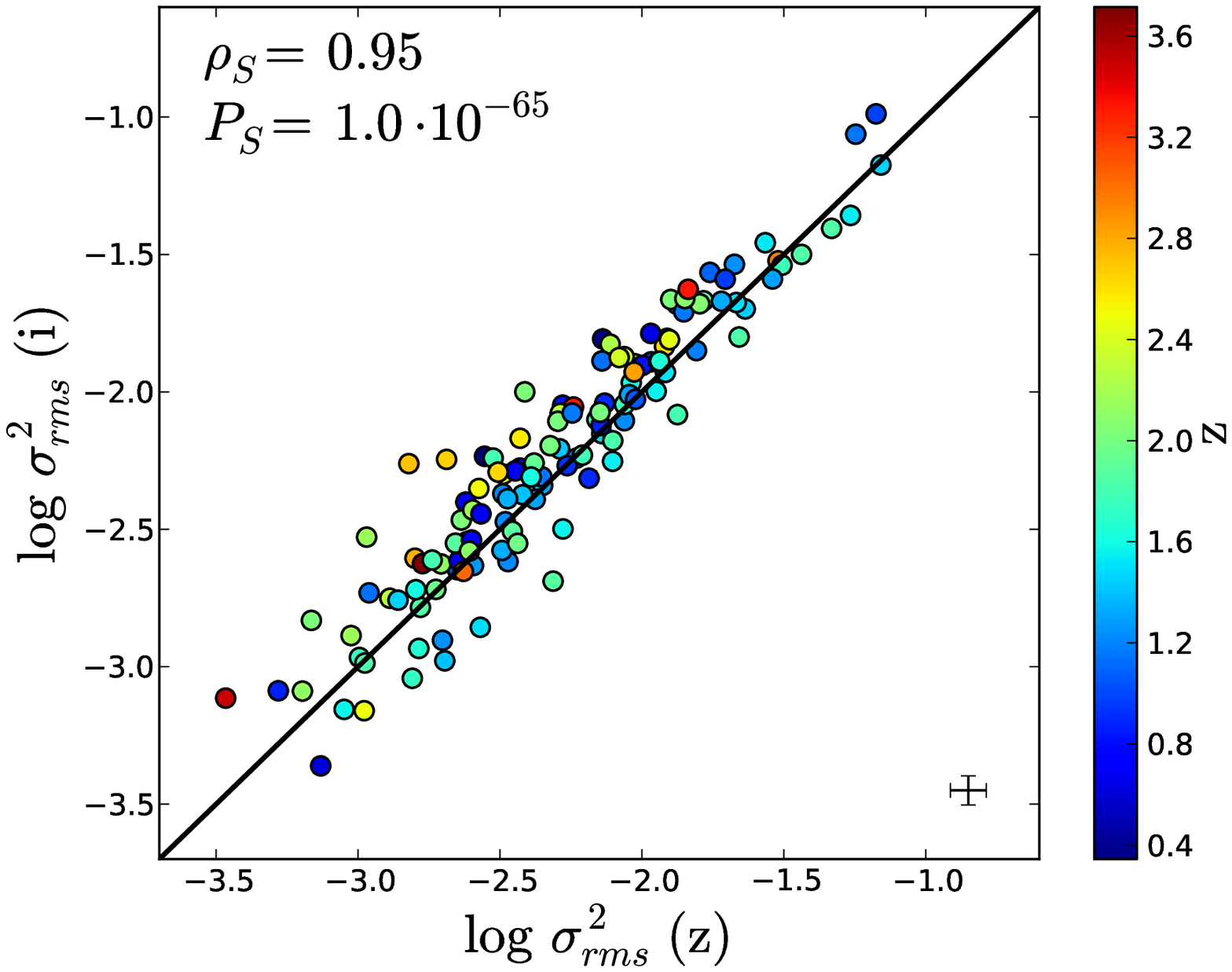}}
\caption{Comparing the excess variance measured on year timescales in the different PS1 bands. The data of all objects from the total sample with variability information in both considered bands are shown. The Spearman correlation coefficient and the respective p-value are reported in each sub-panel. Additionally the redshift is given as a color bar. The black line corresponds to the one to one relation. The black error bars are the average values.}
	\label{fig:cmpnevbands}
\end{figure*}

The multi-band PS1 observations of the MDF04 survey allow for an investigation of the chromatic nature of variability, i.e. the dependence on the radiation wavelength. Fig. \ref{fig:cmpnevbands} shows the excess variances of the total sample (variability timescale of years) for several filter pairs. In detail the intersection of the objects with measured $\sigma_{\mathrm{rms}}^{2}$ values in each of the two considered PS1 bands are plotted. Each sub-panel displays the "bluer" band on the y-axis and the "redder" band on the x-axis, the redshift is given as a color bar. Obviously the $\sigma_{\mathrm{rms}}^{2}$ values are strongly correlated, which is also expressed by the Spearman rank order correlation coefficient $\rho_{\mathrm{S}}$ and the corresponding two-tailed p-value $P_{\mathrm{S}}$, giving the probability that a $\rho_{\mathrm{S}}$ value at least as large as the observed one could arise for an uncorrelated dataset. The $\rho_{\mathrm{S}}$ values quoted on each sub-panel of Fig. \ref{fig:cmpnevbands} are all very close to $+1$ and the respective $P_{\mathrm{S}}$ values are essentially zero. However we observe a systematic trend that the "bluer" bands exhibit larger variability amplitudes than the "redder" bands, as the respective $\sigma_{\mathrm{rms}}^{2}$ values are shifted upwards the one to one relation. The offset gets larger when a specified "blue" band is compared with the series of bands with longer wavelength, i.e. comparing the pairs $\left(g_{\mathrm{P1}},r_{\mathrm{P1}}\right)$, $\left(g_{\mathrm{P1}},i_{\mathrm{P1}}\right)$ and $\left(g_{\mathrm{P1}},z_{\mathrm{P1}}\right)$. Yet the variability amplitudes seem to approach more and more similar values towards the NIR regime. In fact the difference between the $\sigma_{\mathrm{rms}}^{2}$ measurements of the $i_{\mathrm{P1}}$ and $z_{\mathrm{P1}}$ bands is less pronounced than the respective values of the pairs $\left(g_{\mathrm{P1}},r_{\mathrm{P1}}\right)$ and $\left(r_{\mathrm{P1}},i_{\mathrm{P1}}\right)$. We do not recognize any evolution of the aforementioned wavelength dependence with redshift, since there are no regions which are predominantly occupied by high or low redshift sources in any sub-panel. The same trends are observed when using the excess variance values measured on timescales of months. We emphasize that these findings are in agreement with previous studies which observed local and high redshift AGNs to be more variable at shorter wavelength \citep{1990ApJ...359...86E,1991ApJS...75..645K,1994A&A...291...74P,1996ApJ...463..466D,1996MNRAS.282.1191C,2004ApJ...601..692V,2010ApJ...708..927K,2010ApJ...721.1014M,2012ApJ...758..104Z}.     

\subsection{Excess variance versus black hole mass}
\label{sec:rmsvsmbh}

Determining accurate black hole masses for a large number of AGNs across the Universe is observationally expensive. However, recent works probing the high frequency part of the PSD delivered black hole mass estimates with $\sim$0.2--0.4 dex precision based on scaling relations of black hole mass and X-ray variability \citep{2010ApJ...710...16Z,2012A&A...542A..83P,2011ApJ...730...52K,2013ApJ...779..187K}. It is therefore important to know, if  optical variability provides another independent tool to measure black hole masses of AGNs, since massive time-domain optical surveys such as PS1 and LSST would then allow to derive black hole mass estimates for a very large number of quasars, irrespective of the AGN type.

\begin{figure}
\centering
\subfloat{%
	\includegraphics[width=0.50\textwidth]{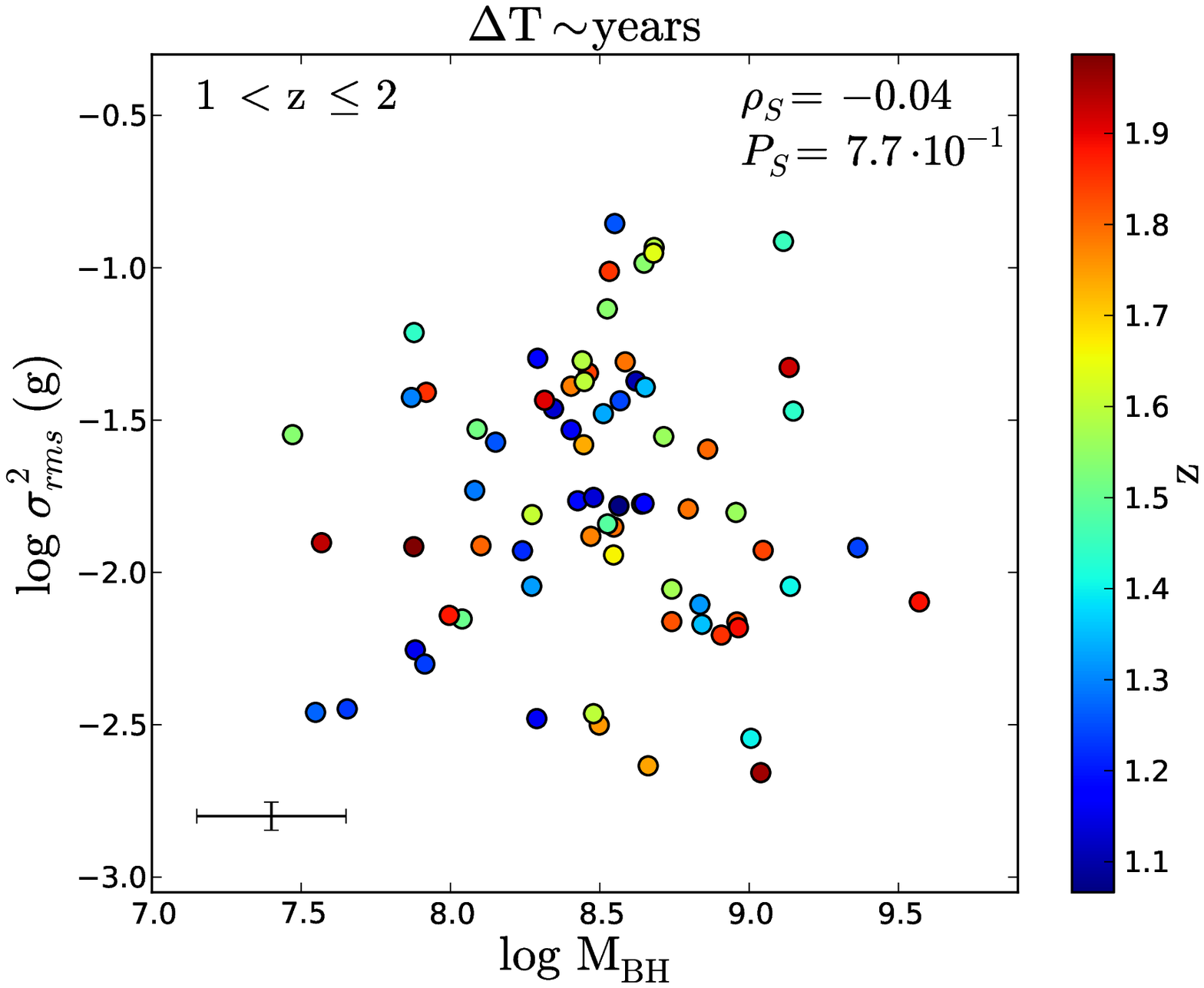}}
		
\subfloat{%
	\includegraphics[width=0.50\textwidth]{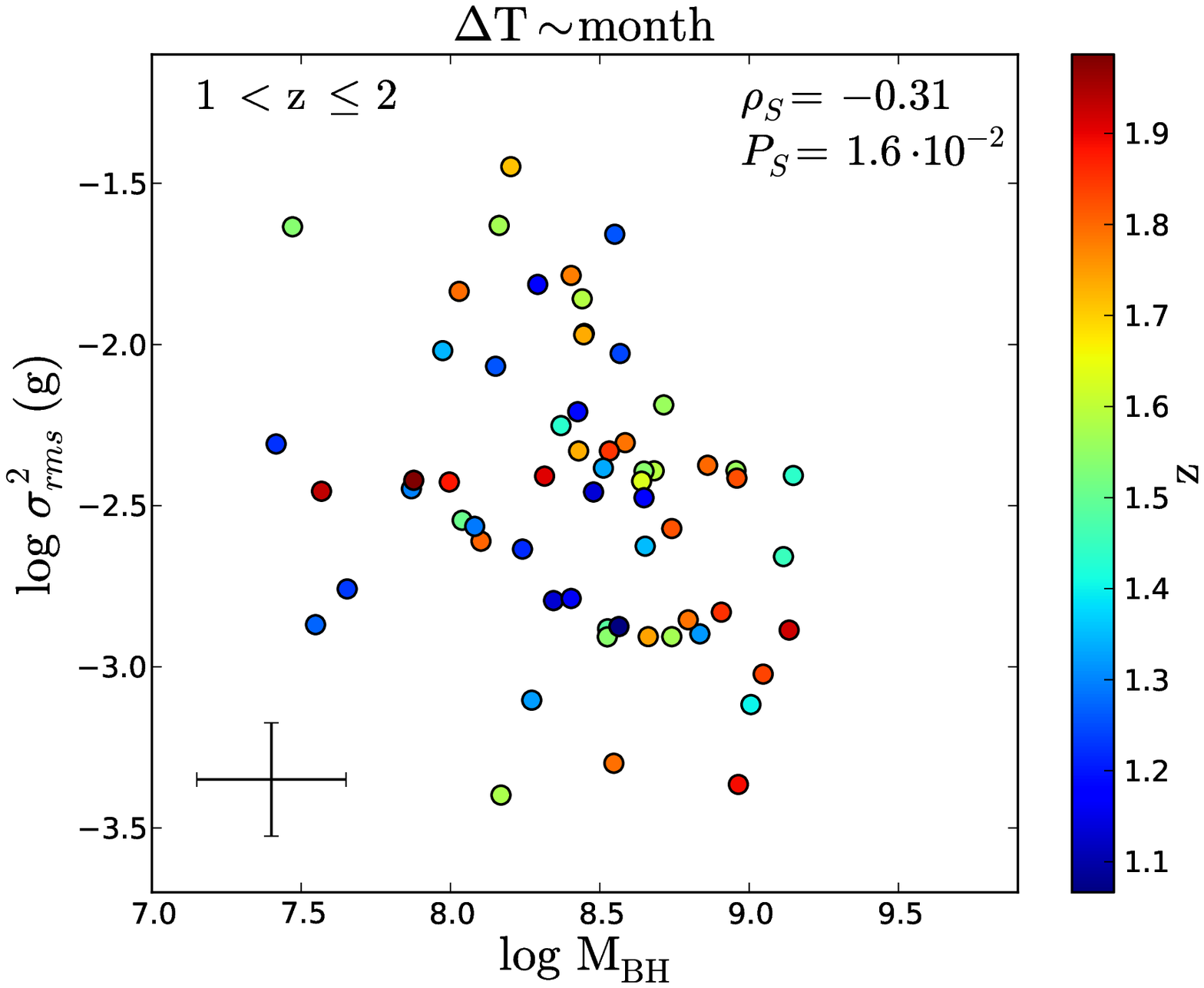}}
\caption{Excess variance ($g_{\mathrm{P1}}$ band) measured on year timescales (\textit{top}) and on month timescales (\textit{bottom}) versus $M_{\mathrm{BH}}$ in units of $M_{\mathrm{\odot}}$  for the 1z2\_MBH sample. Spearman's r and the respective p-value are reported in each sub-panel. Additionally the redshift is given as a color bar. The black error bars correspond to the average values.}
\label{fig:rmsvsmbh}
\end{figure}

In Fig. \ref{fig:rmsvsmbh} we plot the $g_{\mathrm{P1}}$ band excess variance measured on timescales of years and months versus the black hole mass for the 1z2\_MBH sample. Even though the estimated uncertainties of the black hole masses of our sample are large, typically $\sim$0.25 dex, there is little evidence for any correlation between $M_{\mathrm{BH}}$ and $\sigma_{\mathrm{rms}}^{2}$ measured on timescales of years. At least for the $g_{\mathrm{P1}}$ band we observe a weak anti-correlation with $M_{\mathrm{BH}}$ for variability measured on timescales of months with $\rho_{\mathrm{S}}=-0.31$ and $P_{\mathrm{S}}=1.6\cdot 10^{-2}$, but the scatter in the relation is quite large. Moreover we do not find any significant anti-correlation relating $M_{\mathrm{BH}}$ with the month timescale $\sigma_{\mathrm{rms}}^{2}$ values of the remaining PS1 bands. The correlation coefficients and p-values of the 1z2\_MBH sample are summarized in table \ref{tab:rmsvsmbh} for all considered PS1 bands and for both variability timescales. The correlation coefficients of the MBH sample are very similar, which is why we do not report them here. Therefore we conclude that there is no significant anti-correlation between optical variability and black hole mass for the probed variability timescales of our light curves. We stress that other optical variability studies found a correlation of variability and $M_{\mathrm{BH}}$ using different variability estimators, but investigating variability timescales which are similar to the ones of our work. Yet these results are inconsistent in the sense that several works state a positive correlation between the variability amplitude and $M_{\mathrm{BH}}$ \citep[e.g.,][]{2007MNRAS.375..989W,2008MNRAS.383.1232W,2010ApJ...721.1014M}, whereas others report an anti-correlation with $M_{\mathrm{BH}}$ \citep{2009ApJ...698..895K,2013ApJ...779..187K}. Finally, we note that Fig. \ref{fig:rmsvsmbh} shows no obvious dependence on redshift and we do not observe any trend for the other PS1 bands. This is also the case considering the MBH sample. 

\begin{table}
\caption{Spearman correlation coefficient $\rho_{\mathrm{S}}$ and respective p-value $P_{\mathrm{S}}$ of $\sigma_{\mathrm{rms}}^{2}$ and $M_{\mathrm{BH}}$.}
\centering
\begin{tabular}{ccc}
\hline\hline
& \multicolumn{2}{c}{$\Delta\mathrm{T}\sim$years, 1z2\_MBH sample}\\

   Filter & $\rho_{\mathrm{S}}$ & $P_{\mathrm{S}}$ \\
    \hline
    $g_{\mathrm{P1}}$    & -0.04 & $7.7\cdot 10^{-1}$ \\
    $r_{\mathrm{P1}}$    & 0.09 & $4.6\cdot 10^{-1}$ \\
    $i_{\mathrm{P1}}$    & -0.01 & $9.6\cdot 10^{-1}$ \\
    $z_{\mathrm{P1}}$    & -0.06 & $6.7\cdot 10^{-1}$ \\
    \hline
& \multicolumn{2}{c}{$\Delta\mathrm{T}\sim$month, 1z2\_MBH sample}\\

   Filter & $\rho_{\mathrm{S}}$ & $P_{\mathrm{S}}$ \\
    \hline
    $g_{\mathrm{P1}}$   & -0.31 & $1.6\cdot 10^{-2}$ \\
    $r_{\mathrm{P1}}$   &  0.06 & $7.1\cdot 10^{-1}$ \\
    $i_{\mathrm{P1}}$   & -0.03 & $8.9\cdot 10^{-1}$ \\
    $z_{\mathrm{P1}}$   & -0.13 & $5.0\cdot 10^{-1}$ \\
    \hline
\end{tabular}
\tablefoot{The values of the 1z2\_MBH sample are quoted for $\sigma_{\mathrm{rms}}^{2}$ measured on year timescales (\textit{top}) and month timescales (\textit{bottom}), respectively.} 
\label{tab:rmsvsmbh}
\end{table}

\subsection{Excess variance versus luminosity}
\label{sec:rmsvslbol}

\begin{figure}
\centering	
\subfloat{%
	\includegraphics[width=0.50\textwidth]{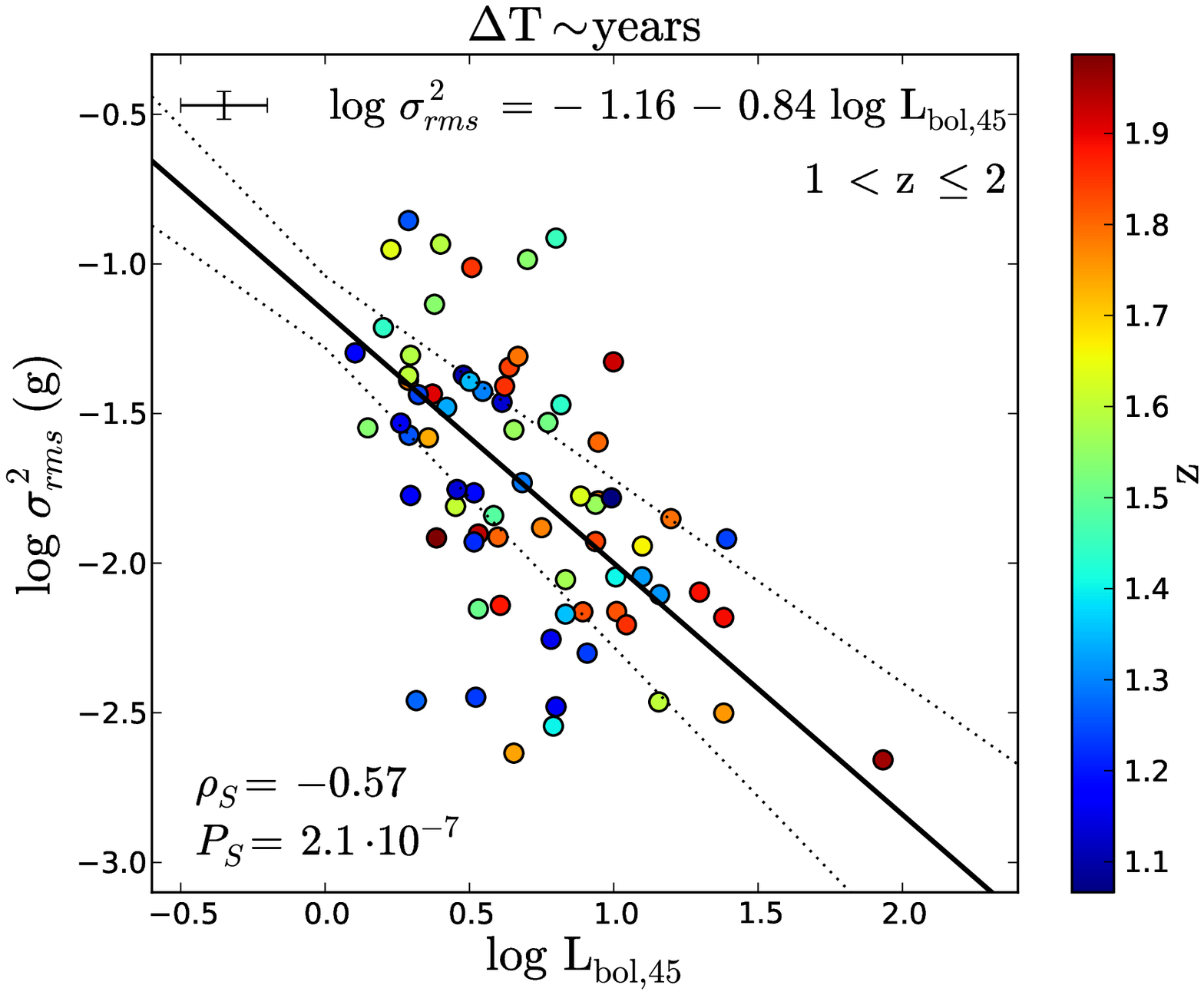}}
		
\subfloat{%
	\includegraphics[width=0.50\textwidth]{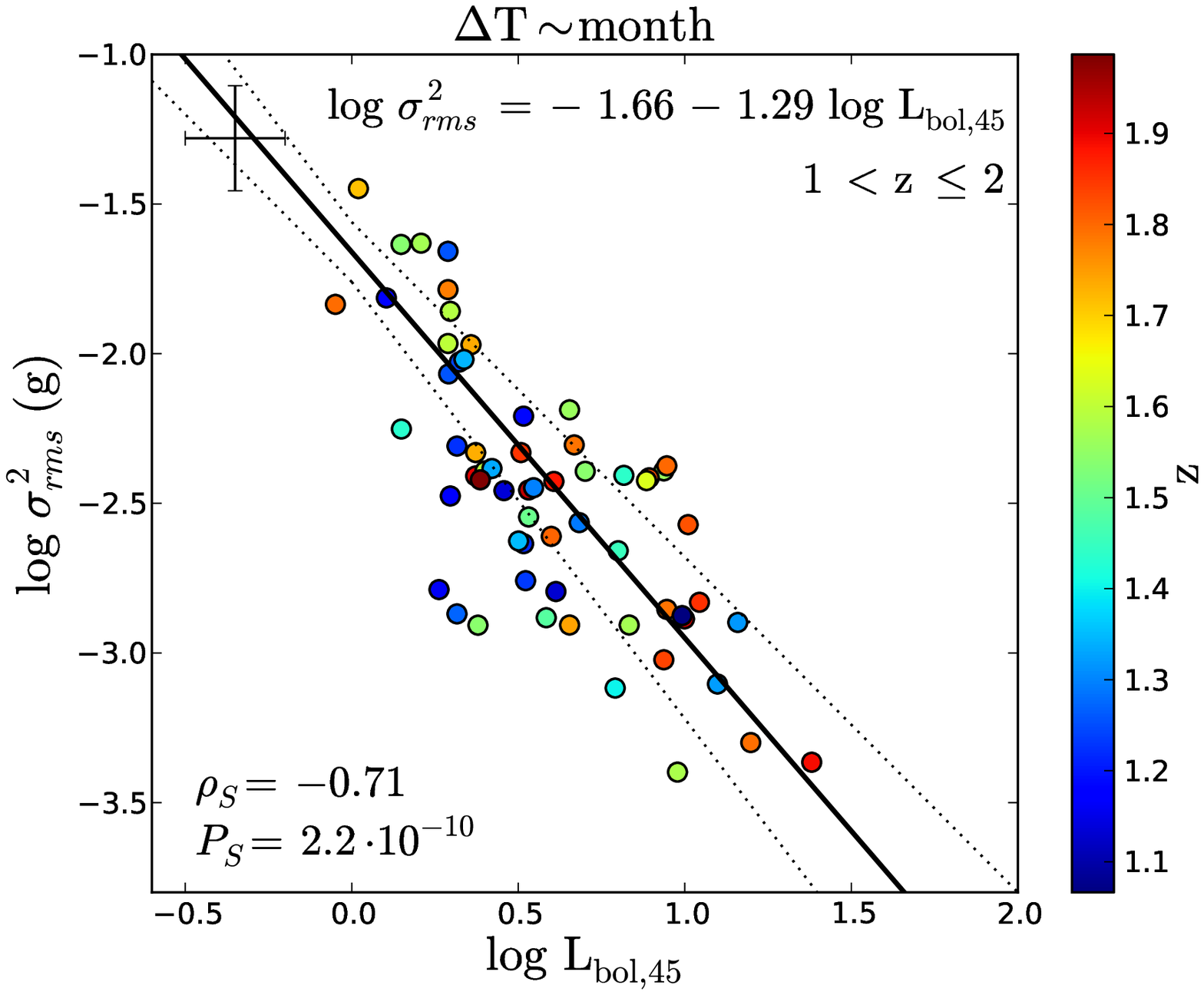}}
\caption{Excess variance ($g_{\mathrm{P1}}$ band) measured on year timescales (\textit{top}) and on month timescales (\textit{bottom}) versus $L_{\mathrm{bol}}$ in units of $10^{45}\,\mathrm{erg\,s^{-1}}$ for the 1z2\_MBH sample. The best power law fit is plotted as black solid line, the dashed lines show the $1\sigma$ errors on the fit parameters. Additionally the redshift is given as a color bar. The black error bars correspond to the average values.}
\label{fig:rmsvslbol}
\end{figure}

The existence of an anti-correlation between optical variability and luminosity has been recognized for many years, but it was often difficult to disentangle the relation from a dependency on redshift. We also observe a strong anti-correlation of the excess variance with bolometric luminosity in our dataset. The respective Spearman correlation coefficients are reported in table \ref{tab:rmsvslbol}. Considering variability on timescales of years the anti-correlation is highly significant in the $g_{\mathrm{P1}}$, $r_{\mathrm{P1}}$ and $i_{\mathrm{P1}}$ bands for the 1z2\_MBH sample. On shorter variability timescales of months the anti-correlation is even stronger and visible in all considered PS1 bands. Furthermore we note that the anti-correlation is generally strongest for the $g_{\mathrm{P1}}$ band and is becoming less significant towards the "redder" bands. We stress that the anti-correlation is also detected with similar significance considering the MBH sample. Fig. \ref{fig:rmsvslbol} presents the $g_{\mathrm{P1}}$ band excess variance as a function of bolometric luminosity for the 1z2\_MBH sample. The figure clearly demonstrates that the anti-correlation with bolometric luminosity is apparent for both probed variability timescales and that the relation is much tighter for shorter timescales of months.  
\begin{table}
\caption{Spearman correlation coefficient $\rho_{\mathrm{S}}$ and respective p-value $P_{\mathrm{S}}$ of $\sigma_{\mathrm{rms}}^{2}$ and $L_{\mathrm{bol}}$.}
\centering
\begin{tabular}{ccc}
\hline\hline
& \multicolumn{2}{c}{$\Delta\mathrm{T}\sim$years, 1z2\_MBH sample}\\

   Filter & $\rho_{\mathrm{S}}$ & $P_{\mathrm{S}}$ \\
    \hline
    $g_{\mathrm{P1}}$    & -0.57 & $2.1\cdot 10^{-7}$ \\
    $r_{\mathrm{P1}}$    & -0.47 & $2.9\cdot 10^{-5}$ \\
    $i_{\mathrm{P1}}$    & -0.49 & $1.6\cdot 10^{-5}$ \\
    $z_{\mathrm{P1}}$    & -0.27 & $4.1\cdot 10^{-2}$ \\
    \hline
& \multicolumn{2}{c}{$\Delta\mathrm{T}\sim$month, 1z2\_MBH sample}\\

   Filter  & $\rho_{\mathrm{S}}$ & $P_{\mathrm{S}}$ \\
    \hline
    $g_{\mathrm{P1}}$    & -0.71 & $2.2\cdot 10^{-10}$ \\
    $r_{\mathrm{P1}}$    & -0.64 & $6.6\cdot 10^{-7}$ \\
    $i_{\mathrm{P1}}$    & -0.64 & $1.6\cdot 10^{-4}$ \\
    $z_{\mathrm{P1}}$    & -0.60 & $3.7\cdot 10^{-4}$ \\
    \hline
\end{tabular}
\tablefoot{The values of the 1z2\_MBH sample are quoted for $\sigma_{\mathrm{rms}}^{2}$ measured on year timescales (\textit{top}) and month timescales (\textit{bottom}), respectively.} 
\label{tab:rmsvslbol}
\end{table}

However, the stronger anti-correlation observed for shorter variability timescales could also just be a selection effect, caused by looking at a particular sub-sample of objects of the larger sample of AGNs which are varying on year timescales. For this reason we additionally searched for the anti-correlation with $L_{\mathrm{bol}}$ selecting the same sub-sample of sources from the 1z2\_MBH sample for both variability timescales. This test revealed that the observed difference regarding the strength of the anti-correlation for the two variability timescales is still present, with $\rho_{\mathrm{S}}=-0.45$, $P_{\mathrm{S}}=7.2\cdot 10^{-4}$ ($g_{\mathrm{P1}}$ band) for variability on year timescales, and $\rho_{\mathrm{S}}=-0.60$, $P_{\mathrm{S}}=1.9\cdot 10^{-6}$ ($g_{\mathrm{P1}}$ band) for variability on month timescales. This finding implies that whichever mechanism causes the anti-correlation of the excess variance with bolometric luminosity must be strongly dependent on the characteristic timescale of variability. 

In order to estimate the functional dependency of $\sigma_{\mathrm{rms}}^{2}$ on $L_{\mathrm{bol}}$ we use the Bayesian linear regression method of \cite{2007ApJ...665.1489K} which considers the measurement uncertainties of both related quantities. In detail we fit the linear model $\log \sigma_{\mathrm{rms}}^{2}=\beta+\alpha\log L_{\mathrm{bol,45}}+\epsilon$ with $L_{\mathrm{bol,45}}=L_{\mathrm{bol}}/10^{45}\,\mathrm{erg\,s^{-1}}$ to the dataset. In addition to the zeropoint $\beta$ and the logarithmic slope $\alpha$ this model also fits the intrinsic scatter $\epsilon$ inherent to the relation. Since the symmetric error of the excess variance given by equation \ref{eq:errnev} becomes asymmetric in log-space, we use a symmetrized error by taking the average of the upper and lower error. Regarding the error of $L_{\mathrm{bol}}$, \cite{2013A&A...560A..72R} observe a rms scatter of 0.11 dex comparing a sub-sample of 63 QSOs with spectra from two different datasets, whereas \cite{2011A&A...534A.110L} find a $1\sigma$ dispersion of 0.2 dex for their SED-fitting method for a larger sample. In this work we perform all fits assuming a conservative average uncertainty of 0.15 dex for each AGN.

The fitted values for the 1z2\_MBH sample are listed in table \ref{tab:fitlbolzbin} for each considered PS1 band and the best-fitting model is also displayed in Fig. \ref{fig:rmsvslbol}. We note that the model fits produce the same logarithmic slopes, at least within the $1\sigma$ errors, for all those PS1 bands showing a significant anti-correlation according to the $\rho_{\mathrm{S}}$ and $P_{\mathrm{S}}$ values. Comparing the two considered variability timescales the determined slopes of the $\sigma_{\mathrm{rms}}^{2}$ values measured on month timescales are systematically steeper. However within one or two standard deviations the fitted slopes are consistent with a value of $\alpha\sim-1$ for both variability timescales, indicating that the relationship may be created by the same physical process\footnote{We found that the assumed x-axis error strongly affects the derived slope for our fitting routine. Performing test fits with the $g_{\mathrm{P1}}$ band data yielded slopes of -0.70, -0.84, -1.00, -1.74 using $\Delta\log L_{\mathrm{bol}}=$0.01, 0.15, 0.2, 0.3, respectively. Larger x-axis errors therefore systematically steepen the fitted slope and this effect is particularly strong for large errors. However, the bulk of data points clearly suggest a value of $\sim$-1.}. We stress that the intrinsic scatter of the relation is only $\sim$0.2--0.25 dex for variability on timescales of months, whereas the scatter is about a factor of two larger for variability on timescales of years. Fitting the linear model to the MBH sample, i.e. including the full redshift range, results in very similar slopes for variability measured on month timescales. Yet the presence of some high redshift outliers in the larger sample with $\sigma_{\mathrm{rms}}^{2}$ measured on year timescales drives the fitting routine towards much flatter slopes of $\alpha\sim-0.5$. Finally, we tested that the anti-correlation between $\sigma_{\mathrm{rms}}^{2}$ and $L_{\mathrm{bol}}$ is also recovered when applying a $3\sigma$ cut in the variability detection (see Eq. \ref{eq:detectnev}). In detail, considering the $g_{\mathrm{P1}}$ band 1z2\_MBH sample we then obtain $\rho_{\mathrm{S}}=-0.58$ and $P_{\mathrm{S}}=1.3\cdot 10^{-7}$ with fitted parameters of $\alpha=-0.85\pm 0.16$, $\beta=-1.15\pm 0.12$, and $\epsilon=0.35\pm 0.04$ for year timescale variability. The corresponding values for month timescale variability read $\rho_{\mathrm{S}}=-0.69$ and $P_{\mathrm{S}}=2.2\cdot 10^{-5}$ with fitted parameters of $\alpha=-1.27\pm 0.22$, $\beta=-1.57\pm 0.14$, and $\epsilon=0.16\pm 0.06$.
\begin{table}
\caption{Scaling of $\sigma_{\mathrm{rms}}^{2}$ with $L_{\mathrm{bol}}$.}
\centering
\begin{tabular}{cccc}
\hline\hline
 & \multicolumn{3}{c}{$\Delta\mathrm{T}\sim$years, 1z2\_MBH sample}\\

   Filter  & $\alpha$ & $\beta$ & $\epsilon$ \\
    \hline
    $g_{\mathrm{P1}}$  & $-0.84\pm 0.16$ & $-1.16\pm 0.12$ & $0.36\pm 0.04$ \\
    $r_{\mathrm{P1}}$  & $-0.74\pm 0.17$ & $-1.43\pm 0.13$ & $0.41\pm 0.04$ \\
    $i_{\mathrm{P1}}$  & $-0.85\pm 0.19$ & $-1.47\pm 0.14$ & $0.43\pm 0.04$ \\
    $z_{\mathrm{P1}}$  & $-0.55\pm 0.23$ & $-1.73\pm 0.19$ & $0.42\pm 0.04$ \\
    \hline
 & \multicolumn{3}{c}{$\Delta\mathrm{T}\sim$month, 1z2\_MBH sample}\\

   Filter & $\alpha$ & $\beta$ & $\epsilon$ \\
    \hline
    $g_{\mathrm{P1}}$  & $-1.29\pm 0.17$ & $-1.66\pm 0.10$ & $0.17\pm 0.05$ \\
    $r_{\mathrm{P1}}$  & $-0.95\pm 0.22$ & $-2.17\pm 0.14$ & $0.22\pm 0.05$ \\
    $i_{\mathrm{P1}}$  & $-1.31\pm 0.35$ & $-2.09\pm 0.20$ & $0.23\pm 0.07$ \\
    $z_{\mathrm{P1}}$  & $-1.37\pm 0.53$ & $-2.12\pm 0.35$ & $0.28\pm 0.08$ \\
    \hline
\end{tabular}
\tablefoot{Fitted values of the relation $\log \sigma_{\mathrm{rms}}^{2}=\beta+\alpha\log L_{\mathrm{bol,45}}+\epsilon$ for each considered PS1 band assuming $\Delta\log L_{\mathrm{bol}}=0.15$. The values for the 1z2\_MBH sample are quoted for $\sigma_{\mathrm{rms}}^{2}$ measured on year timescales (\textit{top}) and month timescales (\textit{bottom}), respectively.} 
\label{tab:fitlbolzbin}
\end{table}

\begin{figure}
\centering	
\includegraphics[width=0.5\textwidth]{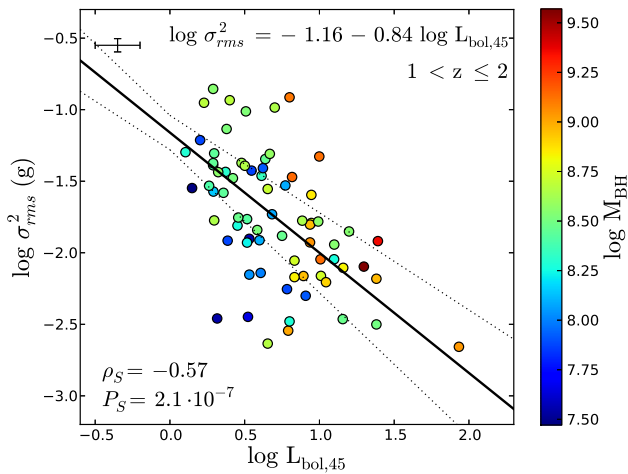}
\caption{Same as Fig. \ref{fig:rmsvslbol} for $\sigma_{\mathrm{rms}}^{2}$ measured on year timescales, but with $M_{\mathrm{BH}}$ as color bar.}
\label{fig:colmbhrmsvslbol}
\end{figure}

Several authors who observed an anti-correlation of $\sigma_{\mathrm{rms}}^{2}$ and luminosity argued that this relation may be a byproduct of a more fundamental anti-correlation of $\sigma_{\mathrm{rms}}^{2}$ and $M_{\mathrm{BH}}$ seen at frequencies above $\nu_{\mathrm{br}}$ in X-ray studies, since the more luminous sources tend to be the more massive systems \citep[e.g.][]{2004MNRAS.348..207P,2012A&A...542A..83P}. This was also proposed by \citet{2014ApJ...781..105L}, studying the low frequency part of the X-ray PSD, and was triggered by the very similar slopes they found for the anti-correlations of $\sigma_{\mathrm{rms}}^{2}$ with $M_{\mathrm{BH}}$ and X-ray luminosity. To test for a similar trend in our data we display the black hole mass as color code in Fig. \ref{fig:colmbhrmsvslbol}, which apart from that shows the same information as the upper panel of Fig. \ref{fig:rmsvslbol}. The rough proportionality of $L_{\mathrm{bol}}$ and $M_{\mathrm{BH}}$ is apparent in the color code as a weak trend that $M_{\mathrm{BH}}$ increases in the x-axis direction. Considering the y-axis direction we observe low and high mass systems at the same level of variability amplitude. This is also the case for $\sigma_{\mathrm{rms}}^{2}$ measured on month timescales (not shown here). However, if the anti-correlation of $\sigma_{\mathrm{rms}}^{2}$ and $L_{\mathrm{bol}}$ would be caused by a hidden anti-correlation with $M_{\mathrm{BH}}$, then the less massive AGNs should predominantly occupy the upper region of the plot and vice versa. Given that $L_{\mathrm{bol}}\propto\dot{M}$, where $\dot{M}$ denotes the mass accretion rate, these findings suggest that the fundamental AGN parameter determining the optical variability amplitude is not the black hole mass, but the accretion rate.  

\subsection{Excess variance versus redshift}
\label{sec:rmsvsz}

\begin{figure}
	\centering
	\includegraphics[width=.5\textwidth]{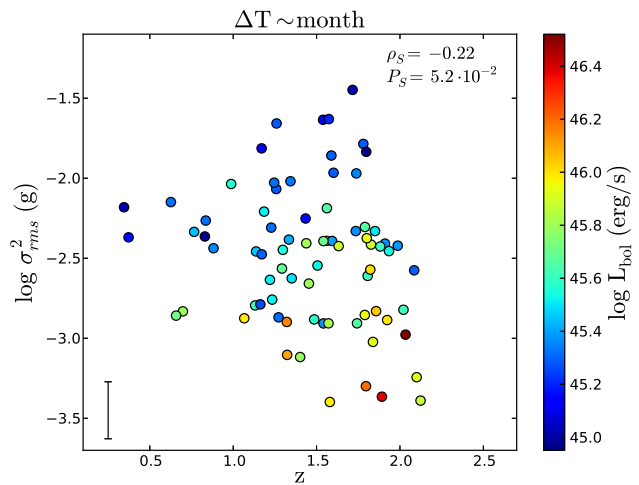}
	\caption{Excess variance ($g_{\mathrm{P1}}$ band) measured on month timescales versus redshift for the MBH sample. The bolometric luminosity is given as a color bar.}
	\label{fig:rmsvsz}
\end{figure}

Considering the relations presented above we do not observe any strong evolution with redshift. In fact correlating the excess variance with the redshift of our AGNs we find no significant dependency in any band, which is summarized in table \ref{tab:rmsvsz}. However we can predict the expected evolution of the variability amplitude with redshift in view of the scaling relations outlined in the previous sections. Since we observe our sources in passbands with a fixed wavelength range, the actual rest-frame wavelength probed by each filter is shifted to shorter wavelength for higher redshift. Towards higher redshift we therefore probe UV variability in the "bluest" PS1 bands, whereas the "redder" bands cover the rest-frame optical variability of the AGNs. Yet we showed in section \ref{sec:rmswave} that the variability amplitude generally decreases with increasing wavelength for our sources. Assuming that the intrinsic variability does not change dramatically from one AGN to another we would therefore expect to observe a positive correlation of the excess variance with redshift for the same band. However we found strong evidence that the intrinsic variability amplitude of AGNs is anti-correlated with bolometric luminosity. Considering the weak selection effect which is apparent in Fig. \ref{fig:physvsz} of appendix \ref{sec:appendixa}, we actually observe the most luminous objects predominantly at higher redshift. From this selection effect alone we would expect an anti-correlation between the excess variance and redshift. The fact that we do not find a dependency of variability on redshift for our AGN sample is likely the result of the superposition of the two aforementioned effects, which are acting in different directions.         
\begin{table}
\caption{Spearman correlation coefficient $\rho_{\mathrm{S}}$ and respective p-value $P_{\mathrm{S}}$ of $\sigma_{\mathrm{rms}}^{2}$ and $z$.}
\centering
\begin{tabular}{ccccc}
\hline\hline
& \multicolumn{2}{c}{$\Delta\mathrm{T}\sim$years} & \multicolumn{2}{c}{$\Delta\mathrm{T}\sim$month}\\

Filter & $\rho_{\mathrm{S}}$ & $P_{\mathrm{S}}$ 
     & $\rho_{\mathrm{S}}$ & $P_{\mathrm{S}}$ \\
    \hline
    $g_{\mathrm{P1}}$    & -0.16  & $1.3\cdot 10^{-1}$  & -0.22 & $5.2\cdot 10^{-2}$ \\
    $r_{\mathrm{P1}}$    & -0.05  & $5.9\cdot 10^{-1}$  & -0.09 & $4.7\cdot 10^{-1}$ \\
    $i_{\mathrm{P1}}$    & -0.14  & $1.8\cdot 10^{-1}$  & -0.20 & $2.0\cdot 10^{-1}$ \\
    $z_{\mathrm{P1}}$    & -0.15  & $1.9\cdot 10^{-1}$  & -0.25 & $1.0\cdot 10^{-1}$ \\
    \hline
\end{tabular}
\tablefoot{The values of the MBH sample are quoted for $\sigma_{\mathrm{rms}}^{2}$ measured on year timescales (\textit{left column}) and month timescales (\textit{right column}), respectively.} 
\label{tab:rmsvsz}
\end{table}
This explanation is in agreement with what we observe in Fig. \ref{fig:rmsvsz}, displaying the excess variance versus redshift and the bolometric luminosity as a color bar. The slight anti-correlation of $\sigma_{\mathrm{rms}}^{2}$ with redshift is counterbalanced by a positive correlation, which is visible in various stripes of constant luminosity showing an increasing variability amplitude. The positive correlation of the variability amplitude with redshift due to the redshift dependent wavelength probed by a given filter was also observed in earlier works \citep{1990AJ....100...56C,1994MNRAS.268..305H,1996MNRAS.282.1191C,1996A&A...306..395C}. Our results are also in accord with recent studies that did not find any significant evolution of variability with redshift or identified an observed correlation to be caused by the aforementioned selection effects \citep{2010ApJ...721.1014M,2012ApJ...758..104Z,2014ApJ...784...92M}. Finally, the low intrinsic scatter in the relation with $L_{\mathrm{bol}}$ suggests that biases due to the broad redshift distribution of our sample are negligible compared to the strong dependence on $L_{\mathrm{bol}}$.     

\subsection{Excess variance versus Eddington ratio}
\label{sec:rmsvsedd}

The last fundamental AGN parameter for which we can probe correlations with variability is the Eddington ratio. The correlation coefficients and p-values suggest an anti-correlation between $\sigma_{\mathrm{rms}}^{2}$ and $\lambda_{\mathrm{Edd}}$ with high significance for both studied variability timescales, considering the MBH sample and the 1z2\_MBH sample. The values for the 1z2\_MBH sample are quoted in table \ref{tab:rmsvsedd}. However the relation is not as tight as the one with bolometric luminosity, but the uncertainty of $\lambda_{\mathrm{Edd}}$ is considerably larger because both the error of $L_{\mathrm{bol}}$ and $M_{\mathrm{BH}}$ contribute to its value. The $1\sigma$ dispersion of the black hole masses is 0.24 dex according to \cite{2013A&A...560A..72R}, yet the actual uncertainty might be even larger due to systematic errors. The anti-correlation is apparent for all considered PS1 bands, though less robust for the $z_{\mathrm{P1}}$ band. Moreover comparing the two variability timescales we find the anti-correlation to be more significant for the $\sigma_{\mathrm{rms}}^{2}$ values measured on timescales of years, in contrast to what is observed in the relation with $L_{\mathrm{bol}}$. However given the comparably large uncertainties of the $\lambda_{\mathrm{Edd}}$ values, this difference should not be over-interpreted. In addition we checked that the $\rho_{\mathrm{S}}$ and $P_{\mathrm{S}}$ values obtained for the same sub-sample of objects are very similar for both variability timescales.      
\begin{table}
\caption{Spearman correlation coefficient $\rho_{\mathrm{S}}$ and respective p-value $P_{\mathrm{S}}$ of $\sigma_{\mathrm{rms}}^{2}$ and $\lambda_{\mathrm{Edd}}$.}
\centering
\begin{tabular}{ccc}
\hline\hline
& \multicolumn{2}{c}{$\Delta\mathrm{T}\sim$years, 1z2\_MBH sample}\\

   Filter & $\rho_{\mathrm{S}}$ & $P_{\mathrm{S}}$ \\
    \hline
    $g_{\mathrm{P1}}$    & -0.52 & $2.6\cdot 10^{-6}$ \\
    $r_{\mathrm{P1}}$    & -0.56 & $2.7\cdot 10^{-7}$ \\
    $i_{\mathrm{P1}}$    & -0.48 & $2.9\cdot 10^{-5}$ \\
    $z_{\mathrm{P1}}$    & -0.25 & $6.5\cdot 10^{-2}$ \\
    \hline
& \multicolumn{2}{c}{$\Delta\mathrm{T}\sim$month, 1z2\_MBH sample}\\

   Filter & $\rho_{\mathrm{S}}$ & $P_{\mathrm{S}}$ \\
    \hline
    $g_{\mathrm{P1}}$    & -0.32 & $1.2\cdot 10^{-2}$ \\
    $r_{\mathrm{P1}}$    & -0.57 & $1.9\cdot 10^{-5}$ \\
    $i_{\mathrm{P1}}$    & -0.47 & $9.4\cdot 10^{-3}$ \\
    $z_{\mathrm{P1}}$    & -0.27 & $1.4\cdot 10^{-1}$ \\
    \hline
\end{tabular}
\tablefoot{The values of the 1z2\_MBH sample are quoted for $\sigma_{\mathrm{rms}}^{2}$ measured on year timescales (\textit{top}) and month timescales (\textit{bottom}), respectively.} 
\label{tab:rmsvsedd}
\end{table}

\begin{figure}
\centering
\subfloat{%
	\includegraphics[width=0.50\textwidth]{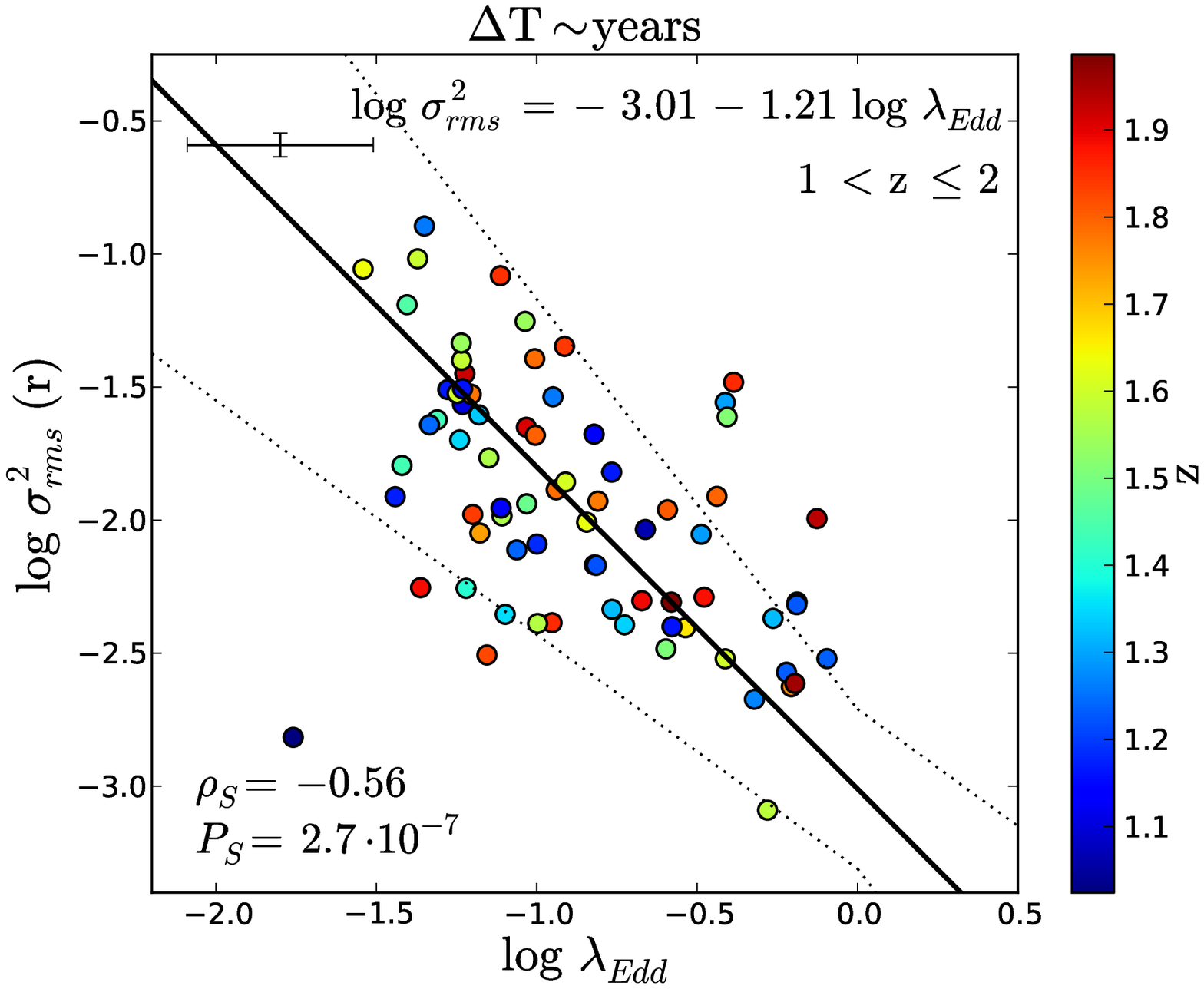}}
		
\subfloat{%
	\includegraphics[width=0.50\textwidth]{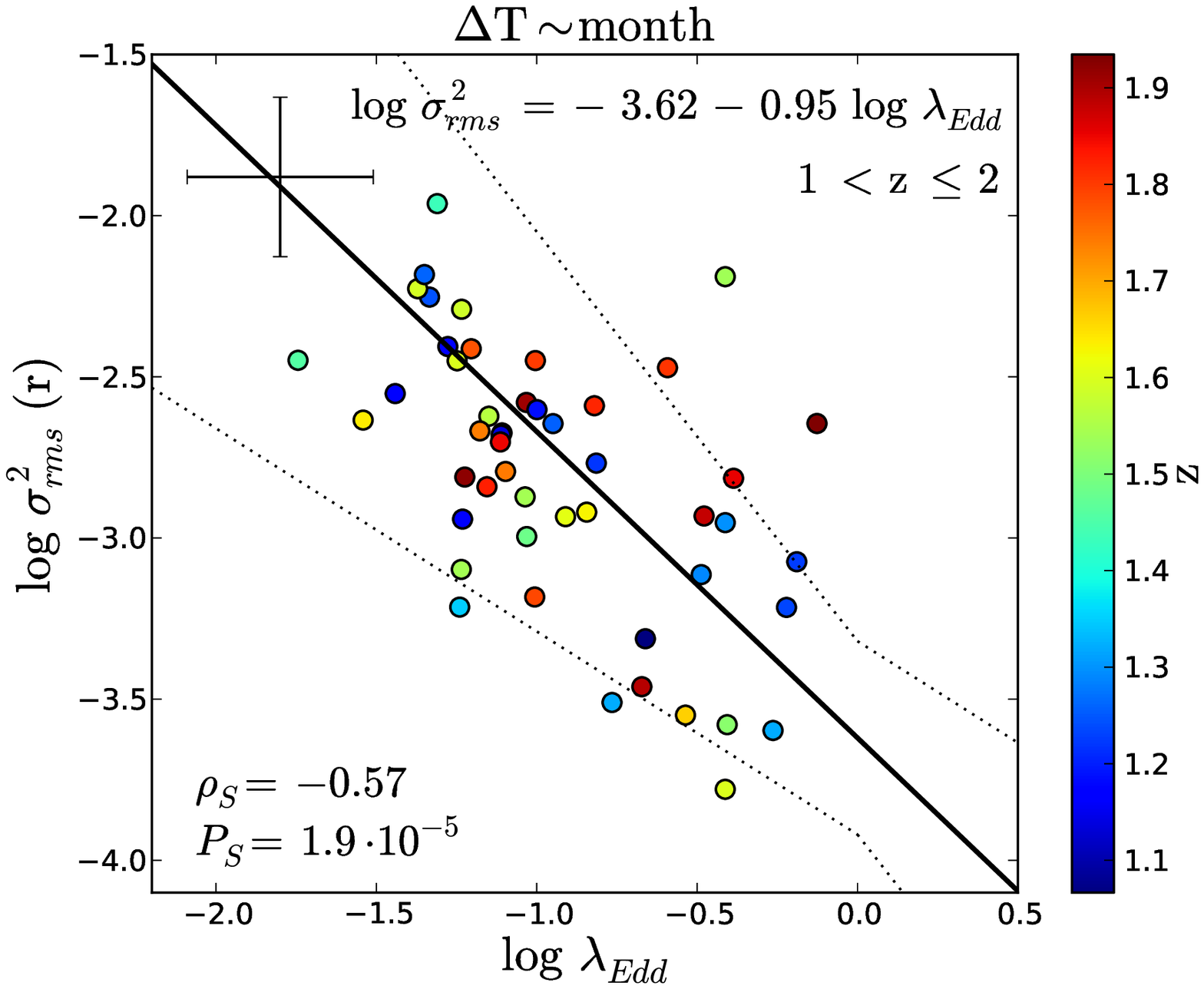}}
\caption{Excess variance ($r_{\mathrm{P1}}$ band) measured on year timescales (\textit{top}) and on month timescales (\textit{bottom}) versus $\lambda_{\mathrm{Edd}}$ for the 1z2\_MBH sample. The best power law fit and other symbols are displayed as in Fig. \ref{fig:rmsvslbol}.}
\label{fig:rmsvsedd}
\end{figure}

We use the same fitting technique as described in section \ref{sec:rmsvslbol} with a power law model of the form $\log \sigma_{\mathrm{rms}}^{2}=\beta+\alpha\log\lambda_{\mathrm{Edd}}+\epsilon$ in order to find the scaling of $\sigma_{\mathrm{rms}}^{2}$ with $\lambda_{\mathrm{Edd}}$. Regarding the error of $\lambda_{\mathrm{Edd}}$ we assume $\Delta\log L_{\mathrm{bol}}=0.15$ and $\Delta\log M_{\mathrm{BH}}=0.25$ for each AGN, added in quadrature\footnote{We also performed the fits using larger uncertainties of $\Delta\log M_{\mathrm{BH}}=$0.3--0.4. However, due to the systematic steepening of the derived slopes for larger x-axis errors we reported in section \ref{sec:rmsvslbol}, these errors lead to slopes that are much steeper than the overall distribution of the data implies.}. The results are listed in table \ref{tab:fiteddzbin} and we show the data with the fitted relation for the $r_{\mathrm{P1}}$ band in Fig. \ref{fig:rmsvsedd} for the 1z2\_MBH sample. We note that owing to the large error bars of the Eddington ratio and the large scatter in the anti-correlation the uncertainties of the fitted parameters are quite large. Considering all PS1 bands exhibiting a significant anti-correlation in view of the correlation coefficients, i.e. the $g_{\mathrm{P1}}$, $r_{\mathrm{P1}}$ and $i_{\mathrm{P1}}$ bands, we find logarithmic slopes very similar to the ones of the $L_{\mathrm{bol}}$ relation with $\alpha\sim -1$ within the $1\sigma$ errors for both variability timescales. The intrinsic scatter of the relation between $\sigma_{\mathrm{rms}}^{2}$ and $\lambda_{\mathrm{Edd}}$ is $\sim$0.2--0.4 dex. 
\begin{table}
\caption{Scaling of $\sigma_{\mathrm{rms}}^{2}$ with $\lambda_{\mathrm{Edd}}$.}
\centering
\begin{tabular}{cccc}
\hline\hline
 & \multicolumn{3}{c}{$\Delta\mathrm{T}\sim$years, 1z2\_MBH sample}\\

   Filter & $\alpha$ & $\beta$ & $\epsilon$ \\
    \hline
    $g_{\mathrm{P1}}$  & $-1.37\pm 0.38$ & $-2.97\pm 0.34$ & $0.28\pm 0.07$ \\
    $r_{\mathrm{P1}}$  & $-1.21\pm 0.33$ & $-3.01\pm 0.30$ & $0.33\pm 0.06$ \\
    $i_{\mathrm{P1}}$  & $-1.16\pm 0.41$ & $-3.12\pm 0.38$ & $0.38\pm 0.07$ \\
    $z_{\mathrm{P1}}$  & $-0.69\pm 0.54$ & $-2.74\pm 0.47$ & $0.41\pm 0.05$ \\
    \hline
 & \multicolumn{3}{c}{$\Delta\mathrm{T}\sim$month, 1z2\_MBH sample}\\

   Filter & $\alpha$ & $\beta$ & $\epsilon$ \\
    \hline
    $g_{\mathrm{P1}}$  & $-1.16\pm 0.89$ & $-3.53\pm 0.85$ & $0.33\pm 0.07$ \\
    $r_{\mathrm{P1}}$  & $-0.95\pm 0.32$ & $-3.62\pm 0.30$ & $0.19\pm 0.07$ \\
    $i_{\mathrm{P1}}$  & $-1.26\pm 0.70$ & $-4.15\pm 0.75$ & $0.28\pm 0.10$ \\
    $z_{\mathrm{P1}}$  & $-0.48\pm 0.61$ & $-3.45\pm 0.55$ & $0.36\pm 0.07$ \\
    \hline
\end{tabular}
\tablefoot{Fitted values of the relation $\log\sigma_{\mathrm{rms}}^{2}=\beta+\alpha\log\lambda_{\mathrm{Edd}}+\epsilon$ for each considered PS1 band assuming $\Delta\log L_{\mathrm{bol}}=0.15$ and $\Delta\log M_{\mathrm{BH}}=0.25$. The values of the 1z2\_MBH sample are quoted for $\sigma_{\mathrm{rms}}^{2}$ measured on year timescales (\textit{top}) and month timescales (\textit{bottom}), respectively.} 
\label{tab:fiteddzbin}
\end{table}

In contrast to the well established anti-correlation of optical variability and luminosity, the actual dependency of the variability amplitude on the Eddington ratio is less clear, but evidence for an anti-correlation was detected in previous investigations \citep{2008MNRAS.383.1232W,2009ApJ...696.1241B,2010ApJ...716L..31A,2010ApJ...721.1014M,2012ApJ...758..104Z,2013ApJ...779..187K}. The highly significant anti-correlations between $\sigma_{\mathrm{rms}}^{2}$ and the quantities $\lambda_{\mathrm{Edd}}$ and $L_{\mathrm{bol}}$ reported in this work strongly support the idea that the accretion rate is the main driver of optical variability.   

\section{Power spectrum analysis}
\label{sec:psdanalysis}

Considering the fact that we do not correct our $\sigma_{\mathrm{rms}}^{2}$ measurements for the range in redshift covered by our sources, but the excess variance depends on the rest-frame time intervals sampled by a light curve, the presented results may be weakly biased although we do not find any strong trend with redshift. Furthermore the individual segments of the MDF04 light curves used in the calculation of the excess variance on timescales of months do not have the same length in general, introducing further biases on these timescales. However, we can perform an independent verification of our results by applying the CARMA modelling of variability described in \citet{2014ApJ...788...33K}, which does not suffer from the latter issues. What is more, this model allows us to study in detail the PSDs of our light curves and therefore provides information about the part of the PSD which is predominantly integrated by our $\sigma_{\mathrm{rms}}^{2}$ measurements.

\subsection{Fitting the CARMA model}
\label{sec:carmafit}

In order to model our light curves as a CARMA(p,q) process we use the software package provided by \citet{2014ApJ...788...33K} involving an adaptive Metropolis MCMC sampler, routines to obtain maximum-likelihood estimates of the CARMA parameters and tools to analyse the output of the MCMC samples. Finding the optimal order of the CARMA process for a given light curve can be difficult and there are several ways to select p and q. Following \citet{2014ApJ...788...33K} we choose the order of the CARMA model invoking the corrected Akaike Information Criterion (AICc) \citep{Akaike1973,HURVICH01061989}. The AICc for a time series of $N$ values $\boldsymbol{y}=y_{1},...,y_{N}$ is defined by
\begin{flalign}
	\label{eq:aicc}	
	AICc\left(p,q\right)=2k-2\log p\left(\boldsymbol{y}|\theta_{\mathrm{mle}},p,q\right)+\frac{2k\left(k+1\right)}{N-k-1}
\end{flalign} 
with $k$ the number of free parameters, $p\left(\boldsymbol{y}|\theta\right)$ the likelihood function of the light curve and $\theta_{\mathrm{mle}}$ the maximum-likelihood estimate of the CARMA model parameters summarized by the symbol $\theta$. The optimal CARMA model for a given light curve is the one that minimizes the AICc. For each pair (p,q) the CARMA software package of \citet{2014ApJ...788...33K} finds the maximum-likelihood estimate $\theta_{\mathrm{mle}}$ by running 100 optimizers with random initial sets of $\theta$ and then selects the order (p,q) that minimizes the AICc for the optimized $\theta_{\mathrm{mle}}$ value.   

Before applying the CARMA model we transform the light curve of each of our objects to the AGN rest-frame according to $t_{i,\mathrm{rest}}=\left(t_{i,\mathrm{obs}}-t_{0,\mathrm{obs}}\right)/\left(1+z\right)$ with $t_{0,\mathrm{obs}}$ denoting the starting point of the light curve. For each source we then find the order (p,q) of the CARMA model by minimizing the AICc on the grid $p=1,..,7$, $q=0,...,p-1$. Given the optimal CARMA(p,q) model we run the MCMC sampler for 75\,000 iterations with the first 25\,000 discarded as burn-in to obtain the PSD of the CARMA process for each of our sources\footnote{We note that running the MCMC sampler without parallel tempering or with 10 parallel chains leads to essentially indistinguishable results for our data.}. This procedure was performed for the flux light curves of the total sample in the four PS1 bands $g_{\mathrm{P1}}$, $r_{\mathrm{P1}}$, $i_{\mathrm{P1}}$ and $z_{\mathrm{P1}}$.    

\subsection{Quantifying the model fit}
\label{sec:assessfit}

As outlined in \citet{2014ApJ...788...33K} the accurateness of the CARMA model fit can be tested by investigating the properties of the standardized residuals $\chi_{i}$. The latter are given by
\begin{flalign}
	\label{eq:residuals}	
	\chi_{i}=\frac{y_{i}-E\left(y_{i}|\boldsymbol{y}_{<i},\theta_{\mathrm{map}}\right)}{\sqrt{Var\left(y_{i}|\boldsymbol{y}_{<i},\theta_{\mathrm{map}}\right)}}	
\end{flalign}
where $\boldsymbol{y}_{<i}=y_{1},...,y_{i-1}$ and $\theta_{\mathrm{map}}$ is the maximum a posteriori value of the CARMA model parameters. The expectation value $E\left(y_{i}|\boldsymbol{y}_{<i},\theta_{\mathrm{map}}\right)$ and variance $Var\left(y_{i}|\boldsymbol{y}_{<i},\theta_{\mathrm{map}}\right)$ of the light curve point $y_{i}$ given all previous values under the CARMA model are calculated using the Kalman filter \citep{JONES01121990}, see also appendix A of \citet{2014ApJ...788...33K}. If the Gaussian CARMA model provides an adequate description of a light curve, then the $\chi_{i}$ should follow a normal distribution with zero mean and unit standard deviation. Moreover the sequence of $\chi_{1},...,\chi_{N}$ should resemble a Gaussian white noise sequence, i.e. the autocorrelation function (ACF) at time lag $\tau$ of the sequence of residuals should be uncorrelated and be normally distributed with zero mean and variance $1/N$. Likewise the sequence of $\chi_{1}^{2},...,\chi_{N}^{2}$ should also be a Gaussian white noise sequence with an ACF distribution of zero mean and variance $1/N$.  

\begin{figure}
\centering
\subfloat{%
	\includegraphics[width=0.50\textwidth]{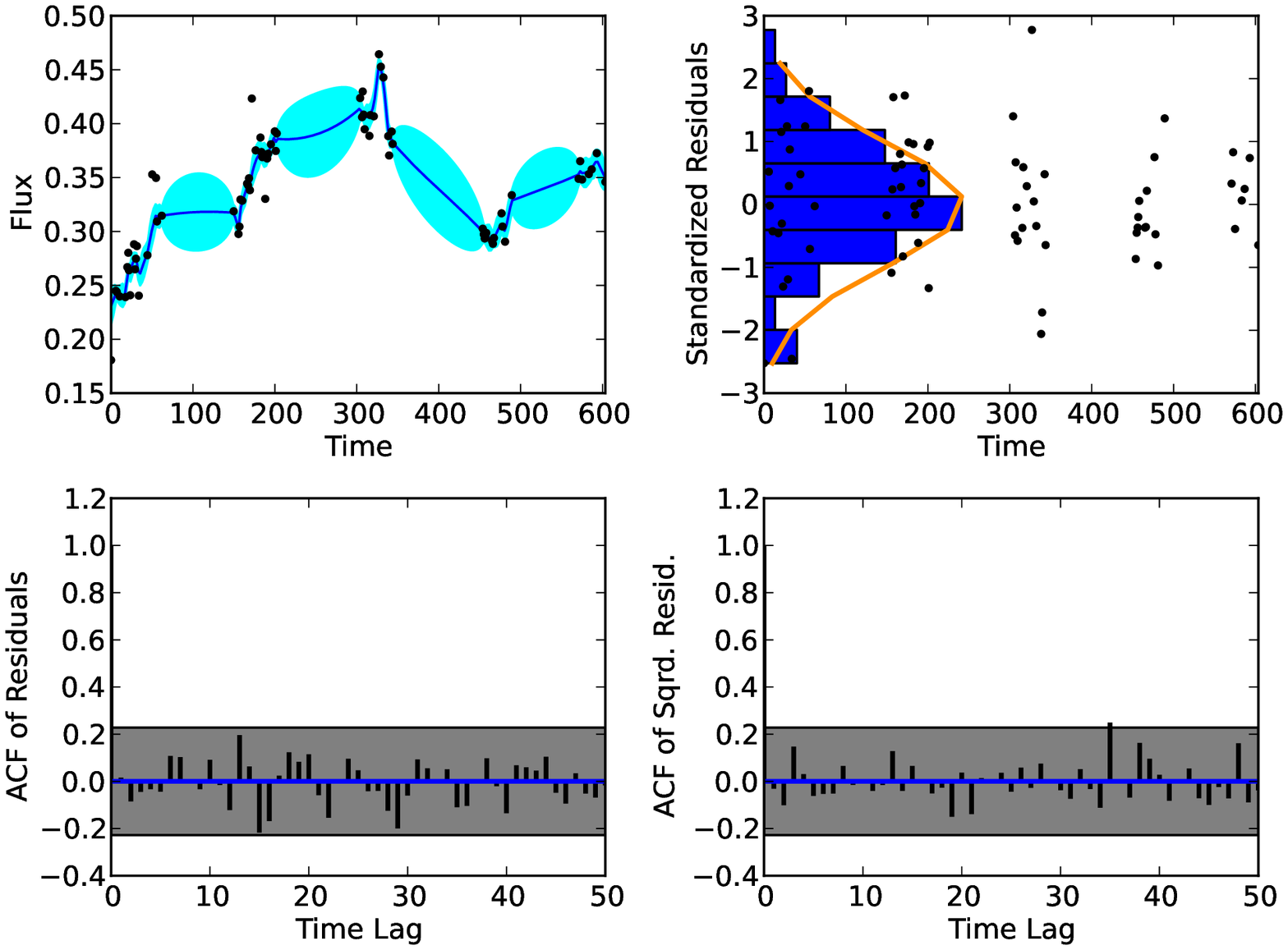}}
		
\subfloat{%
	\includegraphics[width=0.50\textwidth]{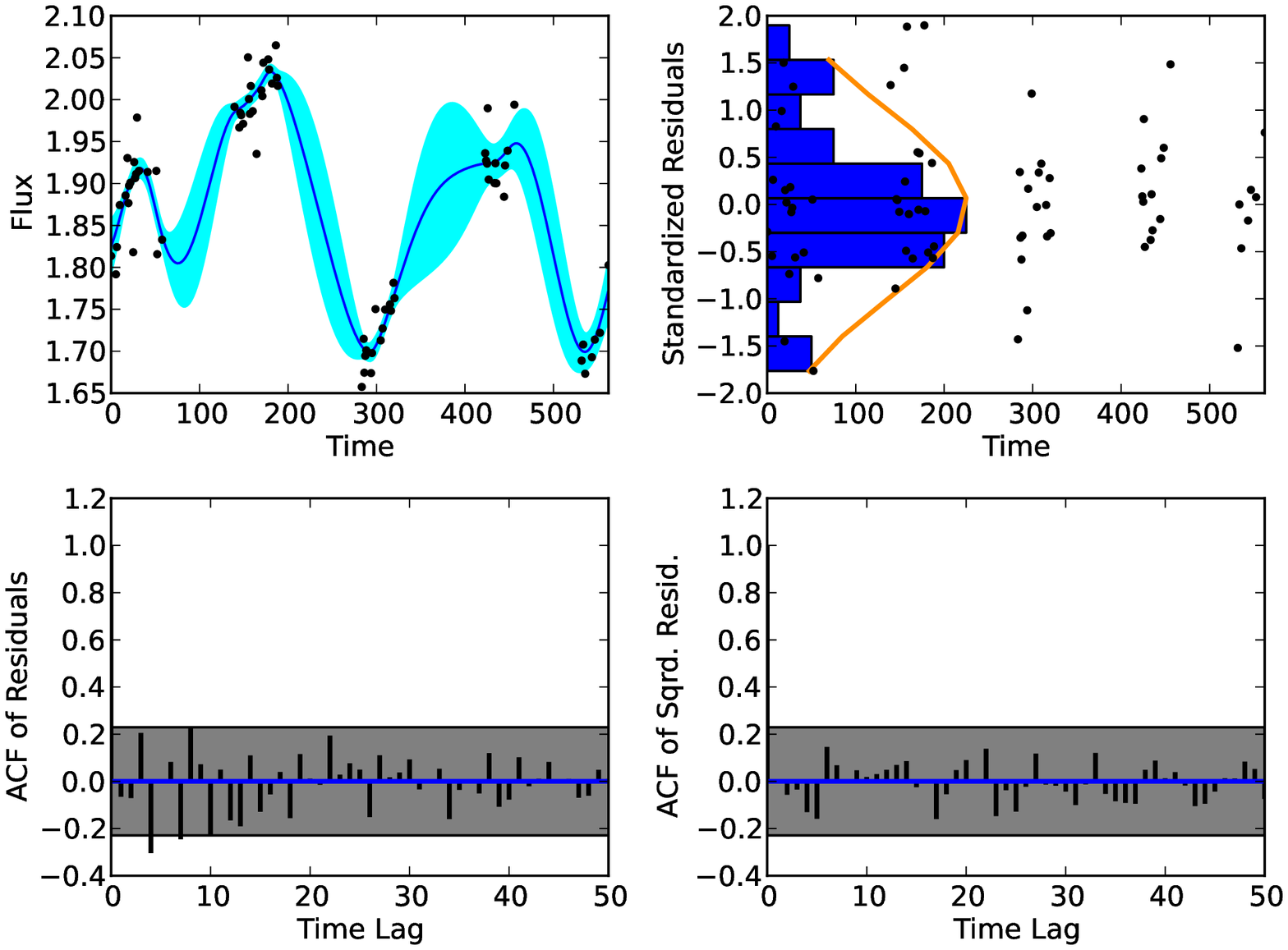}}
\caption{In both subpanels starting from top left: a) $g_{\mathrm{P1}}$ band flux light curve (in units of 3631 Jy times $10^{8}$) with the solid blue line and cyan regions corresponding to the modelled light curve and 1$\sigma$ error bands given the measured data (black points). b) Standardized residuals (black points) and their histogram in blue, over-plotted with the expected standard normal distribution (orange line). c) and d) Autocorrelation functions (ACF) of the standardized residuals (bottom left) and their square (bottom right) with the shaded region displaying the 95\% confidence intervals assuming a white noise process. The top four panels show the data of the AGN with XID 2391 which is best fitted by a CARMA(3,0) process. The bottom four panels show data of the AGN with XID 30 which is best fitted by a CARMA(2,0) process.}
\label{fig:residuals}
\end{figure}

For each of our sources we visually inspected the aforementioned three properties of the residuals. We find that more than 90\% of the AGN light curves of our sample do not exhibit strong deviations from the expected distributions of the residuals in any of the four studied bands. Exemplary, we show the interpolated $g_{\mathrm{P1}}$ band flux light curve, the distribution of the residuals as well as the distributions of the ACF of the sequence of residuals and their square in Fig. \ref{fig:residuals} for two AGNs of our sample. The AGN with XID 2391 (upper panel of Fig. \ref{fig:residuals}) is best modelled by a CARMA(3,0) process according to the minimization of the AICc. There is no evidence for a deviation from a Gaussian CARMA process as the residuals closely follow the expected normal distribution and the sample autocorrelations of the residuals and their square lie well within the 2$\sigma$ interval for all but one time lag. In contrast the AGN with XID 30 (lower panel of Fig. \ref{fig:residuals}), which is best fitted by a CARMA(2,0) process, shows some level of deviation from the expected distribution. Since the histogram of the residuals is significantly narrower than the standard normal, a Gaussian process may not be the best description for this light curve. In fact the light curve indicates weak periodic behaviour which may cause the difference from the normal distribution. However the observed periodicity is likely a coincidence due to the irregular sampling pattern and in fact the PSD of this source does not show any signature of a quasi-periodic oscillation (QPO). Nonetheless the data suggest that the autocorrelation structure is correctly described by the CARMA model for this object.  

In general we observe that the CARMA model performs comparably poor for the light curves of our sample consisting of less than $\sim$40 data points. Additionally, light curves exhibiting a long term trend of rising or falling fluxes that does not turn around within the total length of the observations also show some level of deviations from the Gaussian distribution of the residuals. In the following analysis we exclude sources revealing very strong deviations from a Gaussian white noise process. Finally, we also tested the CARMA model using the magnitude light curves of our AGNs, i.e. modelling the log of the flux. Yet we found that in this case the residuals show stronger deviations from a Gaussian white noise process for a lot more sources than using the flux light curves. Therefore we only present the results obtained with fluxes throughout this work.     

\subsection{Optical PSD shape}
\label{sec:psdshape}

\begin{figure*}
\centering
\subfloat{%
	\includegraphics[width=.48\textwidth]{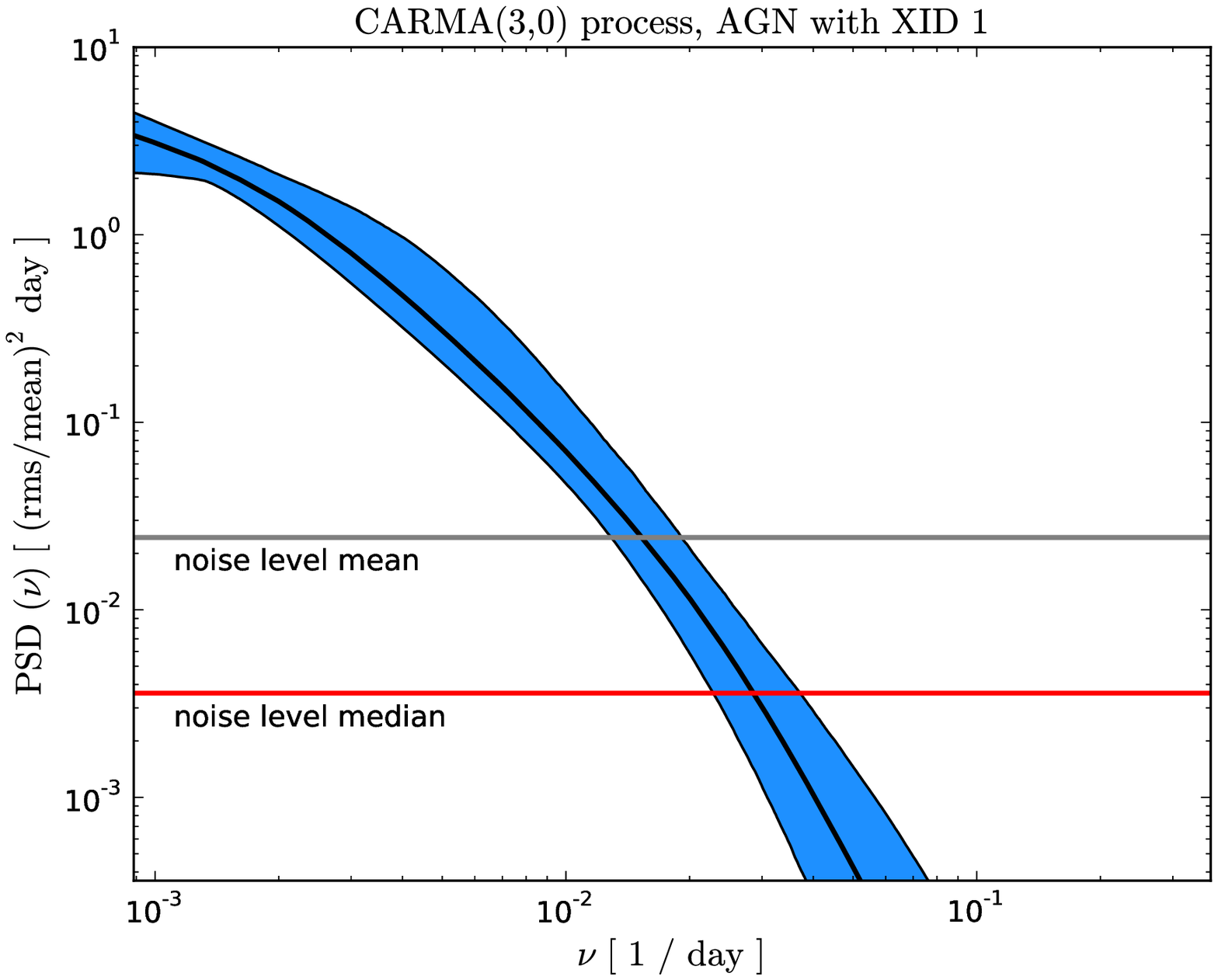}}
\quad
\subfloat{%
	\includegraphics[width=.48\textwidth]{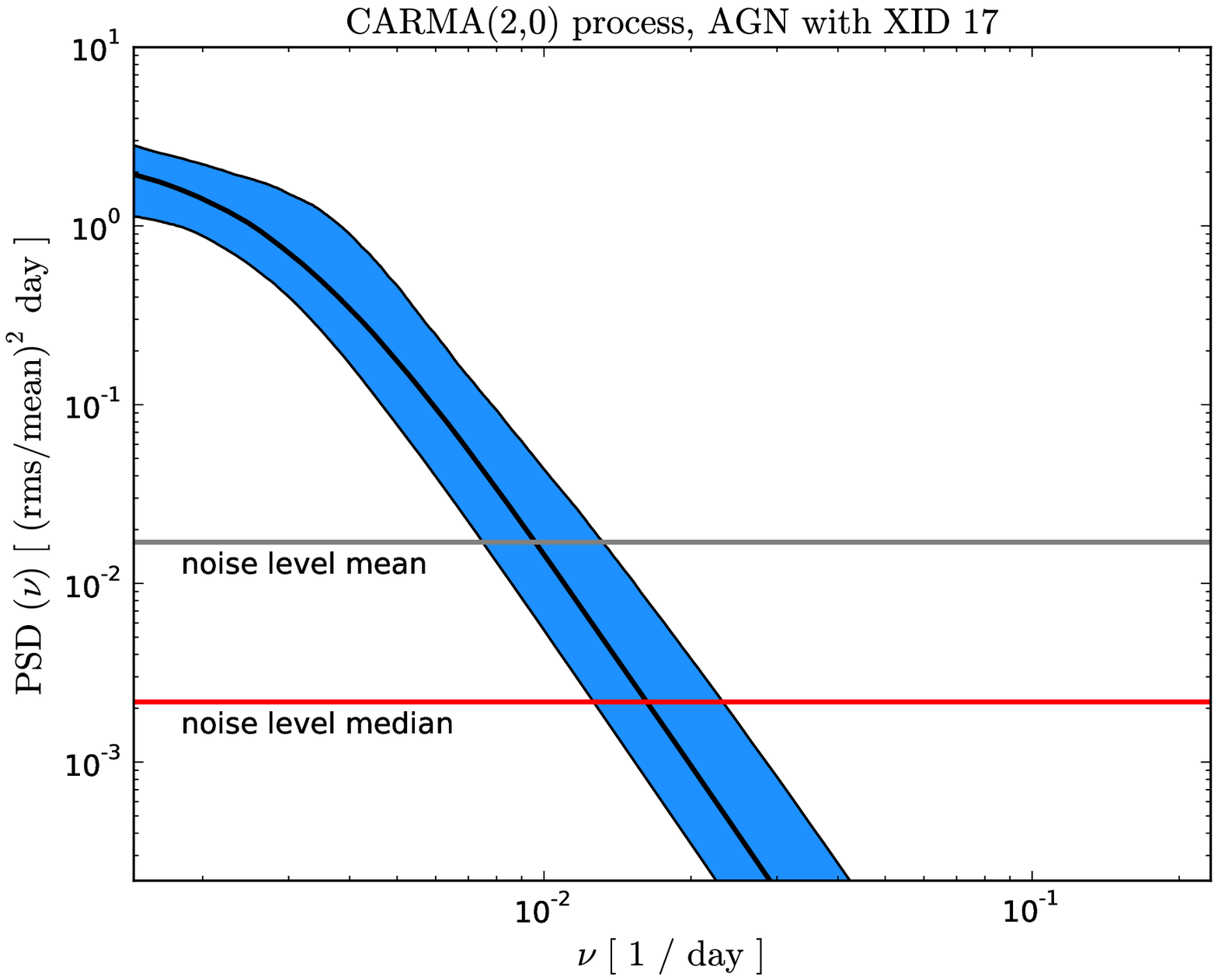}}

\subfloat{%
	\includegraphics[width=.48\textwidth]{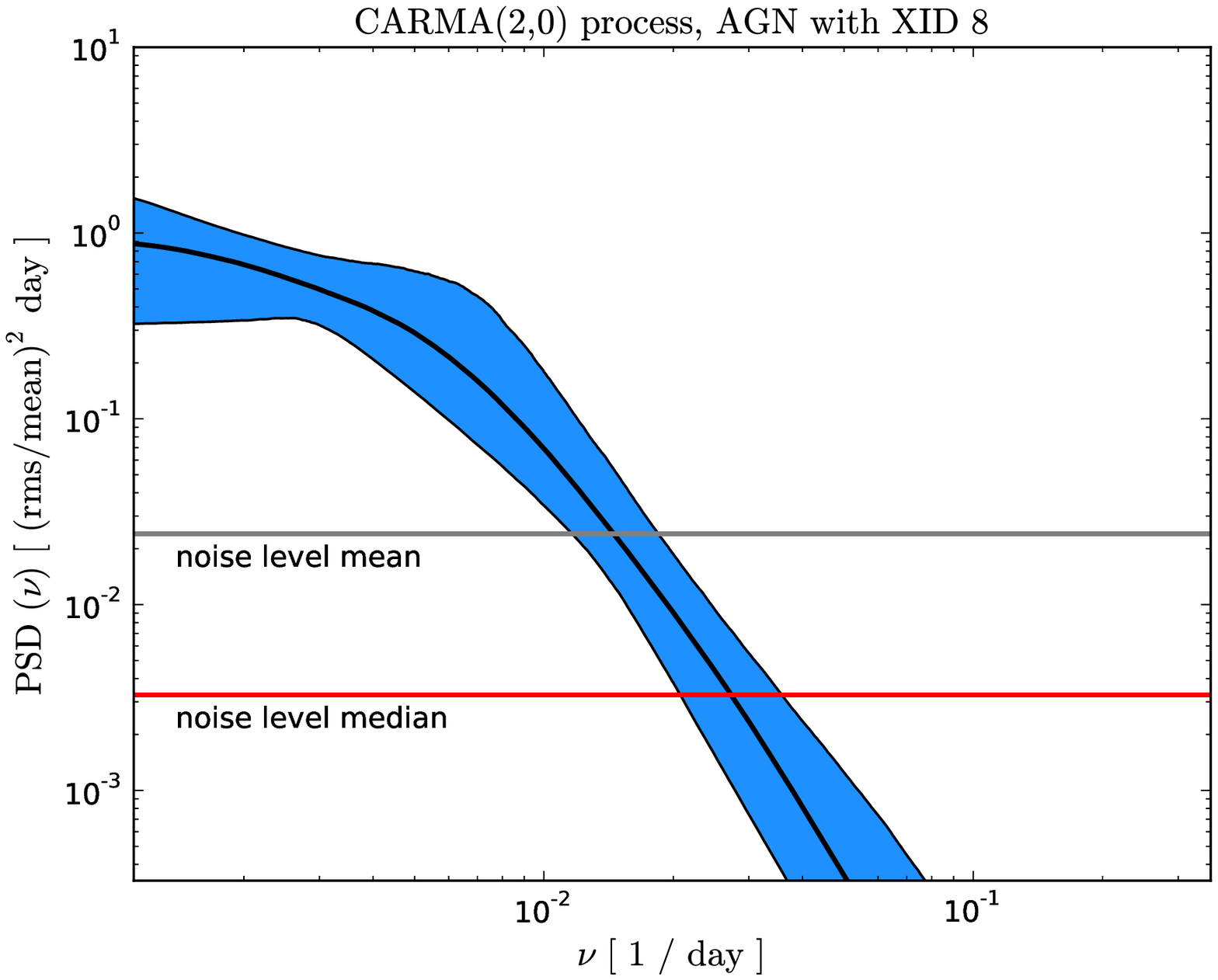}}
\quad
\subfloat{%
	\includegraphics[width=.48\textwidth]{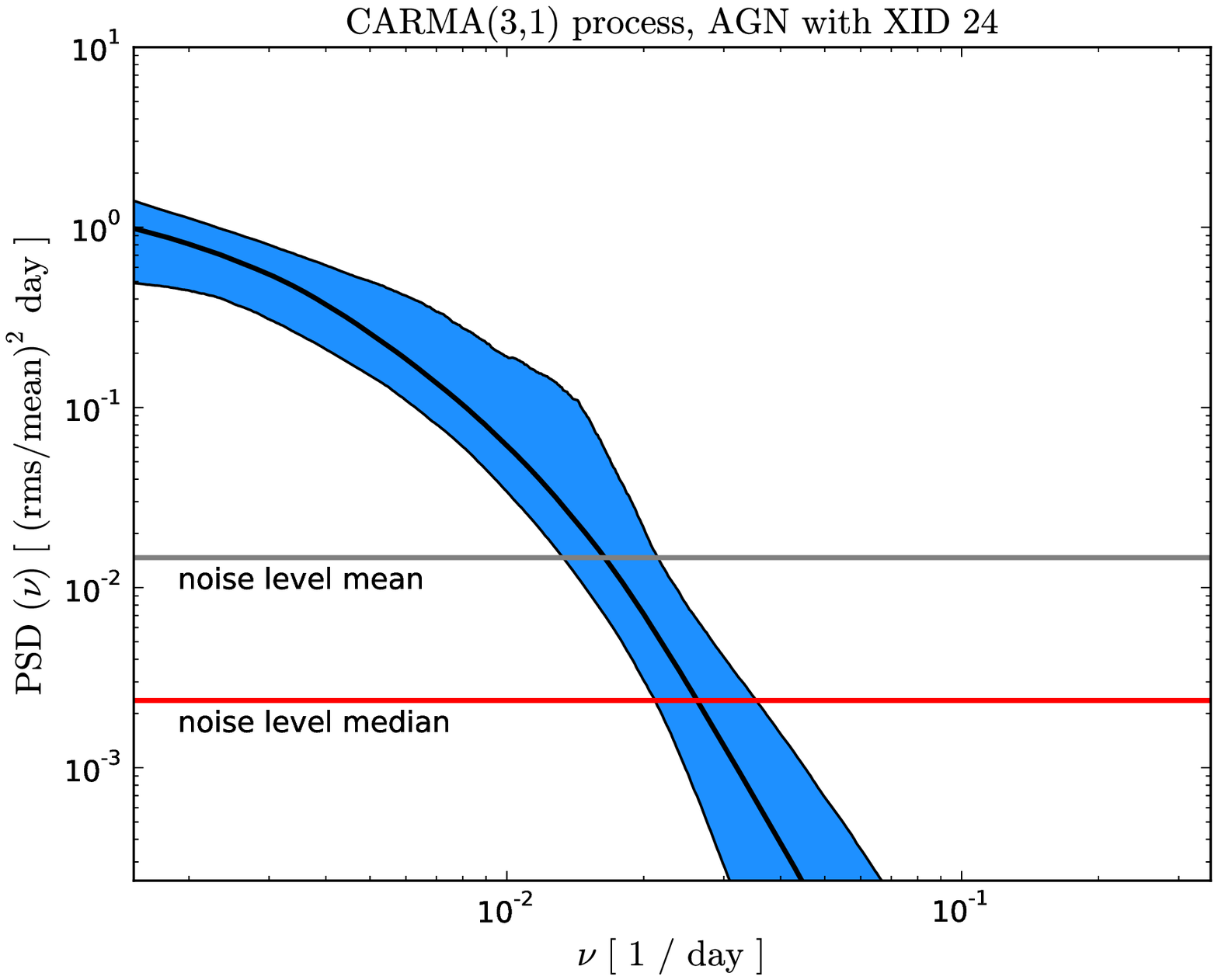}}
\caption{Power spectral densities derived from CARMA model fits to the $g_{\mathrm{P1}}$ band flux light curves for four AGNs of our sample. The solid black line corresponds to the maximum likelihood estimate of the PSD assuming the chosen CARMA model (selected by minimizing the AICc), the blue region shows the 95\% confidence interval. The horizontal lines denote the approximate measurement noise level of the data, estimated by $2\langle\Delta t\rangle\langle\sigma_{\mathrm{y}}^{2}\rangle$ (grey line) and $2median\left(\Delta t\right)median\left(\sigma_{\mathrm{y}}^{2}\right)$ (red line).}
	\label{fig:psdexamples}
\end{figure*}

Following the procedure described in sections \ref{sec:carmafit} and \ref{sec:assessfit} we derived the optical PSDs for the objects of the total sample in four PS1 bands, removing those sources from our sample that exhibit significant deviations from a Gaussian white noise process. Inspecting the modelled PSDs we note that the shape resembles a broken power law for all of our sources. In Fig. \ref{fig:psdexamples} we display four representative $g_{\mathrm{P1}}$ band PSDs of our sample together with the error bounds containing 95\% of the probability on the PSD. Since the modelled PSD should not be evaluated down to arbitrary low variability amplitudes, we show two estimates of the level of measurement noise in our data. The grey line in Fig. \ref{fig:psdexamples} corresponds to the value of $2\langle\Delta t\rangle\langle\sigma_{\mathrm{y}}^{2}\rangle$, where $\langle\Delta t\rangle$ and $\langle\sigma_{\mathrm{y}}^{2}\rangle$ are the average sampling timescale and measurement noise variance. Considering the large gaps between the well sampled segments of our light curves the median may give a better estimate and the red line in Fig. \ref{fig:psdexamples} marks the value of $2median\left(\Delta t\right)median\left(\sigma_{\mathrm{y}}^{2}\right)$, respectively.        

\begin{figure}
	\centering
	\includegraphics[width=.5\textwidth]{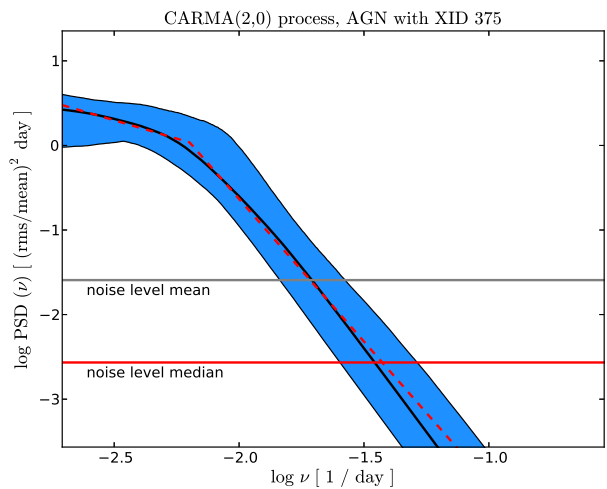}
	\caption{Same as Fig. \ref{fig:psdexamples} in log--space for the AGN with XID 375. The red dashed line is the best-fitting broken power law (equation \ref{eq:psdfit}). Only the values above the red horizontal line were included in the fit.}
	\label{fig:powerlawfit}
\end{figure}

We find that most of our sources are best described by a CARMA(2,0) process (detailed fractions are given below for the final sample we consider during the rest of the paper), meaning that the preferred model PSD is simply given by
\begin{flalign}
	\label{eq:psdcarma20}	
	\mathrm{PSD\left(\nu\right)}=\frac{\sigma^{2}}{|\alpha_{0}+\alpha_{1}\left(2\pi i\nu\right)+\left(2\pi i\nu\right)^{2}|^{2}}
\end{flalign} 
which only depends on the variance of the driving Gaussian white noise process and the first two autoregressive coefficients. One may interpret this PSD in terms of the equivalent expression of a sum of Lorentzian functions, where the roots $r_{k}$ of the autoregressive polynomial determine the widths and centroids of the individual Lorentzians (see \citet{2014ApJ...788...33K} for details). However, in this work we aim to compare our results directly with previous studies parametrizing the PSD as a broken power law of the form  
\begin{flalign}
	\label{eq:psdfit}	
	\text{$\mathrm{PSD\left(\nu\right)}$} =
	\begin{cases}
	\text{$A\left(\frac{\nu}{\nu_{\mathrm{br}}}\right)^{\gamma_{1}}$} & \text{, $\nu\le\nu_{\mathrm{br}}$} \\
	\text{$A\left(\frac{\nu}{\nu_{\mathrm{br}}}\right)^{\gamma_{2}}$} & \text{, $\nu>\nu_{\mathrm{br}}$} 
	\end{cases}
\end{flalign} 
with some amplitude $A$, the break frequency $\nu_{\mathrm{br}}$, a low frequency slope $\gamma_{1}$ and a high frequency slope $\gamma_{2}$. We fit this model to our derived PSDs using the Levenberg-Marquardt-Algorithm. For some of our objects the uncertainties on the PSD are so large that the broken power law fit is very poorly defined. Therefore we visually inspect every power law fit and remove the sources from our sample where the fit failed completely or is of low quality. During the fitting process we only consider the values above the noise level $2median\left(\Delta t\right)median\left(\sigma_{\mathrm{y}}^{2}\right)$. In this way we were able to determine the parameters of equation \ref{eq:psdfit} with acceptable quality for 156 ($g_{\mathrm{P1}}$), 144 ($r_{\mathrm{P1}}$), 124 ($i_{\mathrm{P1}}$), 93 ($z_{\mathrm{P1}}$) sources of the total sample and in the following we refer to this sample as the "PSD sample". For reference we show one of these model fits as red dashed line in Fig. \ref{fig:powerlawfit} for the AGN with XID 375. 

Considering the chosen model order we find that 72\% ($g_{\mathrm{P1}}$), 78\% ($r_{\mathrm{P1}}$), 70\% ($i_{\mathrm{P1}}$), 65\% ($z_{\mathrm{P1}}$) of the AGNs of the PSD sample are best fitted by a CARMA(2,0) process. This may explain why many researchers found that the next simpler model of a CARMA(1,0) process, corresponding to a "damped random walk", provides a very accurate description of optical AGN light curves \citep{2009ApJ...698..895K,2010ApJ...708..927K,2010ApJ...721.1014M,2013A&A...554A.137A}. For 23\% ($g_{\mathrm{P1}}$), 20\% ($r_{\mathrm{P1}}$), 23\% ($i_{\mathrm{P1}}$), 30\% ($z_{\mathrm{P1}}$) the order (3,0) minimized the AICc and the few residual sources of the PSD sample are best described by higher orders of e.g. (3,1), (3,2), (4,1) or even (6,0). 

Among the objects of the PSD sample 89 ($g_{\mathrm{P1}}$), 79 ($r_{\mathrm{P1}}$), 72 ($i_{\mathrm{P1}}$), 55 ($z_{\mathrm{P1}}$) have known black hole masses, bolometric luminosities and Eddington ratios, hereafter termed "PSD\_MBH sample". Fig. \ref{fig:flowchart} summarizes these two samples which are used throughout the PSD analysis.

\begin{figure}
\centering
\subfloat{%
	\includegraphics[width=0.45\textwidth]{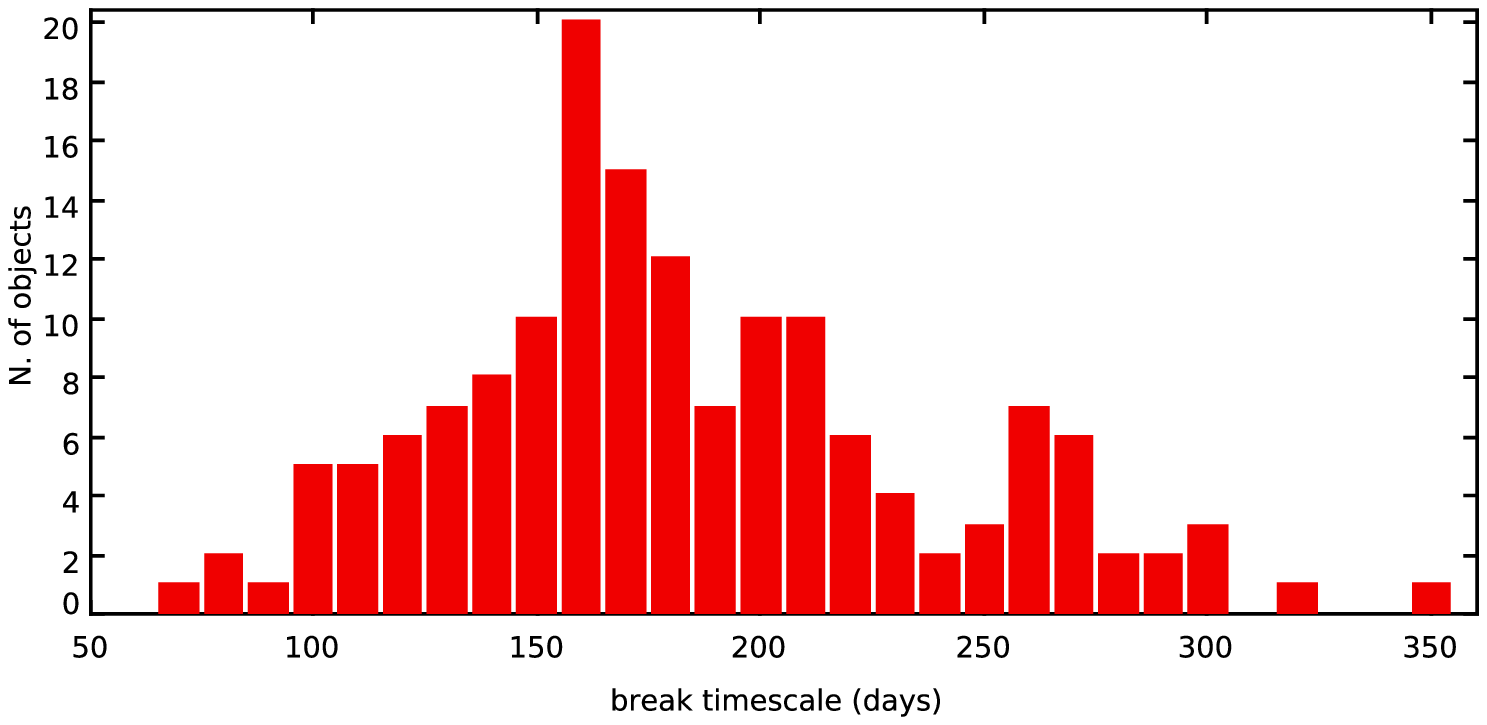}}
	
\subfloat{%
	\includegraphics[width=0.45\textwidth]{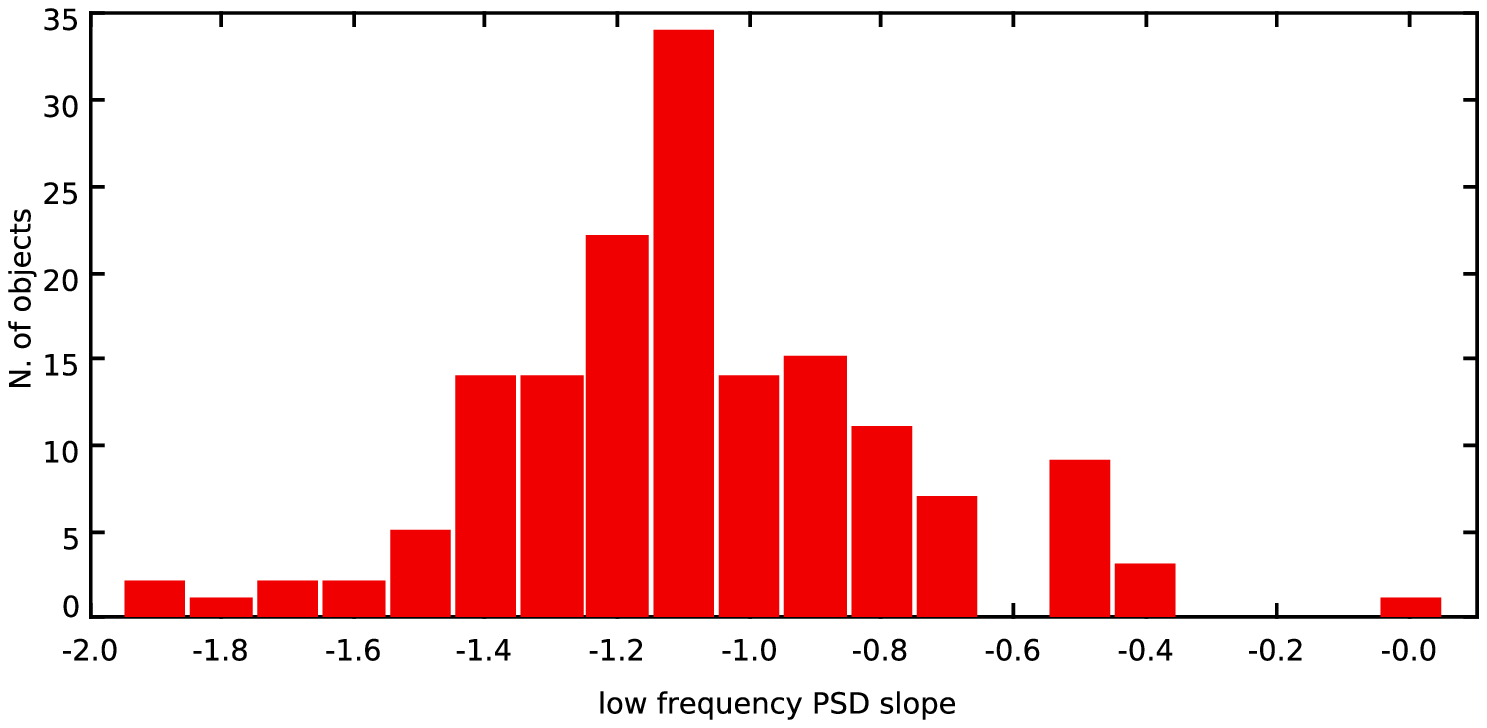}}

\subfloat{%
	\includegraphics[width=0.45\textwidth]{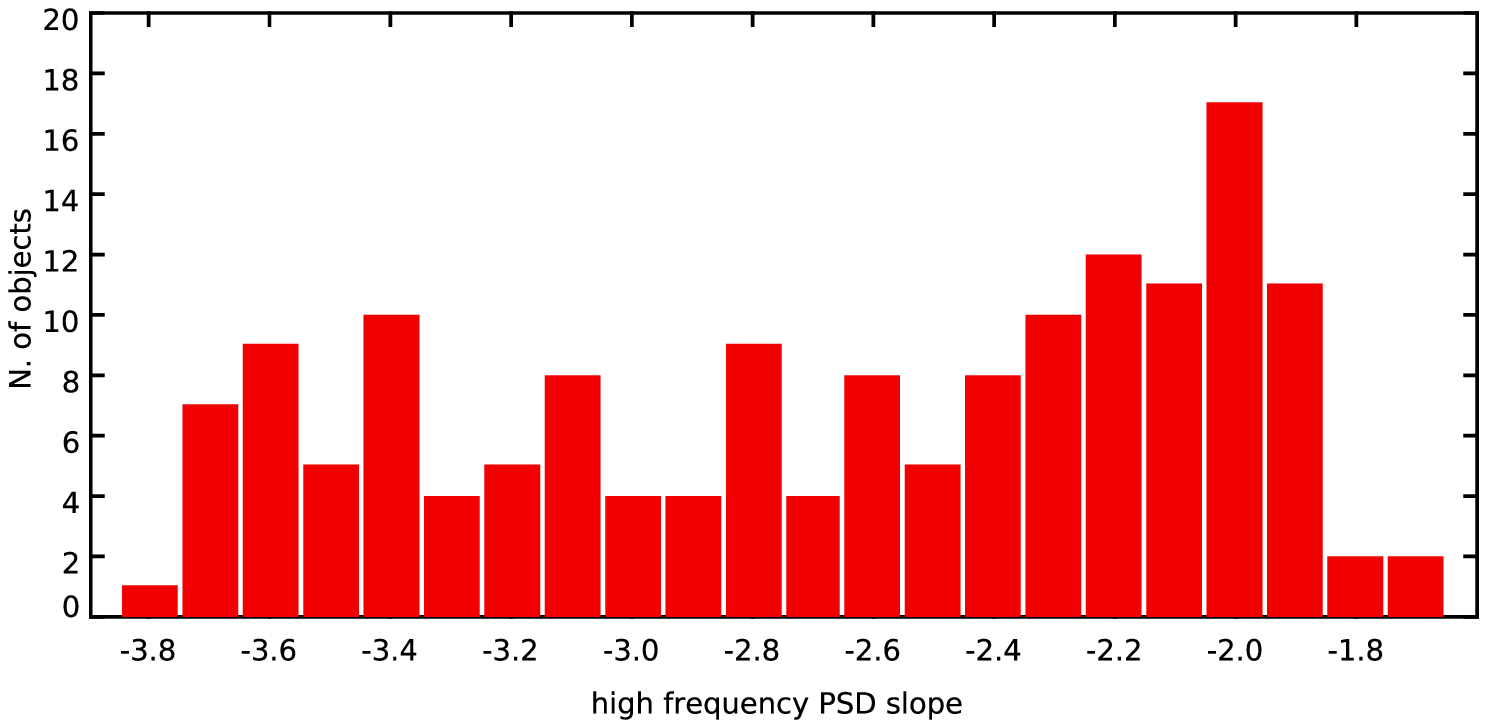}}
\caption{Distributions of the fitted break timescale (\textit{top panel}), the low frequency PSD slope $\gamma_{1}$ (\textit{middle panel}) and the high frequency PSD slope $\gamma_{2}$ (\textit{bottom panel}). The data of the PSD sample, obtained with the $g_{\mathrm{P1}}$ band flux light curves, are shown.}
\label{fig:psdshapehist}
\end{figure}

In Fig. \ref{fig:psdshapehist} we present the distributions of the break timescale $T_{\mathrm{br}}=1/\nu_{\mathrm{br}}$, the low frequency slope $\gamma_{1}$ and the high frequency slope $\gamma_{2}$ for the $g_{\mathrm{P1}}$ band objects of the PSD sample. As it can be seen $T_{\mathrm{br}}$ exhibits a distribution of timescales ranging from about $\sim$100 days to $\sim$300 days with a mean value of 175 days. We note that very similar characteristic timescales for optical quasar light curves have been reported by researchers using the "damped random walk" (DRW) model \citep{2009ApJ...698..895K,2010ApJ...721.1014M}. However, the range of our $T_{\mathrm{br}}$ values is quite narrow, whereas the aforementioned authors also observed characteristic timescales of several 10 days and several years for their objects. In addition we find that the low frequency slope $\gamma_{1}$ is close to a value of $-1$ for most of our sources. The sample average is $-1.08$ ($g_{\mathrm{P1}}$), $-1.11$ ($r_{\mathrm{P1}}$), $-1.17$ ($i_{\mathrm{P1}}$), $-1.21$ ($z_{\mathrm{P1}}$) with a sample standard deviation of 0.31 ($g_{\mathrm{P1}}$), 0.32 ($r_{\mathrm{P1}}$), 0.37 ($i_{\mathrm{P1}}$), 0.33 ($z_{\mathrm{P1}}$). However, in order for the total variability power to stay finite, there must be a second break at lower frequencies after which the PSD flattens to $\gamma_{1}=0$. In fact a flat low frequency PSD is still possible within the $2\sigma$ or $3\sigma$ regions of the maximum likelihood PSD for many of our objects. Furthermore we observe a wide range of high frequency slopes $\gamma_{2}$ showing no clear preference with values between $\sim -2$ and $\sim -4$. This result suggests that optical PSDs of AGNs fall off considerably steeper than the corresponding X-ray PSDs at high frequencies which are typically characterized by a slope of $-2$. This is in accord with recent results obtained with high quality optical Kepler light curves, yielding high frequency slopes of $-2.5$, $-3$ or even $-4$ \citep{2011ApJ...743L..12M,2014ApJ...795....2E,2015MNRAS.451.4328K}. What is more, the distributions of $\gamma_{1}$ and $\gamma_{2}$ reveal significant deviations from the simple DRW model, which is characterized by a flat PSD at low frequencies and a slope of $-2$ at high frequencies. We emphasize that we fit very similar parameters of the broken power law in all four studied PS1 bands. This is consistent with the fact that our light curves vary approximately simultaneous in all PS1 bands with time lags of few days at most. The latter result is supported by a cross-correlation function (CCF) analysis we performed with our light curves using the standard interpolation CCF method \citep{1986ApJ...305..175G,1994PASP..106..879W}. A comparison of the fitted parameters in the different PS1 bands is depicted in appendix \ref{sec:appendixb}.   

\subsection{Comparison of $\sigma_{\mathrm{rms}}^{2}$ and the integrated PSD}
\label{sec:rmsvscarma}

Given that the excess variance is defined to measure the integral of the PSD over the frequency range covered by a light curve (see equation \ref{eq:psdint}), it is interesting to compare the $\sigma_{\mathrm{rms}}^{2}$ measurement with the value of the integrated CARMA PSD for each object. This allows for a consistency test of the two variability methods. 

\begin{figure}
\centering
\includegraphics[width=0.50\textwidth]{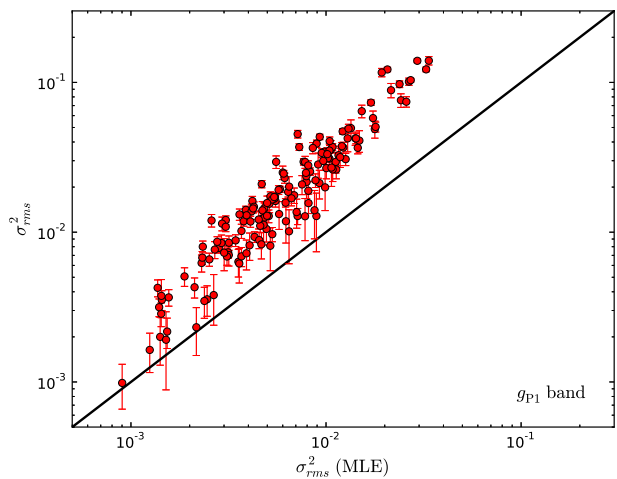}
\caption{Comparison of the integral of the maximum likelihood estimate of the PSD, $\sigma_{\mathrm{rms}}^{2}(\mathrm{MLE})$, with the excess variance $\sigma_{\mathrm{rms}}^{2}$ calculated after equation \ref{eq:nev}. The data for the $g_{\mathrm{P1}}$ band PSD sample are shown. The black line corresponds to the one to one relation.}  
\label{fig:rmsvscarma}
\end{figure}

We perform the integration of each rest-frame PSD within the limits $\nu_{\mathrm{min}}=1/T$ and $\nu_{\mathrm{max}}=1/\left(2median\left(\Delta t\right)\right)$, where $T$ is the rest-frame light curve length and $median\left(\Delta t\right)$ the median rest-frame sampling timescale, respectively. First of all we checked that integrating the maximum likelihood estimate of the PSD (black curve in Fig. \ref{fig:powerlawfit}) and the fitted broken power law PSD (red dashed curve in Fig. \ref{fig:powerlawfit}) yield consistent results. Denoting the integral of the maximum likelihood estimate of the PSD by $\sigma_{\mathrm{rms}}^{2}(\mathrm{MLE})$ and the integral of the fitted broken power law PSD by $\sigma_{\mathrm{rms}}^{2}(\mathrm{FIT})$, we find an average value of $\langle\sigma_{\mathrm{rms}}^{2}(\mathrm{FIT})-\sigma_{\mathrm{rms}}^{2}(\mathrm{MLE})\rangle=1.6\cdot 10^{-4}$ with a standard deviation of $1.2\cdot 10^{-4}$ for the 156 sources of the $g_{\mathrm{P1}}$ band PSD sample. In contrast, as displayed in Fig. \ref{fig:rmsvscarma}, for the same objects there is a systematic offset upwards the one to one relation with a slight tilt with respect to the latter comparing $\sigma_{\mathrm{rms}}^{2}(\mathrm{MLE})$ with the excess variance ($\sigma_{\mathrm{rms}}^{2}$) calculated after equation \ref{eq:nev}. We observe that $\sigma_{\mathrm{rms}}^{2}$ is on average a factor of $\sim$2--3 larger than $\sigma_{\mathrm{rms}}^{2}(\mathrm{MLE})$ for our sources. 

Part of this difference may be explained by noting that the CARMA model light curve fits tend to omit outlier measurements in our light curves (see e.g. Fig. \ref{fig:residuals}), whereas all outliers contribute to the value of the excess variance. Moreover, as shown by \citet{2013ApJ...771....9A}, $\sigma_{\mathrm{rms}}^{2}$ is a biased estimator of the intrinsic normalized source variance. The authors observed that an excess variance measurement of sparsely sampled light curves differs from the intrinsic normalized variance by a bias factor of 1.2, 1.0, 0.6, 0.3, 0.14 for an underlying PSD with a power law slope of $-1$, $-1.5$, $-2$, $-2.5$, $-3$, respectively\footnote{The bias factor is defined by $b=\sigma_{\mathrm{band,norm}}^{2}/\langle\sigma_{\mathrm{rms}}^{2}\rangle$, with the intrinsic "band" normalized variance $\sigma_{\mathrm{band,norm}}^{2}$ (see equation 4 in \citet{2013ApJ...771....9A}). The average value $\langle\sigma_{\mathrm{rms}}^{2}\rangle$ is calculated from "observing" 5000 simulated light curves sampled from the underlying PSD. Therefore multiplying $\sigma_{\mathrm{rms}}^{2}$ with $b$ yields (on average) the unbiased estimate.}. However, they did not study the case of a broken power law PSD. Considering the fact that all of our sources exhibit a broken power law PSD with a low frequency slope of $\sim -1$ and a high frequency slope ranging between $-2$ and $-4$, one may expect an average bias factor somewhere between $\sim0.3$--1.0 for a $\sigma_{\mathrm{rms}}^{2}$ value that is integrating the bend of the PSD for sparsely sampled light curves. This may be another reason for the factor of $\sim 2$--3 difference between our $\sigma_{\mathrm{rms}}^{2}$ measurements and the values suggested by the CARMA PSDs. Finally, we point out that integrating the curve corresponding to the 2$\sigma$ upper error bound of the PSD increases the integral by a factor of $\sim$2 on average. Therefore the excess variance, which does not rely on any statistical property of the light curve, and the rather complex CARMA method yield consistent variability measurements at least within the 95\% error on the PSD.   
 
\subsection{Scaling of the optical break frequency}
\label{sec:nubreak}

Considering the shape of the optical PSD reported in the previous section, the break frequency is the most characteristic feature as it separates two very different variability regimes. For this reason it may be possible to gain insight into the physical system at work, if this characteristic frequency scales with fundamental AGN physical properties. 

\begin{figure}
\centering
\subfloat{%
	\includegraphics[width=0.50\textwidth]{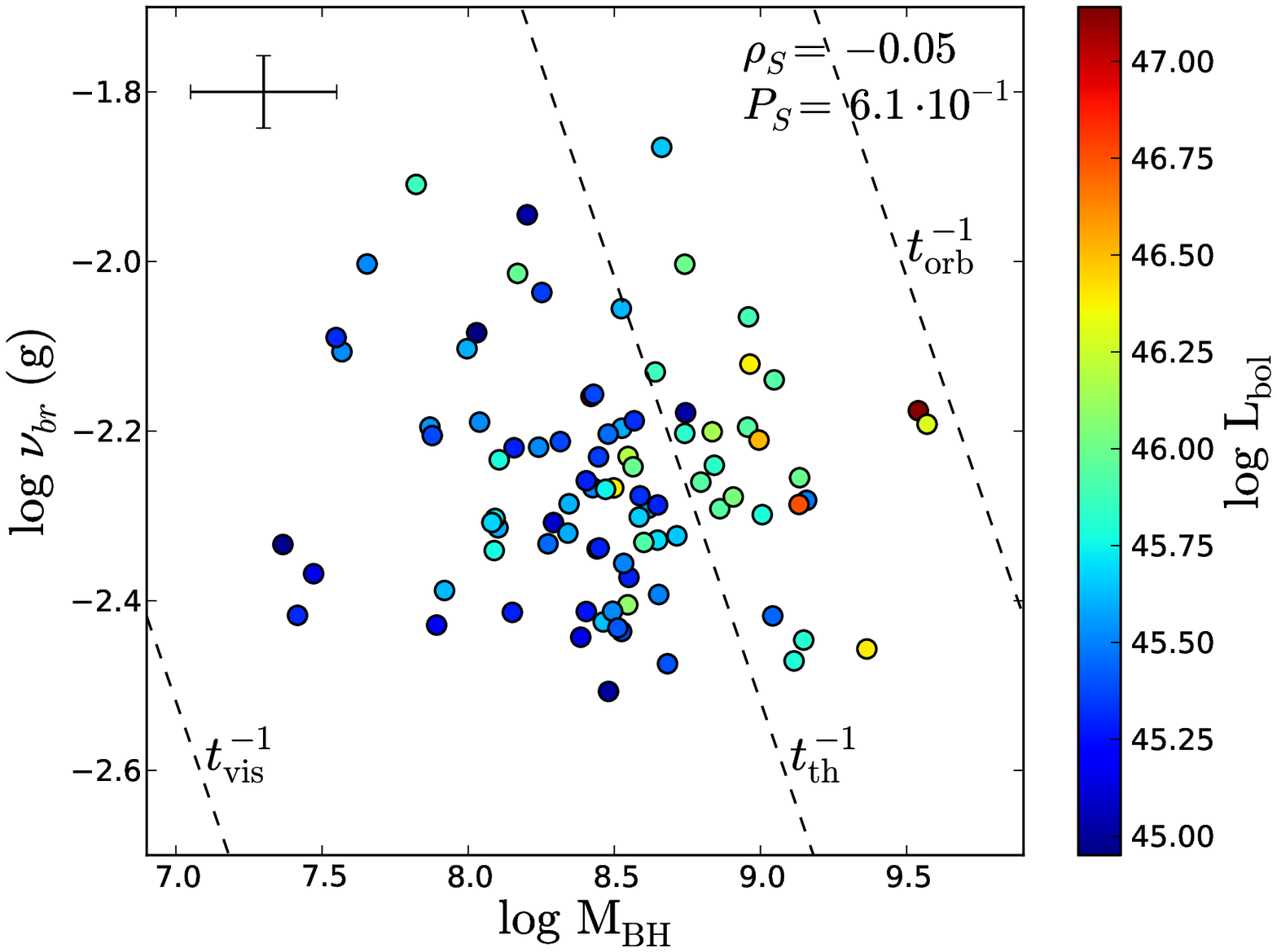}}
		
\subfloat{%
	\includegraphics[width=0.50\textwidth]{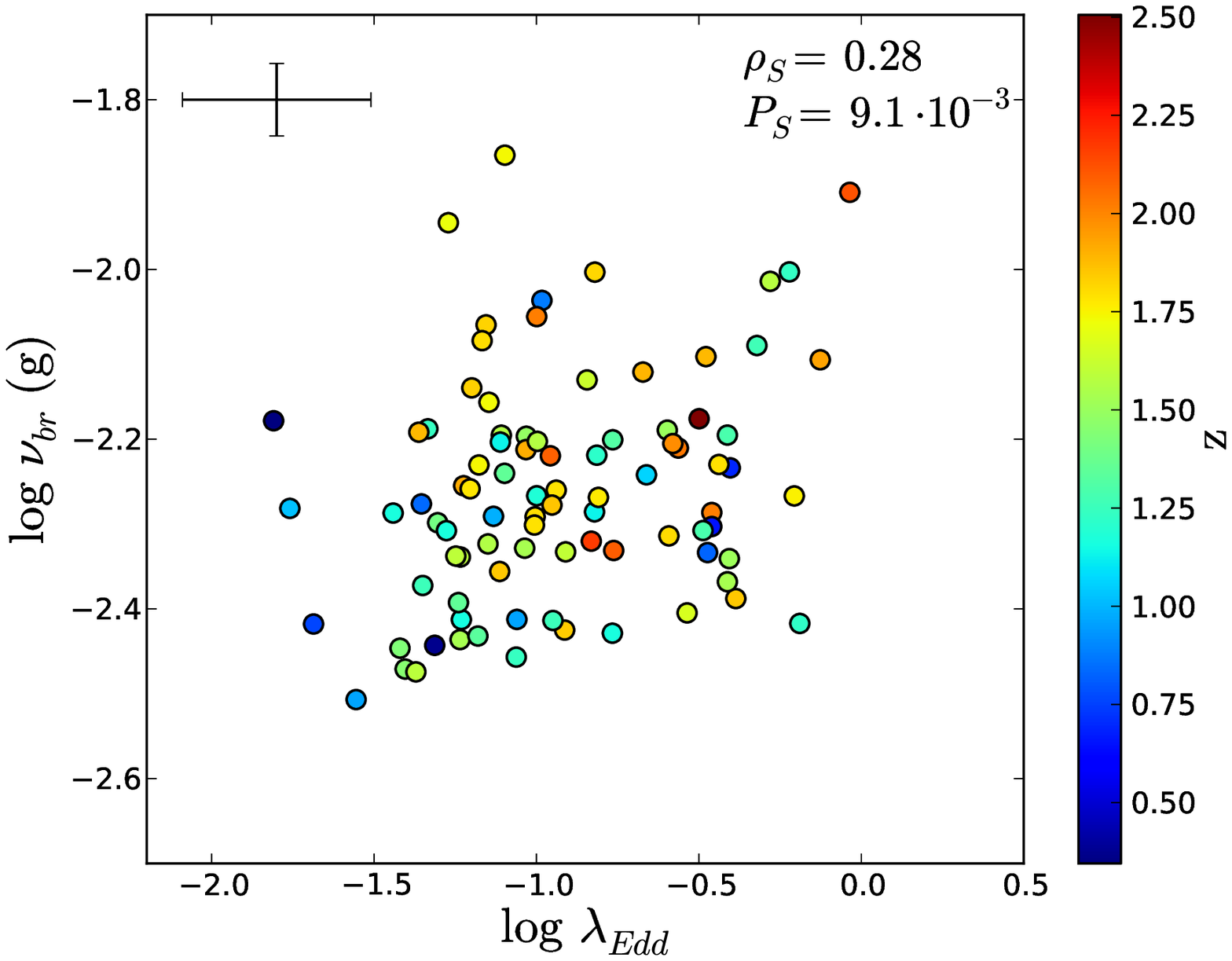}}
\caption{Optical break frequency ($g_{\mathrm{P1}}$ band PSD\_MBH sample) versus $M_{\mathrm{BH}}$, color coded with $L_{\mathrm{bol}}$ (\textit{top}) and $\lambda_{\mathrm{Edd}}$, color coded with redshift (\textit{bottom}). The black error bars are the average values. There is no significant evidence for a correlation with these AGN parameters. The dashed lines in the top panel correspond to the expected scaling of the orbital, thermal and viscous timescales at 10$R_{\mathrm{S}}$, see text for details.}
\label{fig:nubrvsphys}
\end{figure}

Surprisingly we do not find a statistically significant correlation of the measured break frequencies with any of the AGN parameters for the PSD\_MBH sample. The Spearman rank order correlation coefficients of $\nu_{\mathrm{br}}$ and $M_{\mathrm{BH}}$ are $-0.05$ ($g_{\mathrm{P1}}$), $-0.24$ ($r_{\mathrm{P1}}$), $-0.08$ ($i_{\mathrm{P1}}$), $-0.02$ ($z_{\mathrm{P1}}$) with p-values of $0.61$ ($g_{\mathrm{P1}}$), 0.03 ($r_{\mathrm{P1}}$), 0.52 ($i_{\mathrm{P1}}$), 0.91 ($z_{\mathrm{P1}}$). Similarly correlating $\nu_{\mathrm{br}}$ and $L_{\mathrm{bol}}$ gives $\rho_{\mathrm{S}}$ values of 0.23 ($g_{\mathrm{P1}}$), 0.06 ($r_{\mathrm{P1}}$), 0.10 ($i_{\mathrm{P1}}$), 0.12 ($z_{\mathrm{P1}}$) with p-values of 0.03 ($g_{\mathrm{P1}}$), 0.61 ($r_{\mathrm{P1}}$), 0.39 ($i_{\mathrm{P1}}$), 0.38 ($z_{\mathrm{P1}}$). Albeit the "blue" bands exhibit some evidence for a positive correlation between $\nu_{\mathrm{br}}$ and $\lambda_{\mathrm{Edd}}$ with large scatter, the correlation is not significant and not present considering the other bands with  $\rho_{\mathrm{S}}$ values of 0.28 ($g_{\mathrm{P1}}$), 0.28 ($r_{\mathrm{P1}}$), 0.14 ($i_{\mathrm{P1}}$), 0.20 ($z_{\mathrm{P1}}$) with p-values of $9.1\cdot 10^{-3}$ ($g_{\mathrm{P1}}$), 0.01 ($r_{\mathrm{P1}}$), 0.25 ($i_{\mathrm{P1}}$), 0.15 ($z_{\mathrm{P1}}$). In Fig. \ref{fig:nubrvsphys} we plot the break frequency against these AGN parameters for the $g_{\mathrm{P1}}$ band PSD\_MBH sample. Even though it is possible that there might be a hidden correlation within the large uncertainties of the involved quantities, the results obtained with the four PS1 bands suggest that such a correlation must be rather weak. 

These findings are at odds with previous variability studies. For example it is well known that $\nu_{\mathrm{br}}$ scales inversely with $M_{\mathrm{BH}}$ and may also be linearly correlated with $\lambda_{\mathrm{Edd}}$ in the X-ray bands \citep{2006Natur.444..730M,2012A&A...544A..80G}. Furthermore optical variability investigations found evidence that the characteristic timescale of the DRW model is correlated with $M_{\mathrm{BH}}$ and luminosity \citep{2009ApJ...698..895K,2010ApJ...721.1014M}. Finally, if the break timescale is associated with a characteristic physical timescale of the system, we would expect a positive correlation with $M_{\mathrm{BH}}$. This follows from the fact that relevant timescales like the light crossing time, the gas orbital timescale as well as the thermal and viscous timescales of the accretion disc all increase with $M_{\mathrm{BH}}$ \citep[see e.g.][]{1988PASP..100..427T}. For reference the dashed lines in the top panel of Fig. \ref{fig:nubrvsphys} show the frequency scaling with $M_{\mathrm{BH}}$ for the orbital $t_{\mathrm{orb}}\sim 3.3\left(R/10R_{\mathrm{S}}\right)^{3/2}(M_{\mathrm{BH}}/10^{8}M_{\odot})$, thermal $t_{\mathrm{th}}=\alpha^{-1}t_{\mathrm{orb}}$ and viscous $t_{\mathrm{vis}}=(H/R)^{2}t_{\mathrm{th}}$ timescale assuming a viscosity parameter of $\alpha=0.1$ and a disc scale height to radius ratio of $H/R=0.1$ at a distance of 10$R_{\mathrm{S}}$, where $R_{\mathrm{S}}$ denotes the Schwarzschild radius. Although the derived $T_{\mathrm{br}}$ values of our sample seem to be uncorrelated with $M_{\mathrm{BH}}$, the magnitude of the timescales are roughly consistent with $t_{\mathrm{th}}$ at some 10$R_{\mathrm{S}}$ (remember that the $g_{\mathrm{P1}}$ band data shown in Fig. \ref{fig:nubrvsphys} are actually rest-frame UV data for the majority of objects). We stress however that the parameter space of our sample covers only a small range in frequencies and it may be possible that a correlation appears for a much larger sample of objects spanning a wide range of values.    

\subsection{Scaling of the optical PSD amplitude}
\label{sec:psdamp}

Another important characteristic of the PSD is its normalization, stating the amplitude of the PSD for each source. We test for correlations of the PSD amplitude, which is given by $\mathrm{PSD_{amp}}=A\nu_{\mathrm{br}}$, with the fundamental AGN parameters $M_{\mathrm{BH}}$, $L_{\mathrm{bol}}$ and $\lambda_{\mathrm{Edd}}$ using the PSD\_MBH sample. As was observed for our excess variance measurements, there is no significant correlation between the variability amplitude and the black hole mass. Spearman's r for $\mathrm{PSD_{amp}}$ and $M_{\mathrm{BH}}$ reads $-0.02$ ($g_{\mathrm{P1}}$), $0.07$ ($r_{\mathrm{P1}}$), $0.08$ ($i_{\mathrm{P1}}$), $0.00$ ($z_{\mathrm{P1}}$) with p-values of $0.82$ ($g_{\mathrm{P1}}$), 0.56 ($r_{\mathrm{P1}}$), 0.50 ($i_{\mathrm{P1}}$), 1.0 ($z_{\mathrm{P1}}$). However, we find very significant evidence that $\mathrm{PSD_{amp}}$ is anti-correlated with $L_{\mathrm{bol}}$ and $\lambda_{\mathrm{Edd}}$ for all of the four studied PS1 bands. The $\rho_{\mathrm{S}}$ and $P_{\mathrm{S}}$ values of $\mathrm{PSD_{amp}}$ and $L_{\mathrm{bol}}$ are $-0.39$ ($g_{\mathrm{P1}}$), $-0.41$ ($r_{\mathrm{P1}}$), $-0.47$ ($i_{\mathrm{P1}}$), $-0.41$ ($z_{\mathrm{P1}}$) and $1.3\cdot 10^{-4}$ ($g_{\mathrm{P1}}$), $1.8\cdot 10^{-4}$ ($r_{\mathrm{P1}}$), $3.2\cdot 10^{-5}$ ($i_{\mathrm{P1}}$), $2.1\cdot 10^{-3}$ ($z_{\mathrm{P1}}$), respectively. The anti-correlation is even more significant for $\mathrm{PSD_{amp}}$ and $\lambda_{\mathrm{Edd}}$ with $\rho_{\mathrm{S}}$ values of $-0.39$ ($g_{\mathrm{P1}}$), $-0.45$ ($r_{\mathrm{P1}}$), $-0.50$ ($i_{\mathrm{P1}}$), $-0.45$ ($z_{\mathrm{P1}}$) and p-values of $1.5\cdot 10^{-4}$ ($g_{\mathrm{P1}}$), $2.7\cdot 10^{-5}$ ($r_{\mathrm{P1}}$), $7.5\cdot 10^{-6}$ ($i_{\mathrm{P1}}$), $6.5\cdot 10^{-4}$ ($z_{\mathrm{P1}}$). We point out that these results represent an entirely independent verification of the correlations we found using the excess variance as variability estimator.
  
\begin{figure}
\centering
\subfloat{%
	\includegraphics[width=0.50\textwidth]{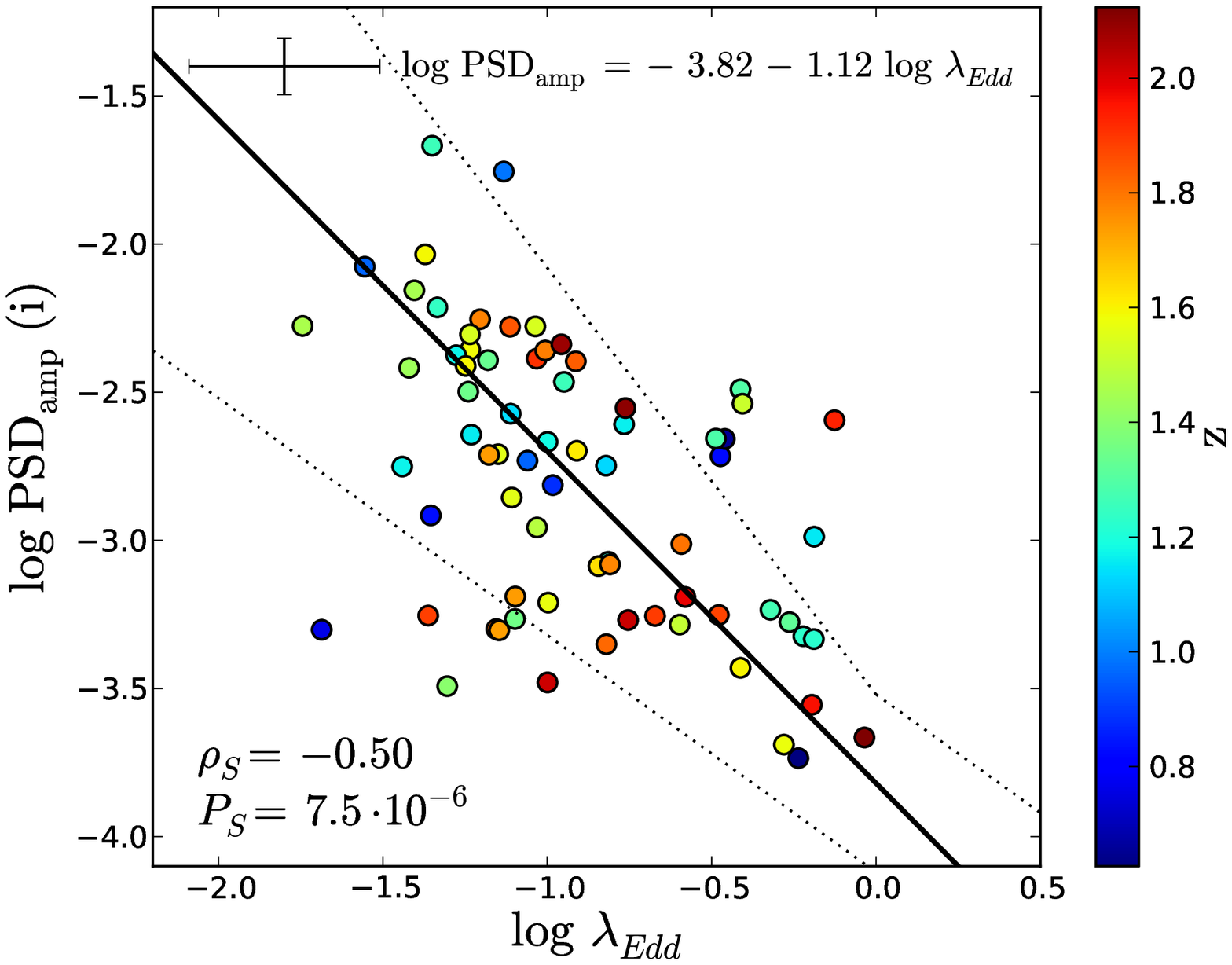}}
		
\subfloat{%
	\includegraphics[width=0.50\textwidth]{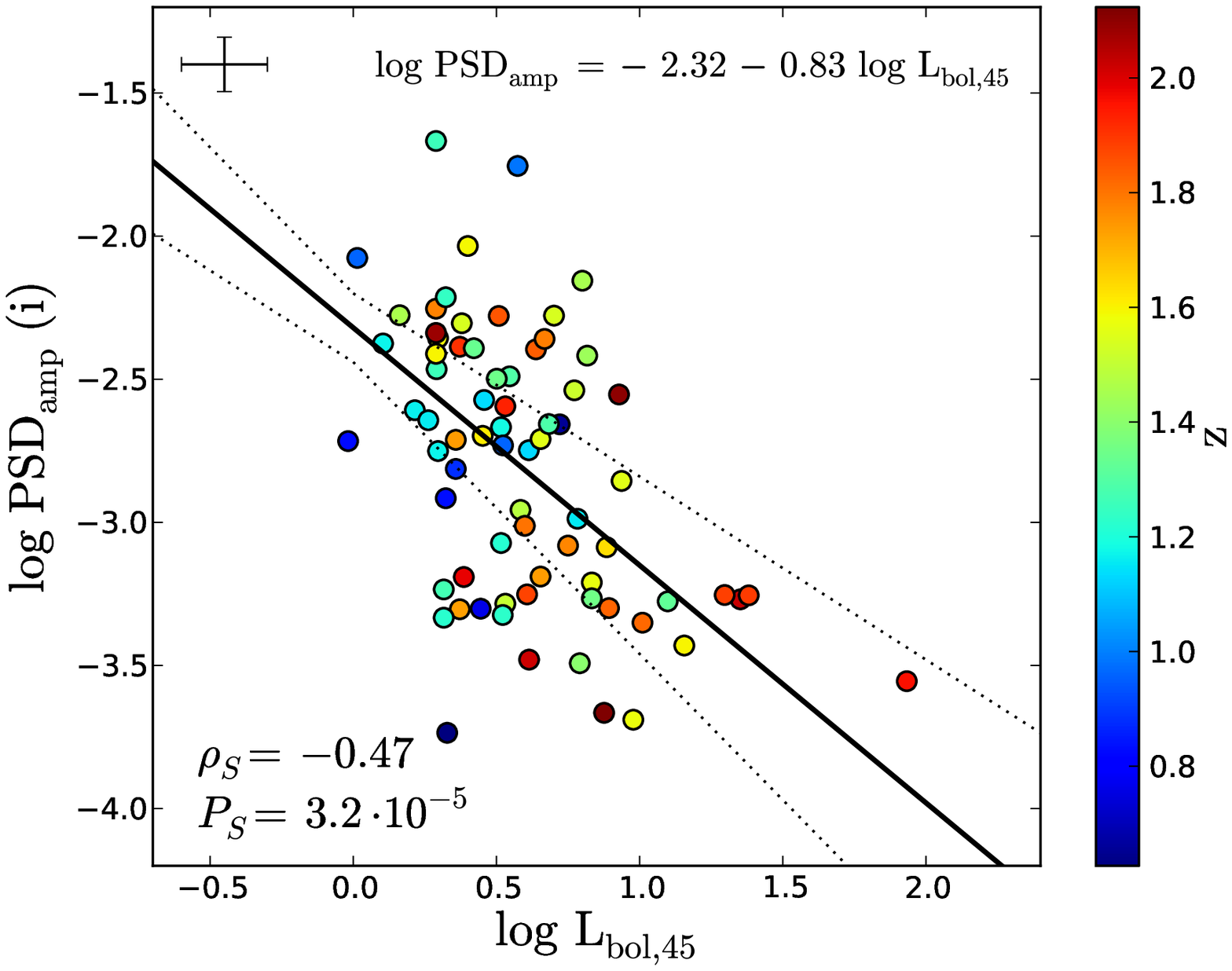}}
\caption{PSD amplitude ($i_{\mathrm{P1}}$ band PSD\_MBH sample) versus $\lambda_{\mathrm{Edd}}$ (\textit{top}) and $L_{\mathrm{bol}}$ (\textit{bottom}). The redshift is given as a color bar. The best power law fit and other symbols are displayed as in Fig. \ref{fig:rmsvslbol}.}
\label{fig:psdampvsphys}
\end{figure}

In the same way as done for the excess variance, we perform a linear regression fit of the form $\log\mathrm{PSD_{amp}}=\beta+\alpha\log x+\epsilon$ for each PS1 band with $x=L_{\mathrm{bol}},\lambda_{\mathrm{Edd}}$. The fitted values of the slope, zeropoint and intrinsic scatter are summarized in table \ref{tab:fitpsdamp}. The linear regressions obtained for the $i_{\mathrm{P1}}$ band are displayed in Fig. \ref{fig:psdampvsphys} together with the data. We note that our data suffer from few fatal outliers, showing significant deviations from the bulk of the data points. These are preferentially associated with high redshift sources that may have low quality $L_{\mathrm{bol}}$ and $\lambda_{\mathrm{Edd}}$ measurements or with objects whose residuals indicate some level of deviations from a Gaussian white noise process. However we found that the presence of these few fatal outliers generally causes the slope of our fitted correlations to flatten. Therefore the slopes of the $g_{\mathrm{P1}}$, $r_{\mathrm{P1}}$ and $z_{\mathrm{P1}}$ band relations of $\mathrm{PSD_{amp}}$ and $L_{\mathrm{bol}}$ listed in table \ref{tab:fitpsdamp} are considerably shallower than the $i_{\mathrm{P1}}$ band slope, because the data of the latter are less affected by outliers. This is also true for the slopes of the $g_{\mathrm{P1}}$ and $r_{\mathrm{P1}}$ band considering the scaling of $\mathrm{PSD_{amp}}$ with $\lambda_{\mathrm{Edd}}$. Once we remove the few fatal outliers from our sample the fitted slopes are more similar for the different PS1 bands. In detail the slopes then read $-0.45\pm 0.13$ ($g_{\mathrm{P1}}$), $-0.73\pm 0.18$ ($r_{\mathrm{P1}}$), $-1.01\pm 0.23$ ($i_{\mathrm{P1}}$), $-0.94\pm 0.31$ ($z_{\mathrm{P1}}$) for the $\mathrm{PSD_{amp}}$--$L_{\mathrm{bol}}$ relation and $-1.18\pm 0.35$ ($g_{\mathrm{P1}}$), $-1.27\pm 0.24$ ($r_{\mathrm{P1}}$), $-1.45\pm 0.35$ ($i_{\mathrm{P1}}$), $-1.11\pm 0.38$ ($z_{\mathrm{P1}}$) for the $\mathrm{PSD_{amp}}$--$\lambda_{\mathrm{Edd}}$ relation, respectively. We stress that these values are consistent with the slopes obtained by relating the excess variance with these quantities (see table \ref{tab:fitlbolzbin} and table \ref{tab:fiteddzbin}) and suggest a common value of $\alpha\sim -1$. 
\begin{table}
\caption{Scaling of $\mathrm{PSD_{amp}}$ with $L_{\mathrm{bol}}$ and $\lambda_{\mathrm{Edd}}$.}
\centering
\begin{tabular}{cccc}
\hline\hline
 & \multicolumn{3}{c}{$\mathrm{PSD_{amp}}$ versus $L_{\mathrm{bol}}$}\\

   Filter & $\alpha$ & $\beta$ & $\epsilon$ \\
    \hline
    $g_{\mathrm{P1}}$ & $-0.35\pm 0.10$ & $-2.28\pm 0.08$ & $0.33\pm 0.03$ \\
    $r_{\mathrm{P1}}$ & $-0.53\pm 0.16$ & $-2.32\pm 0.11$ & $0.41\pm 0.04$ \\
    $i_{\mathrm{P1}}$ & $-0.83\pm 0.19$ & $-2.32\pm 0.12$ & $0.41\pm 0.04$ \\
    $z_{\mathrm{P1}}$ & $-0.55\pm 0.18$ & $-2.54\pm 0.14$ & $0.41\pm 0.05$ \\
    \hline
 & \multicolumn{3}{c}{$\mathrm{PSD_{amp}}$ versus $\lambda_{\mathrm{Edd}}$}\\

   Filter & $\alpha$ & $\beta$ & $\epsilon$ \\
    \hline
    $g_{\mathrm{P1}}$ & $-0.56\pm 0.18$ & $-3.01\pm 0.17$ & $0.31\pm 0.03$ \\
    $r_{\mathrm{P1}}$ & $-0.63\pm 0.19$ & $-3.23\pm 0.18$ & $0.39\pm 0.04$ \\
    $i_{\mathrm{P1}}$ & $-1.12\pm 0.32$ & $-3.82\pm 0.30$ & $0.35\pm 0.06$ \\
    $z_{\mathrm{P1}}$ & $-0.84\pm 0.30$ & $-3.60\pm 0.24$ & $0.36\pm 0.06$ \\
    \hline
\end{tabular}
\tablefoot{Fitted values of the relation $\log\mathrm{PSD_{amp}}=\beta+\alpha\log x+\epsilon$ for each considered PS1 band of the PSD\_MBH sample with $x=L_{\mathrm{bol}},\lambda_{\mathrm{Edd}}$ assuming $\Delta\log L_{\mathrm{bol}}=0.15$ and $\Delta\log M_{\mathrm{BH}}=0.25$.} 
\label{tab:fitpsdamp}
\end{table}

\subsection{Scaling of the high frequency PSD slope}
\label{sec:highnu}

\begin{figure}
	\centering
	\includegraphics[width=.5\textwidth]{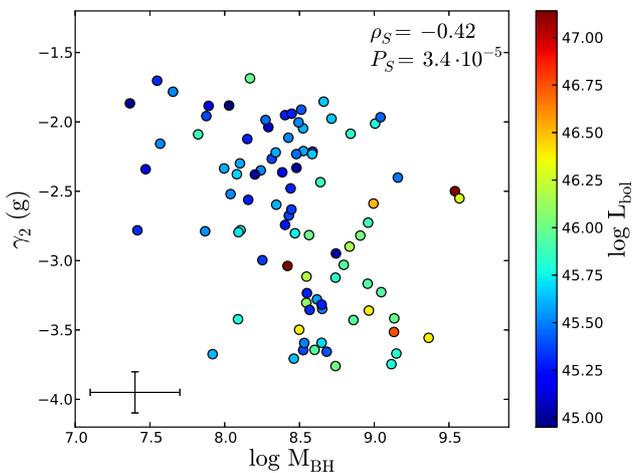}
	\caption{High frequency PSD slope $\gamma_{2}$ ($g_{\mathrm{P1}}$ band PSD\_MBH sample) versus black hole mass. The bolometric luminosity is given as a color bar. The black error bars are the average values.}
	\label{fig:g2vsmbh}
\end{figure}

Finally, we observe a weak tendency for the high frequency slope $\gamma_{2}$ of the optical PSD to scale inversely with $M_{\mathrm{BH}}$ and $L_{\mathrm{bol}}$. This would imply that high frequency variability is increasingly suppressed in higher mass systems. We show the scaling of the $g_{\mathrm{P1}}$ band slope $\gamma_{2}$ with $M_{\mathrm{BH}}$ in Fig. \ref{fig:g2vsmbh}. Obviously there are many outliers and the scatter in the relation is very large. Furthermore these anti-correlations are statistically significant only in the "blue" PS1 bands and essentially disappear in the "red" PS1 bands. Spearman's r of $\gamma_{2}$ and $M_{\mathrm{BH}}$ for the PSD\_MBH sample is given by $-0.42$ ($g_{\mathrm{P1}}$), $-0.36$ ($r_{\mathrm{P1}}$), $-0.30$ ($i_{\mathrm{P1}}$), $-0.11$ ($z_{\mathrm{P1}}$) with p-values of $3.4\cdot 10^{-5}$ ($g_{\mathrm{P1}}$), $1.1\cdot 10^{-3}$ ($r_{\mathrm{P1}}$), $1.1\cdot 10^{-2}$ ($i_{\mathrm{P1}}$), $0.42$ ($z_{\mathrm{P1}}$). The corresponding values of $\rho_{\mathrm{S}}$ for $\gamma_{2}$ and $L_{\mathrm{bol}}$ are $-0.41$ ($g_{\mathrm{P1}}$), $-0.26$ ($r_{\mathrm{P1}}$), $-0.02$ ($i_{\mathrm{P1}}$), $-0.10$ ($z_{\mathrm{P1}}$) with p-values of $7.1\cdot 10^{-5}$ ($g_{\mathrm{P1}}$), $2.0\cdot 10^{-2}$ ($r_{\mathrm{P1}}$), $0.85$ ($i_{\mathrm{P1}}$), $0.47$ ($z_{\mathrm{P1}}$). We note that the high frequency part of the PSD mostly covers low variability amplitudes close to the estimated noise level. For this reason the uncertainty on the high frequency PSD is quite large for many of our sources, leading to poorly defined $\gamma_{2}$ values. Therefore it may well be that these anti-correlations are just a coincidence and may not be present for a much larger sample. 

\begin{figure}
	\centering
	\includegraphics[width=.5\textwidth]{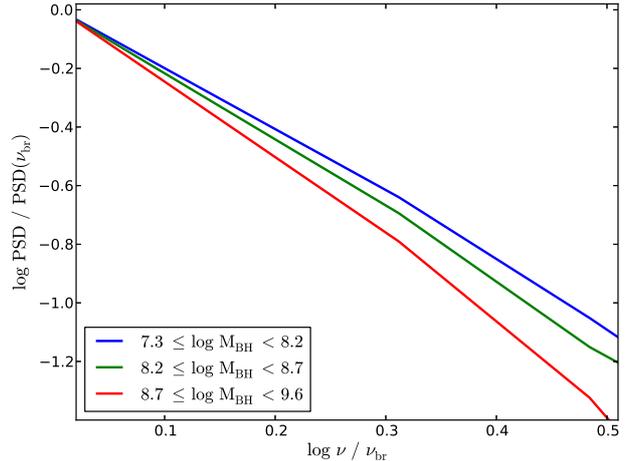}
	\caption{High frequency PSD ($g_{\mathrm{P1}}$ band PSD\_MBH sample) for three bins of black hole mass. Shown is the sample median for each $M_{\mathrm{BH}}$ bin after normalizing the PSD to the values at $\nu_{\mathrm{br}}$ in order to transform the PSD of each AGN to the same unit scale.}
	\label{fig:g2mbhbin}
\end{figure}

Nevertheless, the anti-correlation between $\gamma_{2}$ and $M_{\mathrm{BH}}$ is clearly apparent when we graphically compare the high frequency PSD for three different bins of black hole mass, as shown in Fig. \ref{fig:g2mbhbin}. The latter figure presents the median high frequency PSD for each $M_{\mathrm{BH}}$ bin after properly scaling the PSD to the value at $\nu_{\mathrm{br}}$. In this way the PSD of each object is mapped to the same unit scale and we can take the median at every frequency to obtain a typical PSD for each black hole mass bin. As it can be seen from Fig. \ref{fig:g2mbhbin} the slope of the median PSD is systematically steeper for higher mass systems, as suggested by the distribution of the fitted slopes displayed in Fig. \ref{fig:g2vsmbh}. Future studies, utilizing much larger samples, may be able to validate the observed anti-correlation.      

\section{Discussion}
\label{sec:discussion}

The power spectrum analysis performed in this work indicates that the optical PSD of AGNs can be described by a broken power law with a low frequency slope of $\gamma_{1}\sim -1$ and a high frequency slope $\gamma_{2}$ ranging from $-2$ to $-4$ with no preferred value. The characteristic bend in the PSD occurs at a break timescale between $\sim$100 and $\sim$300 days for our objects. We note that these values are larger than the expected X-ray break timescales of our sources. According to the scaling relation $\nu_{\mathrm{br}}=0.003\,\lambda_{\mathrm{Edd}}\left(M_{\mathrm{BH}}/10^{6}M_{\mathrm{\odot}}\right)^{-1}$ reported by \citet{2006Natur.444..730M} the X-ray break timescales for the sample average of $\langle\lambda_{\mathrm{Edd}}\rangle=$0.15 and $\log M_{\mathrm{BH}}=$7, 8, 9 are $\sim$0.3, 3, 30 days, whereas for the sample average mass $\langle\log M_{\mathrm{BH}}\rangle=8.5$ and $\lambda_{\mathrm{Edd}}=$0.01, 0.1, 1 they read $\sim$120, 12, 1.2 days, respectively. Furthermore we observe that the PSD amplitude scales inversely with $L_{\mathrm{bol}}\propto\dot{M}$ with a logarithmic slope of $\sim -1$. Assuming the power law PSD of equation \ref{eq:psdfit} we can predict the expected scaling of the excess variance by performing the integral according to equation \ref{eq:psdint}, giving three different cases depending on the integration limits   
\begin{flalign}
	\label{eq:rmspsd}
	\text{$\sigma_{\mathrm{rms}}^{2}\propto$}
	\begin{cases}
	\text{$\dot{M}^{-1}\frac{\nu_{\mathrm{br}}^{-\left(\gamma_{2}+1\right)}}{\gamma_{2}+1}\left(\nu_{\mathrm{max}}^{\gamma_{2}+1}-\nu_{\mathrm{min}}^{\gamma_{2}+1}\right)$} & \text{, $\nu_{\mathrm{min}}>\nu_{\mathrm{br}}$} \\
	\text{$\dot{M}^{-1}\left[\ln\left(\frac{\nu_{\mathrm{br}}}{\nu_{\mathrm{min}}}\right)+\frac{1}{\gamma_{2}+1}\left(\left(\frac{\nu_{\mathrm{max}}}{\nu_{\mathrm{br}}}\right)^{\gamma_{2}+1}-1\right)\right]$} & \text{, $\nu_{\mathrm{min}}<\nu_{\mathrm{br}}<\nu_{\mathrm{max}}$} \\
	\text{$\dot{M}^{-1}\ln\left(\frac{\nu_{\mathrm{max}}}{\nu_{\mathrm{min}}}\right)$} & \text{, $\nu_{\mathrm{max}}<\nu_{\mathrm{br}}$.}
	\end{cases}
\end{flalign}
Given the observed break timescales of our objects this means that measuring $\sigma_{\mathrm{rms}}^{2}$ on timescales of 1--3 years corresponds to the middle case of equation \ref{eq:rmspsd}, whereas calculating $\sigma_{\mathrm{rms}}^{2}$ on timescales of 1--3 months we are probing the high frequency part of the PSD (upper case of equation \ref{eq:rmspsd}). Note that due to $\mathrm{PSD_{amp}}\propto L_{\mathrm{bol}}^{-1}$ we expect an anti-correlation between $\sigma_{\mathrm{rms}}^{2}$ and $L_{\mathrm{bol}}$ irrespective of the frequencies sampled by the light curve. Moreover, since we do not find any significant correlation between $\nu_{\mathrm{br}}$ and the AGN parameters, the different scaling of $\sigma_{\mathrm{rms}}^{2}$ with $\nu_{\mathrm{br}}$ for each of the three cases of equation \ref{eq:rmspsd} does not introduce further dependencies on $M_{\mathrm{BH}}$, $L_{\mathrm{bol}}$ or $\lambda_{\mathrm{Edd}}$. Therefore the observed relations for the excess variance are consistent with the results obtained in our PSD analysis. Yet these findings are very different from the results of previous X-ray variability studies suggesting that the observed anti-correlation between the X-ray excess variance and $M_{\mathrm{BH}}$ is introduced by the relation $\nu_{\mathrm{br}}\propto M_{\mathrm{BH}}^{-1}$ \citep{2006Natur.444..730M}. We stress however that \citet{2012A&A...542A..83P} proposed $\mathrm{PSD_{amp}}\propto\dot{M}^{\alpha}$ with $\alpha\sim -0.8$, remarkably similar to the value derived in this work, in order to explain their X-ray excess variance results. Yet we note that directly comparing our findings with the results of X-ray variability studies requires to assume that there is a strong temporal coupling between the variability of both spectral regimes. A fair comparison would only be possible by analysing simultaneous X-ray and UV/optical light curves.              

Many researchers found correlations between the optical variability properties and physical parameters of AGNs that are similar to the ones reported in this work. However, to date there is no self-consistent and physically motivated explanation for all the observed scalings. Although it is possible to predict correlations between the variability amplitude and the AGN parameters (also an anti-correlation with $\lambda_{\mathrm{Edd}}$) within the standard $\alpha$--disc prescription \citep{1973A&A....24..337S} by assuming global fluctuations of $\dot{M}$ \citep[e.g.][]{2012ApJ...758..104Z,2013A&A...560A.104M}, the resulting scaling relations are much flatter than the observed ones. What is more, the typical timescales of optical variability are much shorter than the time needed for global changes of the mass accretion rate in the entire disc, associated with the sound crossing or viscous timescale \citep{1991A&A...248..389C,2014SSRv..183..453U,2014ApJ...783..105R,2015MNRAS.449...94K}. For this reason it is unlikely that global accretion rate changes are the sole driver of optical variability.   

A promising alternative may be the strongly inhomogeneous disc model proposed by \citet{2011ApJ...727L..24D} in which many independent local temperature fluctuations account for the flux variability. Within this toy model the accretion disc is subdivided in $N$ varying regions which are allowed to fluctuate according to a "damped random walk" with an amplitude $\sigma_{\mathrm{T}}$ about the mean temperature of the standard geometrically thin optically thick accretion disc. The total variance is proportional to $N^{-1}$ and \citet{2011ApJ...727L..24D} were able to explain the observed 10\%--20\% amplitudes of optical variability for $N\sim 10^{2}$--$10^{3}$ and      $\sigma_{\mathrm{T}}\sim$0.35--0.5 dex. However, \citet{2015MNRAS.449...94K} argues that the inhomogeneous disc model can not adequately explain the tight inter-band flux-flux correlations of optical variability, which are also present in our light curves. 

Although both global mass accretion rate changes and localized temperature fluctuations can not predict all characteristics of AGN variability, it may well be that both mechanisms act in parallel. Slowly changing values of $\dot{M}$, occurring on timescales of thousands to millions of years, regulate the global, long term, accretion state of an AGN, whereas a large number of localized disc inhomogeneities may account for the observed short term variability. In fact disc inhomogeneities are likely to develop in a turbulent accretion flow due to the thermal, magnetorotational or Parker instabilities. In addition, considering that our sources are luminous quasars, the accretion discs are likely radiation pressure dominated and therefore the radiation pressure instability may also play an important role \citep{2014SSRv..183...21B,2014SSRv..183..453U}. We observed that the variability amplitude is anti-correlated with the bolometric luminosity and the Eddington ratio. These anti-correlations may be qualitatively explained by assuming that more luminous quasars with a higher mass accretion rate develop a greater number $N$ of disc inhomogeneities due to an enhanced radiation pressure instability, leading to a smaller total variance in flux according to the inhomogeneous disc model. 

In view of the fact that the fundamental accretion disc timescales such as the orbital, thermal or viscous timescale all depend on $M_{\mathrm{BH}}$, it is somewhat surprising that the optical break timescale $T_{\mathrm{br}}$ derived in this work seems to be uncorrelated with the latter and also with the other AGN physical parameters. For a standard thin $\alpha$-disc the characteristic radius for emission at wavelength $\lambda$ is governed by $M_{\mathrm{BH}}$ and $\lambda_{\mathrm{Edd}}$ according to $R_{\lambda}\propto M_{\mathrm{BH}}^{2/3}\lambda_{\mathrm{Edd}}^{1/3}\lambda^{4/3}$ \citep[see][]{2002apa..book.....F}. Under the assumption that most of the variable flux of a given wavelength band is emitted at $R_{\lambda}$ and associating $T_{\mathrm{br}}$ with e.g. the thermal timescale $t_{\mathrm{th}}\propto\alpha^{-1} R^{3/2}M_{\mathrm{BH}}^{-1/2}$ at $R_{\lambda}$ we find $T_{\mathrm{br}}\propto M_{\mathrm{BH}}^{1/2}\lambda_{\mathrm{Edd}}^{1/2}\lambda^{2}\propto L_{\mathrm{bol}}^{1/2}\lambda^{2}$. Obviously the broad redshift distribution of our sample means that a variety of different radii contribute to the radiation in a fixed broadband filter and an investigation of the scaling of $T_{\mathrm{br}}$ is only meaningful considering small bins of $M_{\mathrm{BH}}$, $\lambda_{\mathrm{Edd}}$ and $z$. Otherwise any correlation can be smeared out by the range of parameters. However our sample is not large enough to perform such a binning in an appropriate way. On the other hand, comparing the $T_{\mathrm{br}}$ values derived in different PS1 bands, i.e. at fixed $M_{\mathrm{BH}}$, $\lambda_{\mathrm{Edd}}$ and $z$ for each source, a possible dependence on $\lambda$ would be visible. Yet, as shown in appendix \ref{sec:appendixb}, $T_{\mathrm{br}}$ seems to be on average the same for each PS1 band. Eventually, if quasar accretion discs indeed consist of many small localized regions of different temperature, then the radiation at a given rest-frame wavelength originates from a large range of radii, naturally attenuating scalings with $M_{\mathrm{BH}}$, $\lambda_{\mathrm{Edd}}$ and $\lambda$. Therefore it remains unclear which physical process defines the characteristic optical break timescale. The tight correlation of the X-ray break timescale with $M_{\mathrm{BH}}$ is generally interpreted to reflect the size scale of the X-ray emitting region \citep{2011ApJ...730...52K}. Since most of the X-ray luminosity is likely released within a compact region of few gravitational radii in the close vicinity of the black hole, either in a roughly spherical optically thin hot corona or at the base of a relativistic jet, the effect outlined above that may smooth out a dependence of $T_{\mathrm{br}}$ on the size scale in the optical is probably not an issue in the X-rays. 

Considering the observed anti-correlation of the rms variability amplitude with $L_{\mathrm{bol}}$ and $\lambda_{\mathrm{Edd}}$, it is tempting to interpret these results as a representation of different accretion states in AGNs. In fact it is well known that black hole X-ray binary (BHXRB) systems undergo state changes which are believed to be connected to different accretion flow geometries. In states with high $\lambda_{\mathrm{Edd}}$ an optically thick disc is thought to extend down close to the black hole, giving rise to a significant thermal component in the spectrum (soft state). The soft state is characterized by low rms variability amplitudes, typically less than $\sim$5\%. On the contrary, in states with low $\lambda_{\mathrm{Edd}}$ the disc may be replaced by an optically thin hot medium at some truncation radius, generating the observed power law hard X-ray emission (hard state) \citep{2004astro.ph..9254C,2011ApJ...730...52K}. During the hard state the rms variability is large, with amplitudes up to $\sim$30\%--40\% \citep{2011MNRAS.410..679M}. In the transition region between these two "canonical" states the picture is less clear and different intermediate states have been defined, such as the very high state (VHS) exhibiting a spectrum of intermediate hardness and a very high X-ray flux \citep[see e.g.][for a review]{2010LNP...794.....B}. Since the timescales associated with state transitions are assumed to increase with black hole mass, a cycle through the states is expected to last thousands or millions of years for systems with supermassive black holes. For this reason different states may only be visible for a large sample of objects and to date it is unclear whether AGNs show the same accretion states as BHXRBs, however similarities have been observed \citep[see e.g.][]{2006MNRAS.372.1366K}. Interestingly, our sample contains AGNs showing high values of $L_{\mathrm{bol}}$ and $\lambda_{\mathrm{Edd}}$ with rms amplitude of $\sim$5\%, but also sources having low values of $L_{\mathrm{bol}}$ and $\lambda_{\mathrm{Edd}}$ with rms amplitude of $\sim$30\%. However, our objects are mostly luminous quasars characterized by a dominant disc component and therefore it is unlikely that some of these are in a hard state, which may rather be associated with low luminosity AGNs \citep{2004astro.ph..9254C,2006MNRAS.372.1366K,2014SSRv..183...21B}. Nevertheless, the trends observed in this work may indicate that the less variable AGNs populate a state similar to the soft state, whereas the highly variable AGNs may be in an intermediate state between the soft and hard state.  

\section{Conclusions}
\label{sec:conclu}

We studied correlations between the rest-frame UV/optical variability amplitude, expressed by the excess variance, and the fundamental AGN parameters for about 90 quasars covering a wide redshift range (0.3 to 2.5) from the XMM-COSMOS survey. The excess variance is measured from the multi-epoch light curves of the Pan-STARRS1 Medium Deep Field 04 survey in the four bands $g_{\mathrm{P1}}$, $r_{\mathrm{P1}}$, $i_{\mathrm{P1}}$, $z_{\mathrm{P1}}$ and on two different timescales of 1--3 months and 1--3 years, depending on the redshift of each source. We searched for scalings of the excess variance computed on these two timescales with wavelength, redshift, black hole mass, bolometric luminosity and Eddington ratio. Additionally, we performed a power spectrum analysis of our optical light curves in the AGN rest-frame by using the CARMA model prescription of the PSD introduced by \citet{2014ApJ...788...33K}. We also tested for relations between the derived PSD parameters and the aforementioned AGN physical properties. Our main results can be summarized as follows: 
\begin{enumerate}
\item The excess variances calculated in various PS1 bands are highly correlated. This is in accord with the fact that our sources vary approximately simultaneous in the different bands. The variability amplitude is observed to generally decrease with wavelength as found in many previous studies. 
\item We find no significant correlation between the variability amplitude and the black hole mass, neither on timescales of years nor on timescales of months. In contrast we observe a very strong anti-correlation between the excess variance and the bolometric luminosity for both probed variability timescales and in all PS1 bands. The logarithmic slope of the anti-correlation is consistent with a value of $-1$ for both variability timescales.
\item The variability amplitude is also strongly anti-correlated with the Eddington ratio in all PS1 bands. The relation with $\lambda_{\mathrm{Edd}}$ exhibits the same logarithmic slope of $-1$ as observed for the bolometric luminosity. 
\item In all of the aforementioned correlations there is no significant evolution with redshift. That is understood as the result of two counter-directed selection effects related to the wavelength dependence of the variability amplitude and the anti-correlation with luminosity.  
\item The optical PSD of all of our sources resembles a broken power law with break timescales between $\sim$100 and $\sim$300 days. The break timescales seem to be uncorrelated with the black hole mass, the bolometric luminosity, the Eddington ratio and the radiation wavelength. This lack of correlation indicates that the optical break timescale is not associated with any of the characteristic physical timescales of the accretion disc. The low frequency slope of the PSD is roughly consistent with a value of $-1$, similar to the value observed in X-ray PSDs. However, the high frequency slope exhibits a broad distribution of values between $-2$ and $-4$, generally steeper than the high frequency slopes of X-ray PSDs. The observed shape of the optical PSD suggests significant deviations from the PSD of the DRW model. Finally we observe a weak trend that the high frequency optical PSD slope may decrease with increasing black hole mass.  
\item The PSD amplitude is anti-correlated with the bolometric luminosity and the Eddington ratio. The anti-correlations are seen in all PS1 bands and the fitted slopes for the relations with $L_{\mathrm{bol}}$ and $\lambda_{\mathrm{Edd}}$ suggest a common value of $-1$ once fatal outliers are removed from the sample, as observed for the excess variance. We detect no correlation between the PSD amplitude and the black hole mass. Therefore the observed correlations between the excess variance and the AGN physical parameters are consistent with the relations found in the PSD analysis. The observed scalings are in favour of the accretion rate being the fundamental AGN parameter driving the optical variability amplitude. 
\end{enumerate}

Although our studies are based on a well defined statistical sample of QSOs, there are still limitations in view of the probed parameter space of the AGN physical quantities. Therefore it may well be that some of the less significant correlations reported in this work will change for a more complete AGN sample. The upcoming eROSITA mission \citep{2007SPIE.6686E..17P} will deliver an unparalleled large sample of X-ray selected AGNs, allowing for a thorough validation of the observed correlations when combined with massive time-domain optical surveys such as LSST \citep{2006AAS...209.8602I}. 

\begin{acknowledgements}
The Pan-STARRS1 Surveys (PS1) have been made possible through contributions of the Institute for Astronomy, the University of Hawaii, the Pan-STARRS Project Office, the Max-Planck Society and its participating institutes, the Max Planck Institute for Astronomy, Heidelberg and the Max Planck Institute for Extraterrestrial Physics, Garching, The Johns Hopkins University, Durham University, the University of Edinburgh, Queen's University Belfast, the Harvard-Smithsonian Center for Astrophysics, the Las Cumbres Observatory Global Telescope Network Incorporated, the National Central University of Taiwan, the Space Telescope Science Institute, the National Aeronautics and Space Administration under Grant No. NNX08AR22G issued through the Planetary Science Division of the NASA Science Mission Directorate, the National Science Foundation under Grant No. AST-1238877, the University of Maryland, and Eotvos Lorand University (ELTE) and the Los Alamos National Laboratory. We thank the anonymous referee for very beneficial comments. T.S. thanks Andrea Merloni, Phil Uttley, Brandon Kelly, Jason Dexter, Simone Scaringi and Barbara De Marco for many helpful discussions and comments. 
\end{acknowledgements}

\bibliographystyle{aa} 
\bibliography{bibliographie} 

\begin{appendix}
\section{Scaling of the AGN parameters in our sample}
\label{sec:appendixa}

\begin{figure*}
\centering
\subfloat{%
	\includegraphics[width=.48\textwidth]{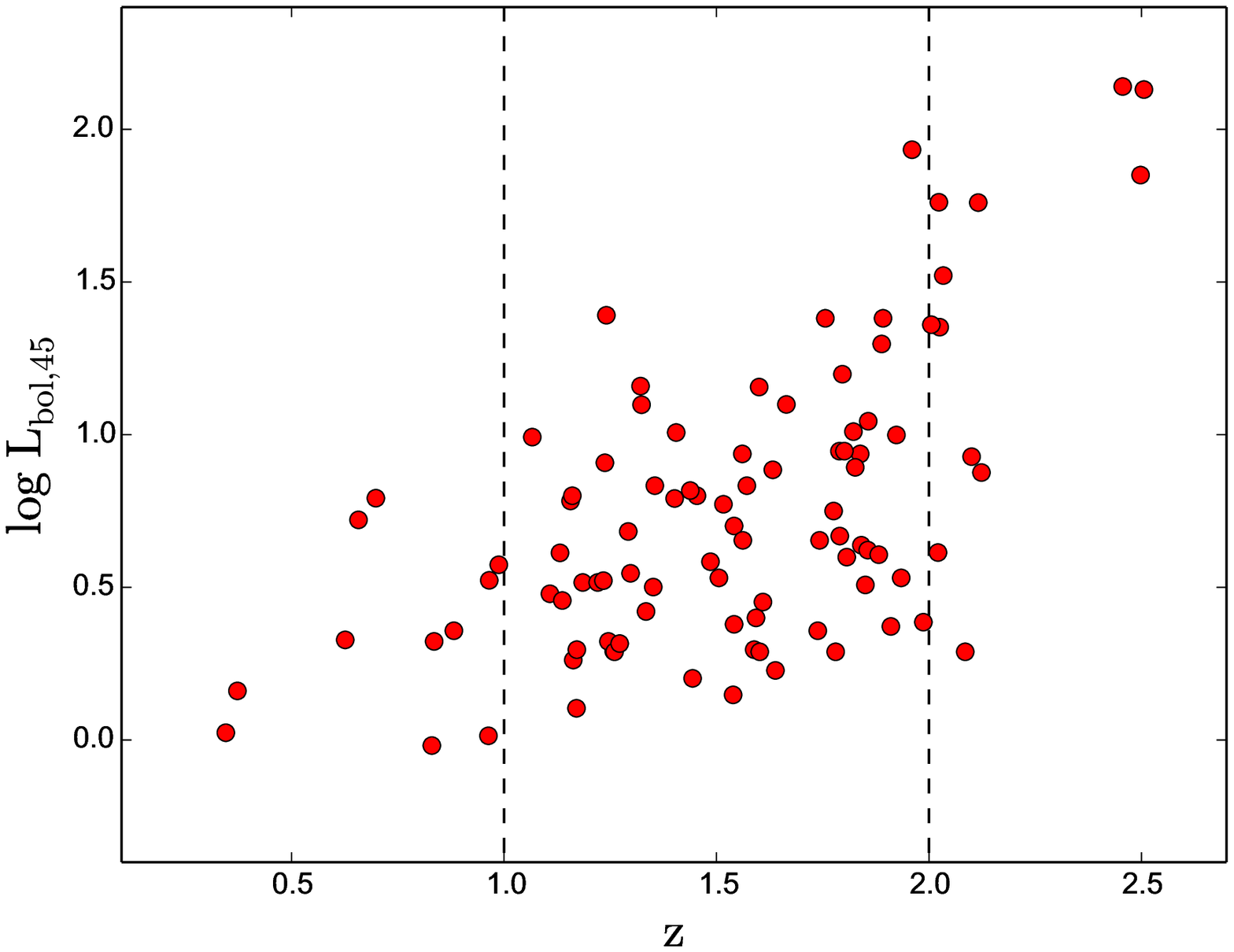}}
\quad
\subfloat{%
	\includegraphics[width=.48\textwidth]{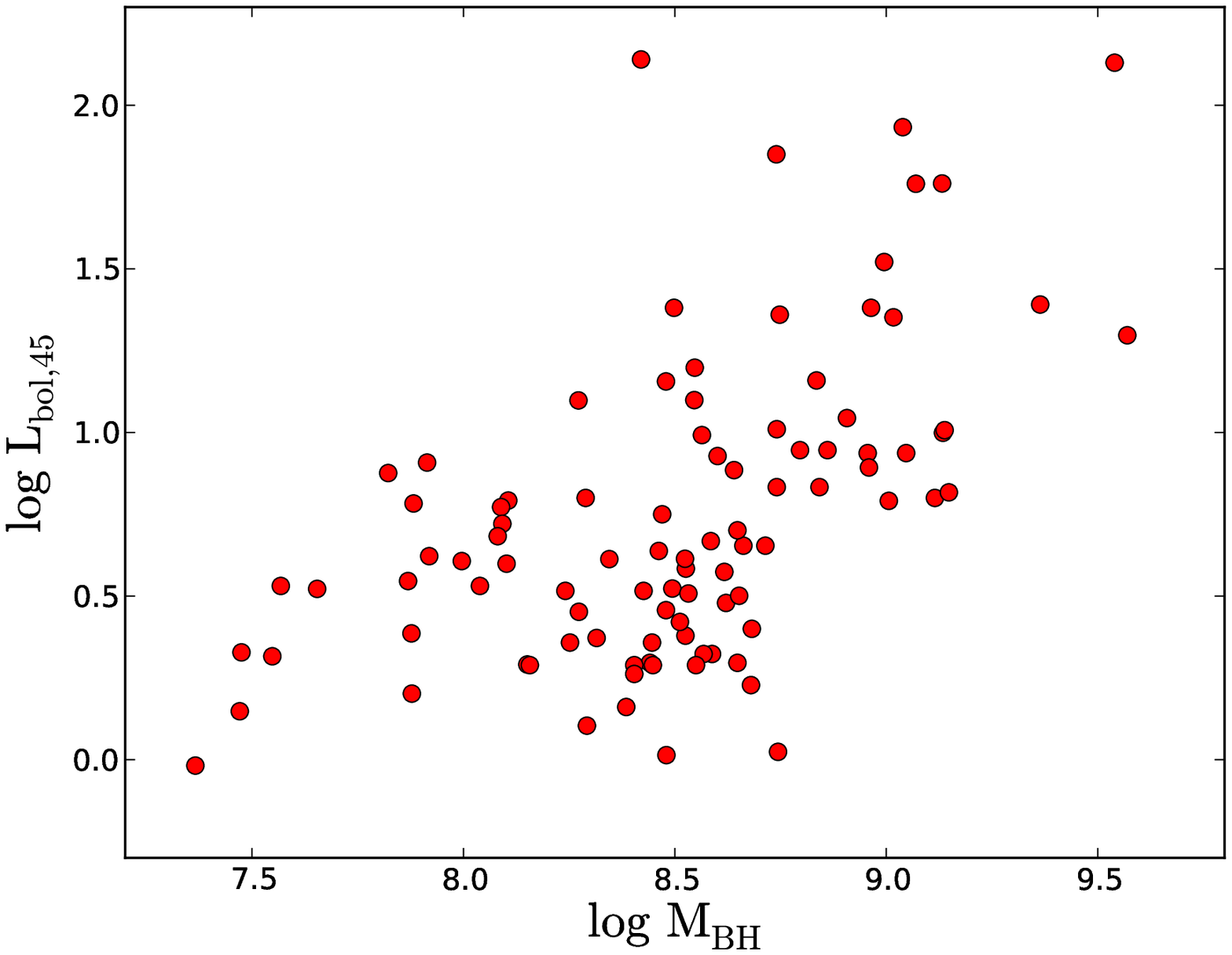}}

\subfloat{%
	\includegraphics[width=.48\textwidth]{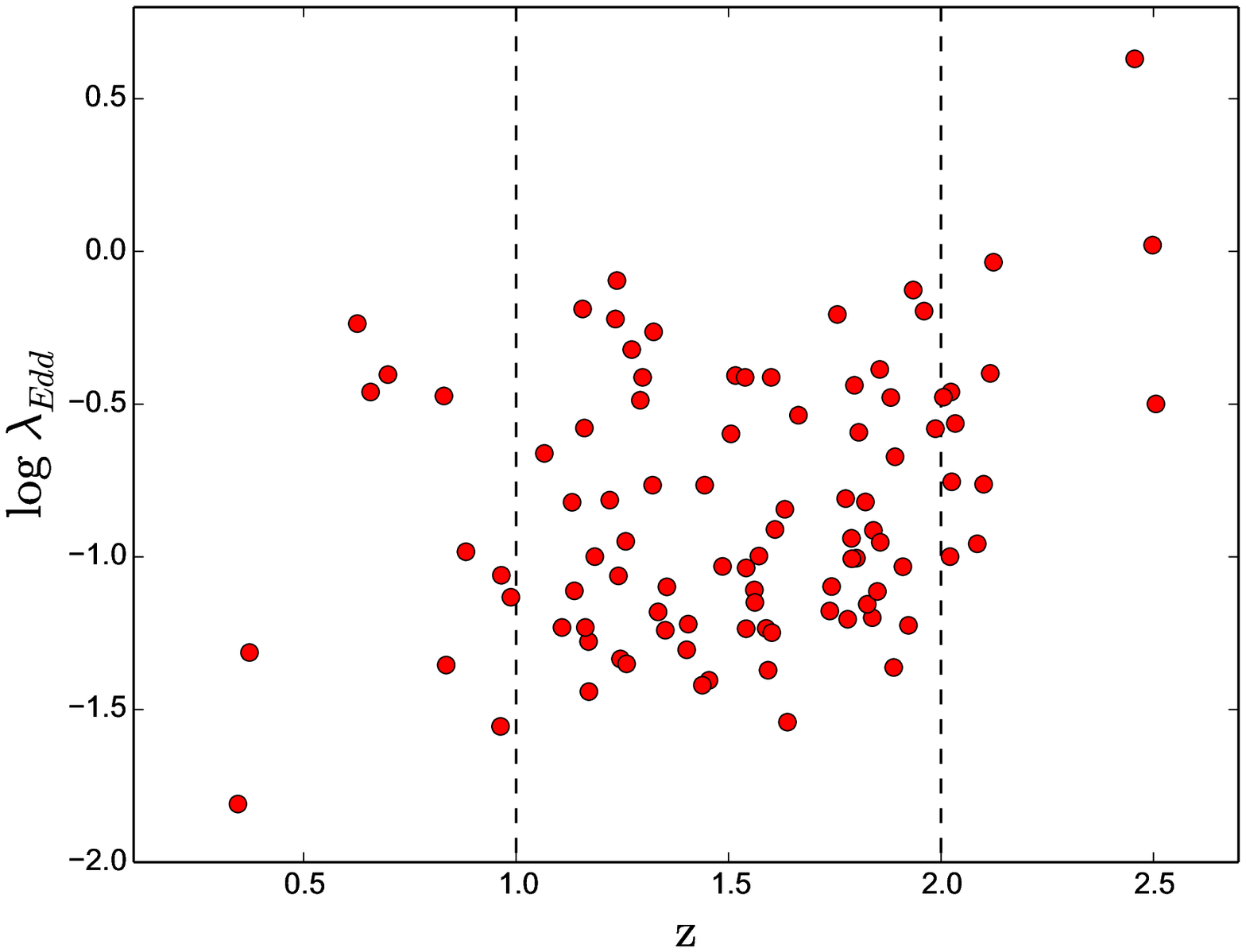}}
\quad
\subfloat{%
	\includegraphics[width=.48\textwidth]{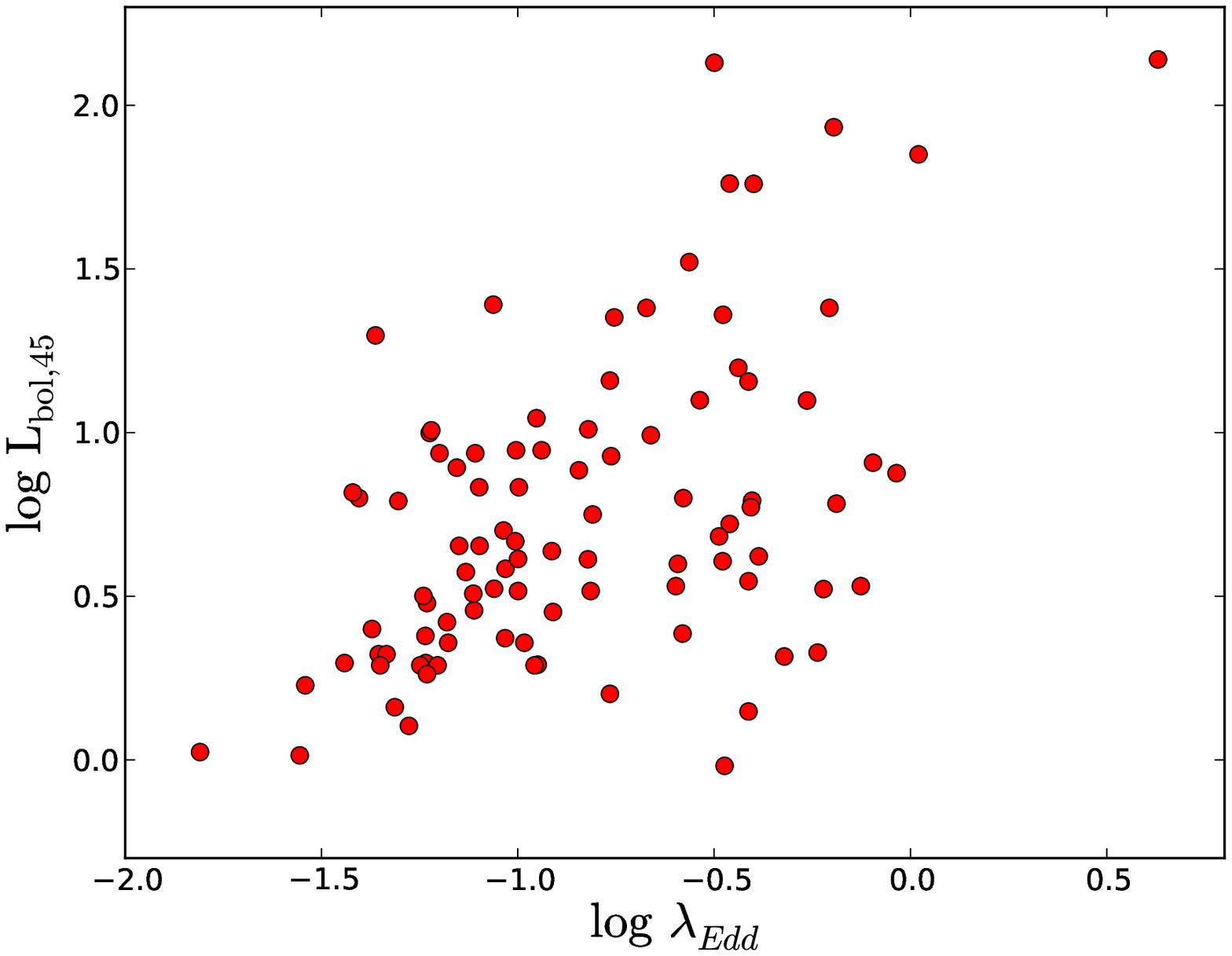}}

\subfloat{%
	\includegraphics[width=.48\textwidth]{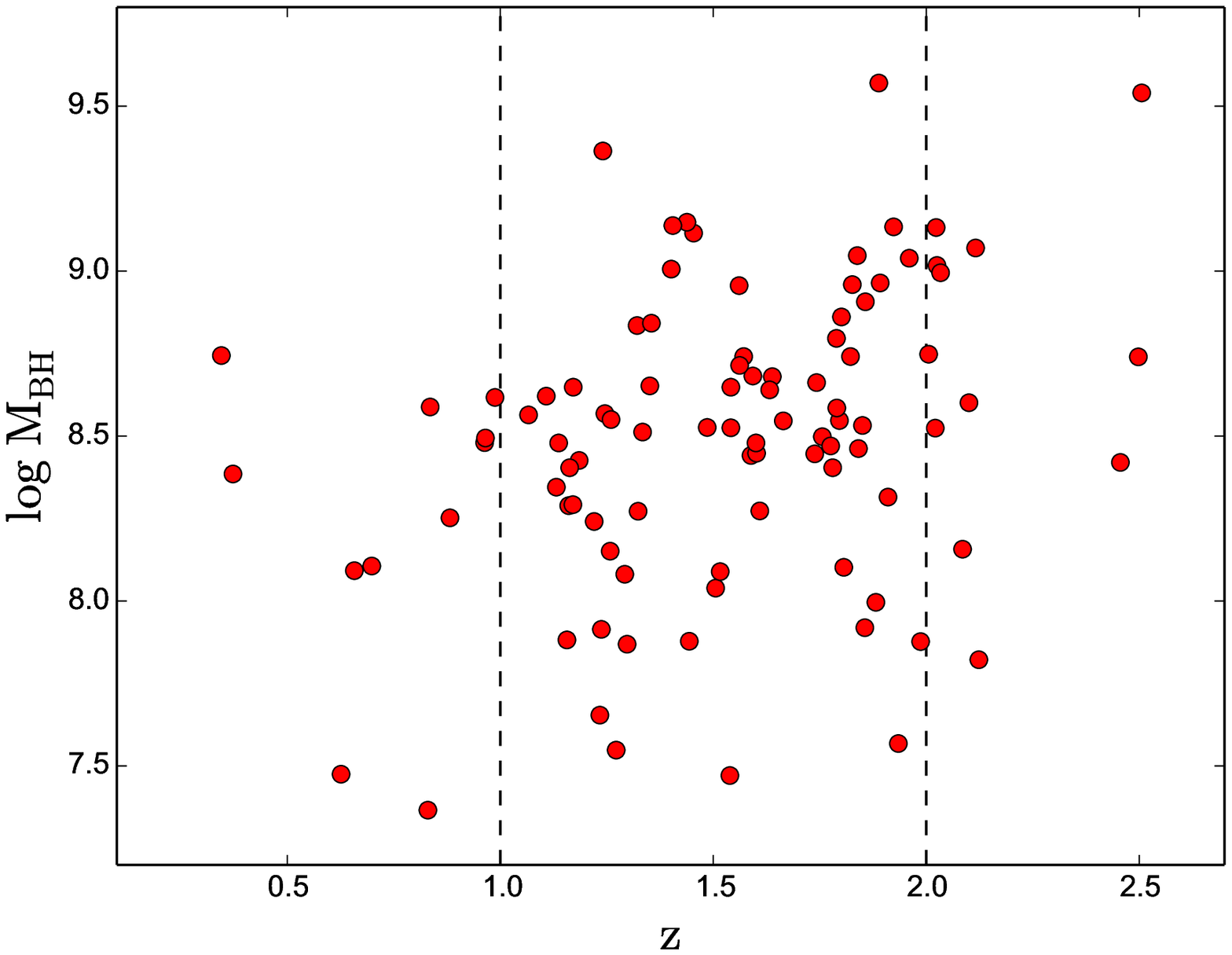}}
\quad
\subfloat{%
	\includegraphics[width=.48\textwidth]{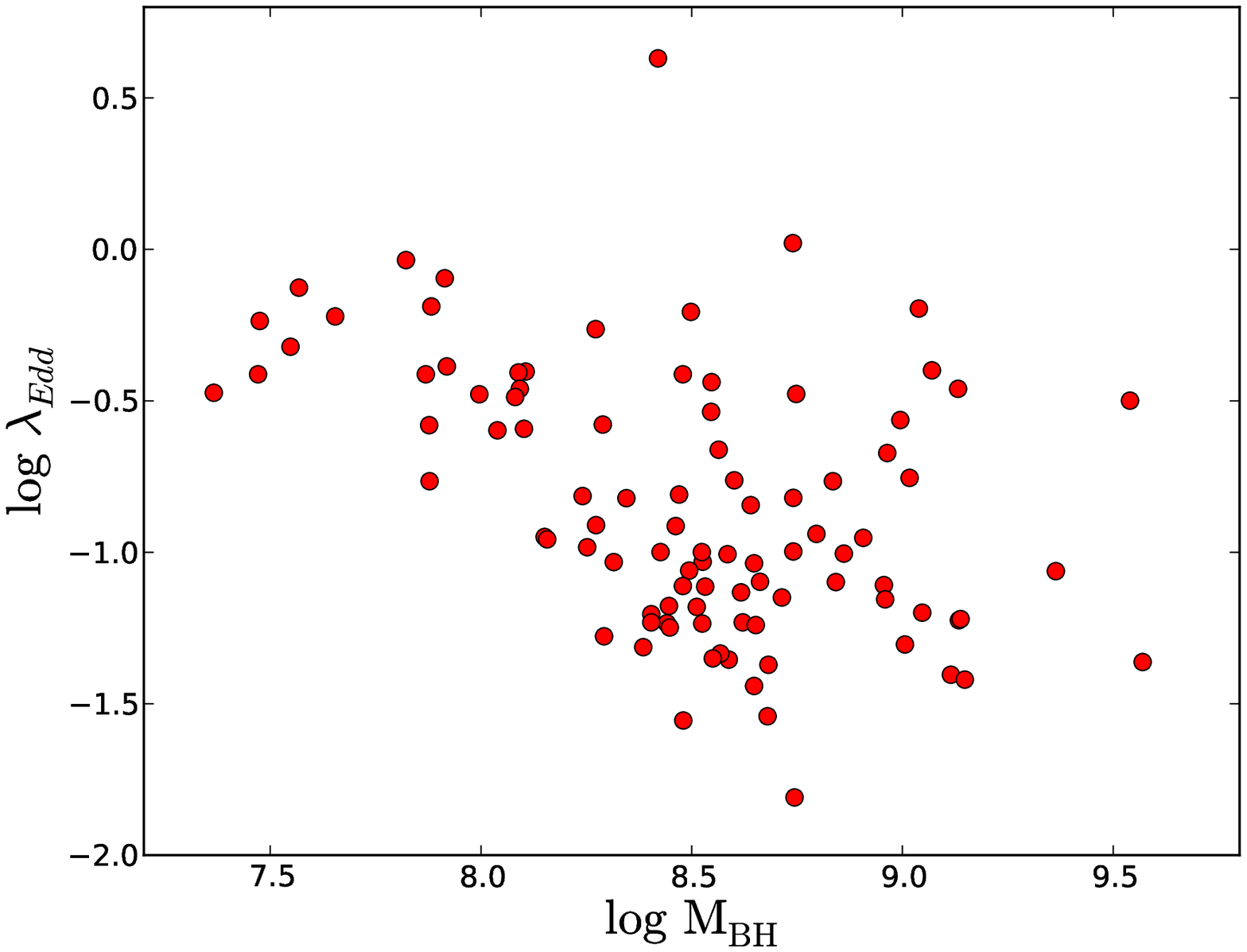}}
\caption{Dependence of the quantities $L_{\mathrm{bol}}$, $\lambda_{\mathrm{Edd}}$ and $M_{\mathrm{BH}}$ on redshift (\textit{left column}) and scalings between the physical quantities (\textit{right column}). The object data of the $g_{\mathrm{P1}}$ band MBH sample are shown. The dashed vertical lines in the left column enclose the sources of the 1z2\_MBH sample.}
	\label{fig:physvsz}
\end{figure*}

In order to characterize the used sample in view of the AGN physical properties Fig. \ref{fig:physvsz} shows the dependence of the fundamental AGN parameters on redshift and the relations between the physical properties for the MBH sample. As it can be seen the quantities $L_{\mathrm{bol}}$, $\lambda_{\mathrm{Edd}}$ and $M_{\mathrm{BH}}$ do not strongly depend on the redshift, meaning that our sample does not suffer from significant selection effects. There is only a weak trend that we preferentially select the more luminous objects at higher redshifts. We note that $L_{\mathrm{bol}}$ is roughly proportional to $M_{\mathrm{BH}}$, whereas there is only a very weak positive correlation with $\lambda_{\mathrm{Edd}}$. The Eddington ratio itself is anti-correlated with $M_{\mathrm{BH}}$. However, all these dependencies show large scatter and resemble the trends also observed in other quasar samples used to study optical variability e.g. \citet{2010ApJ...721.1014M,2012ApJ...758..104Z}. Therefore the considered sample is not strongly biased by selection effects and covers about two orders of magnitude in the AGN physical parameters.

\section{Comparing the PSD power law fit parameters in various PS1 bands}
\label{sec:appendixb}

\begin{figure*}
\centering
\subfloat{%
	\includegraphics[width=.4\textwidth]{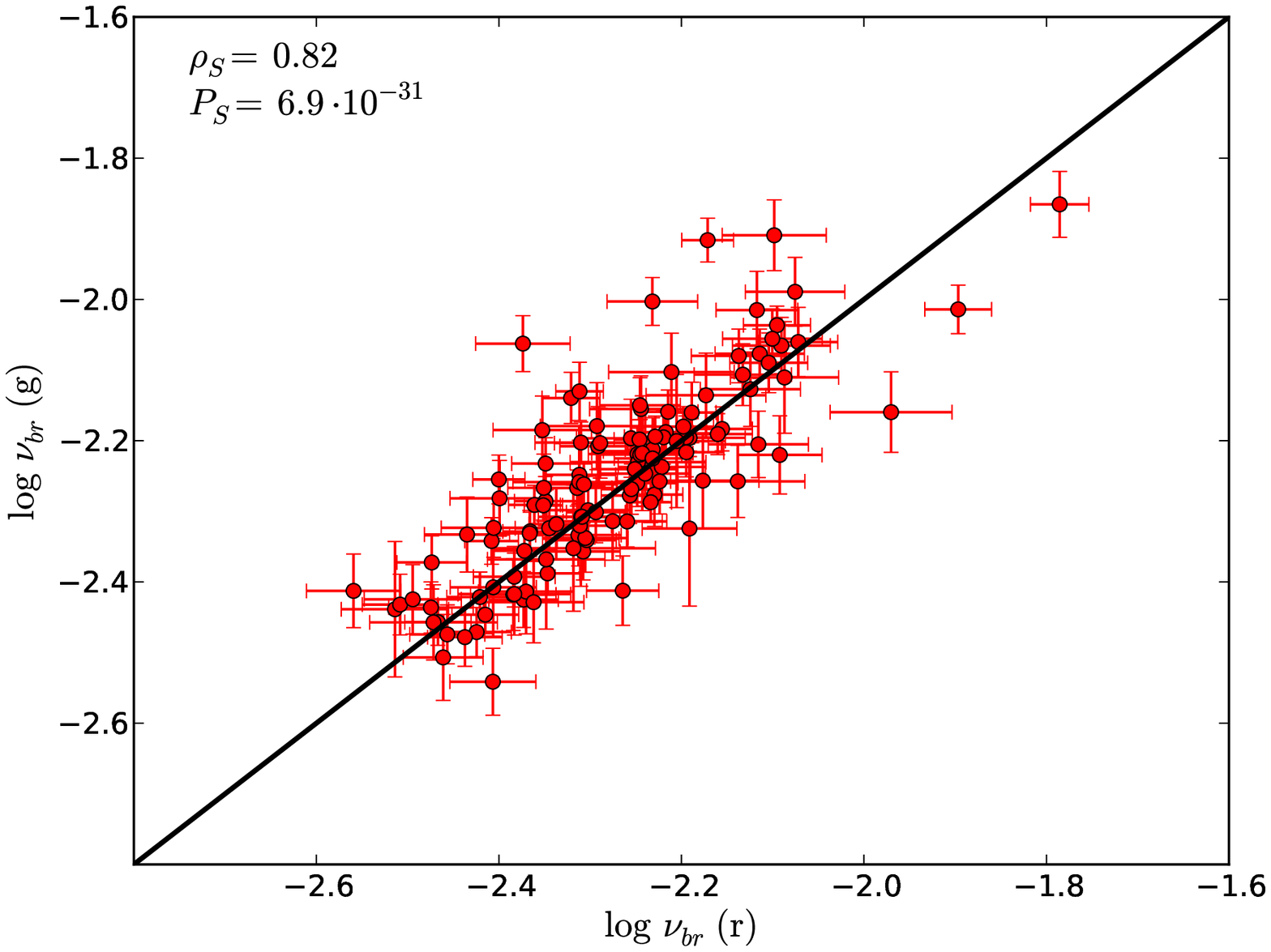}}
\quad
\subfloat{%
	\includegraphics[width=.4\textwidth]{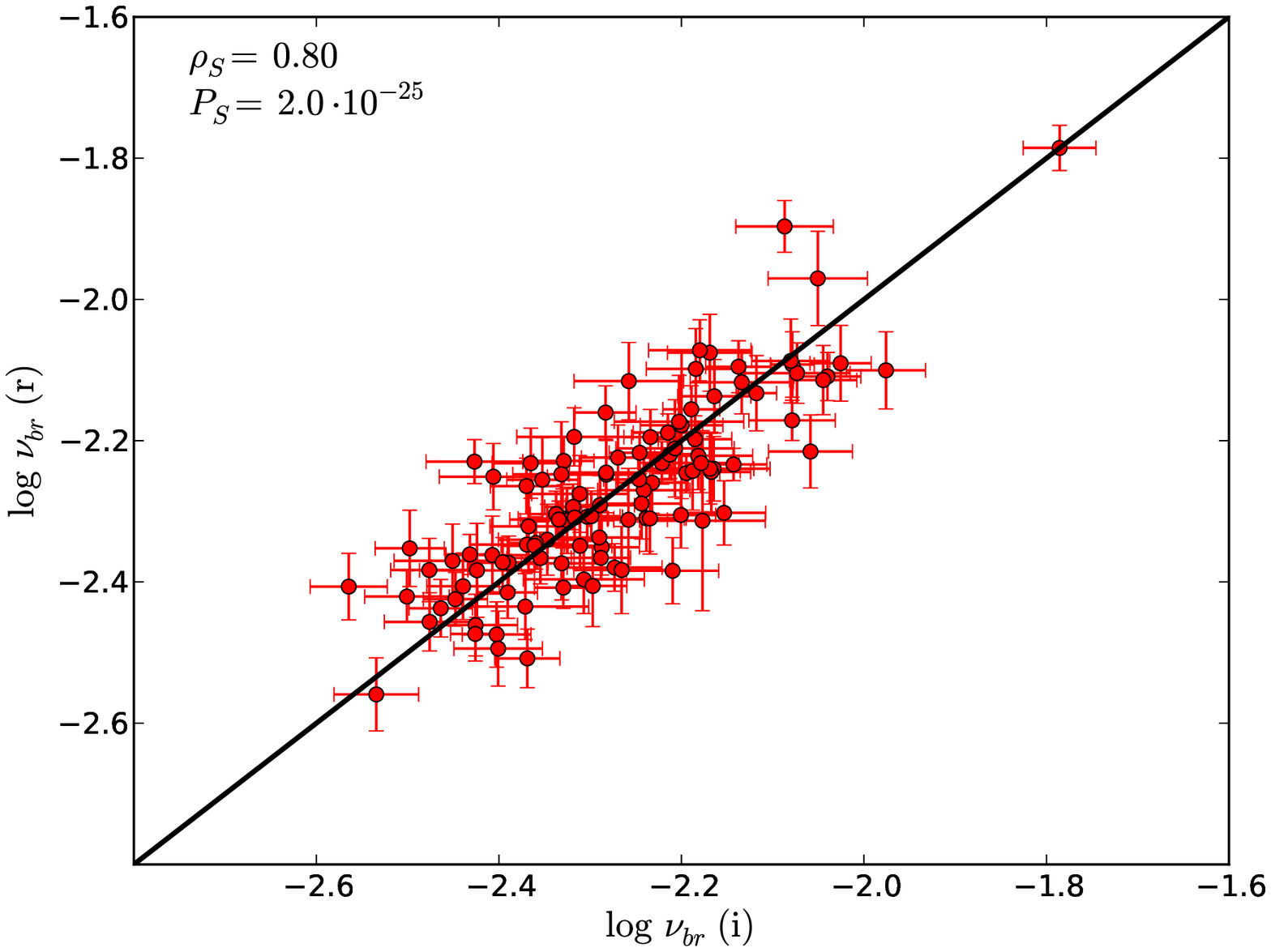}}

\subfloat{%
	\includegraphics[width=.4\textwidth]{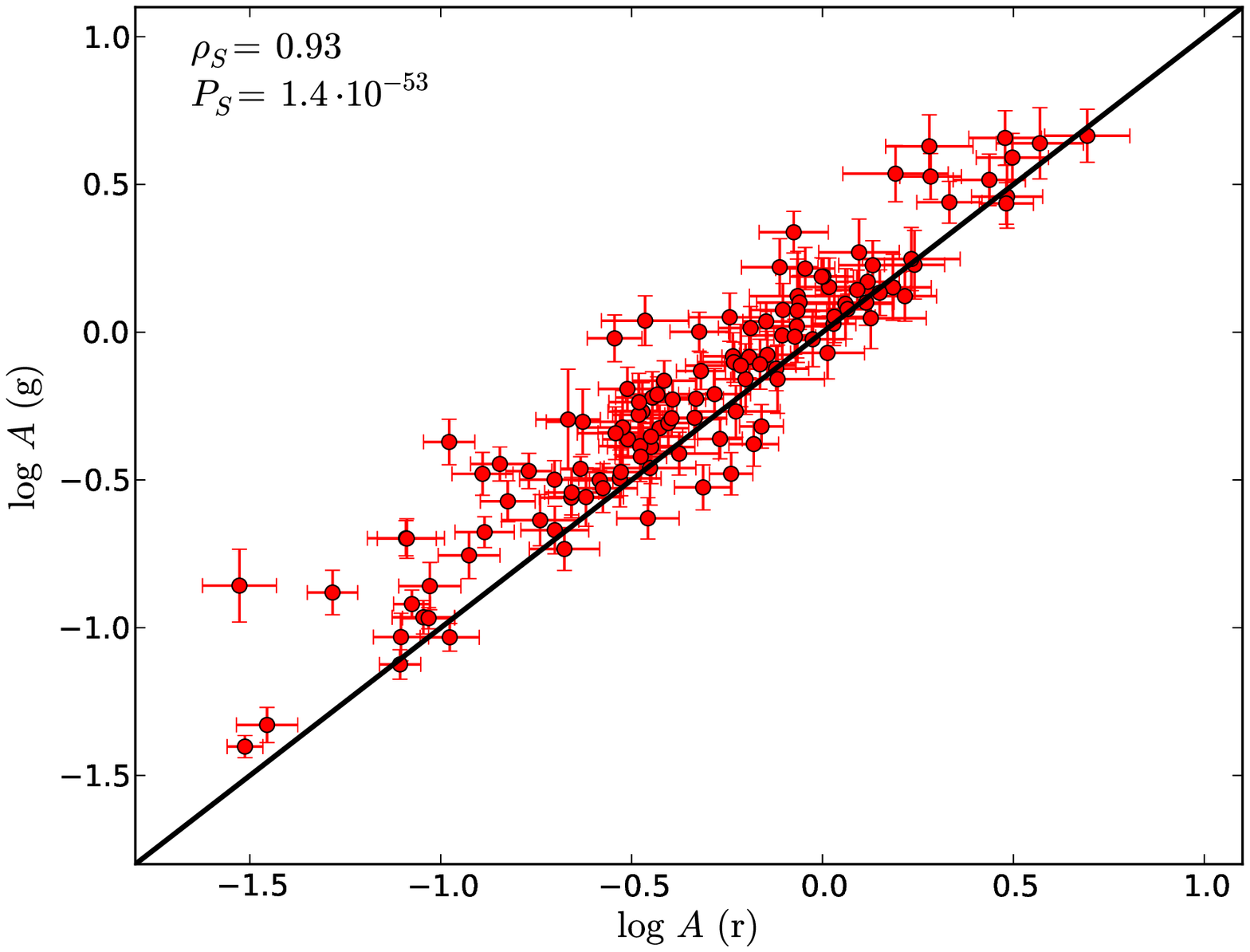}}
\quad
\subfloat{%
	\includegraphics[width=.4\textwidth]{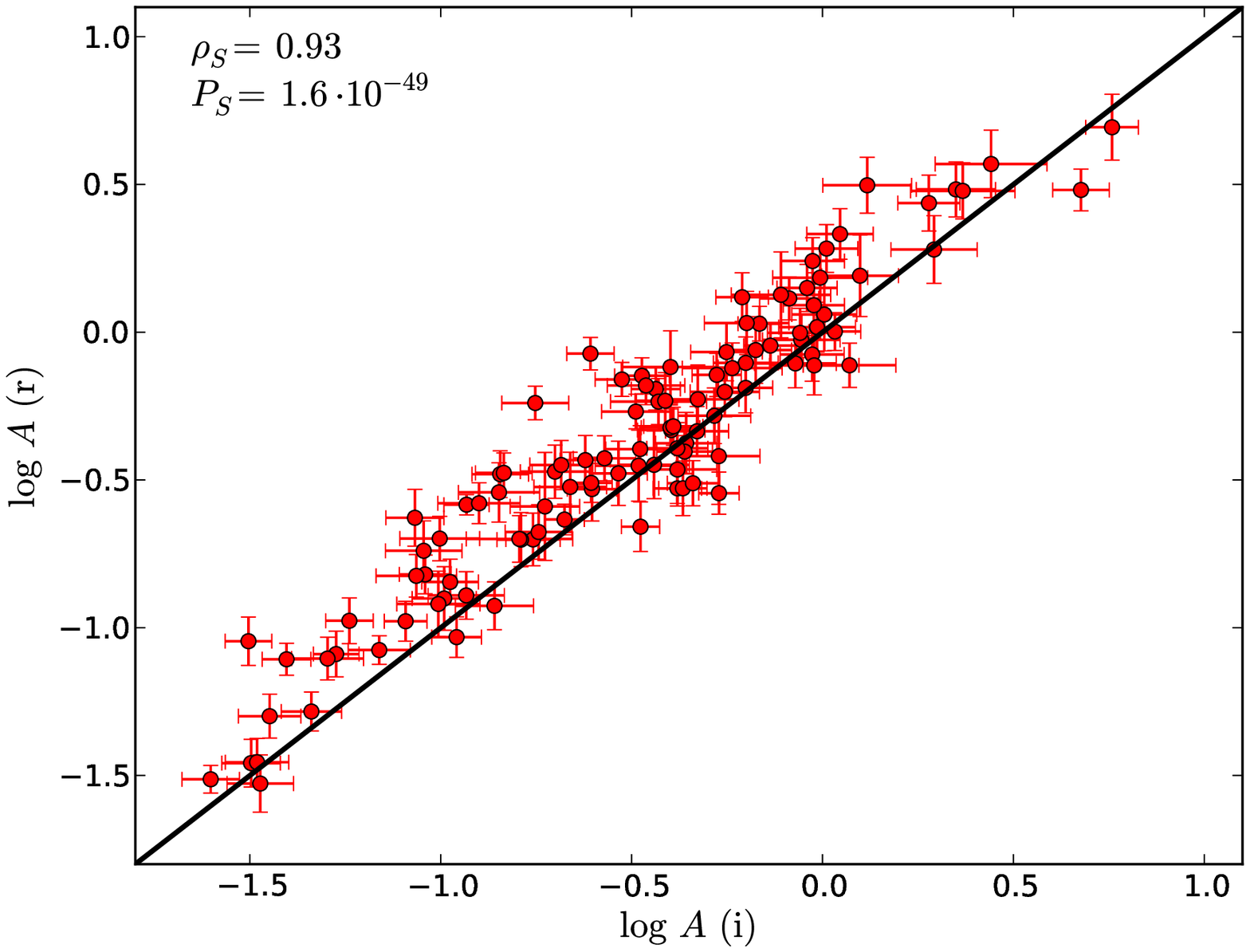}}

\subfloat{%
	\includegraphics[width=.4\textwidth]{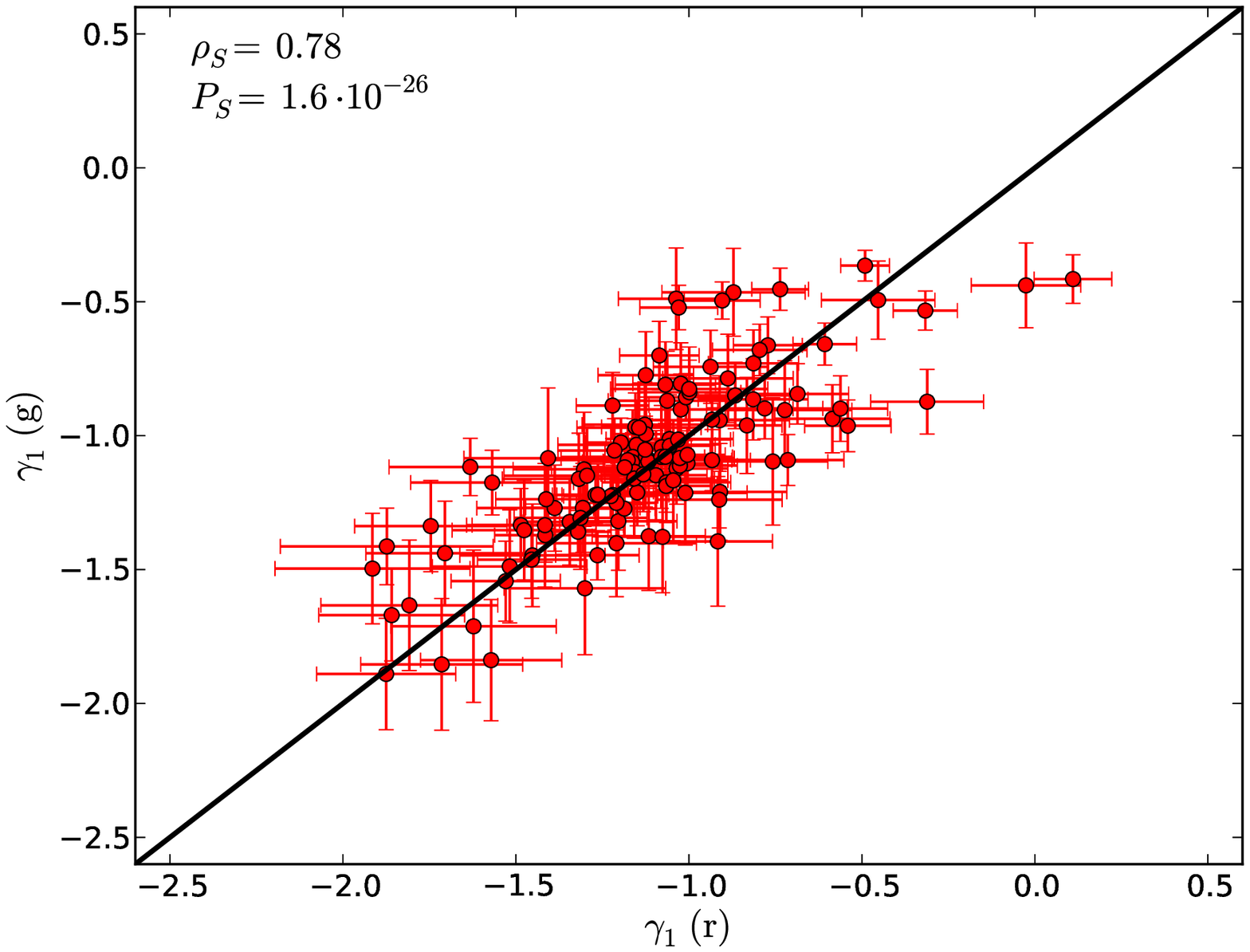}}
\quad
\subfloat{%
	\includegraphics[width=.4\textwidth]{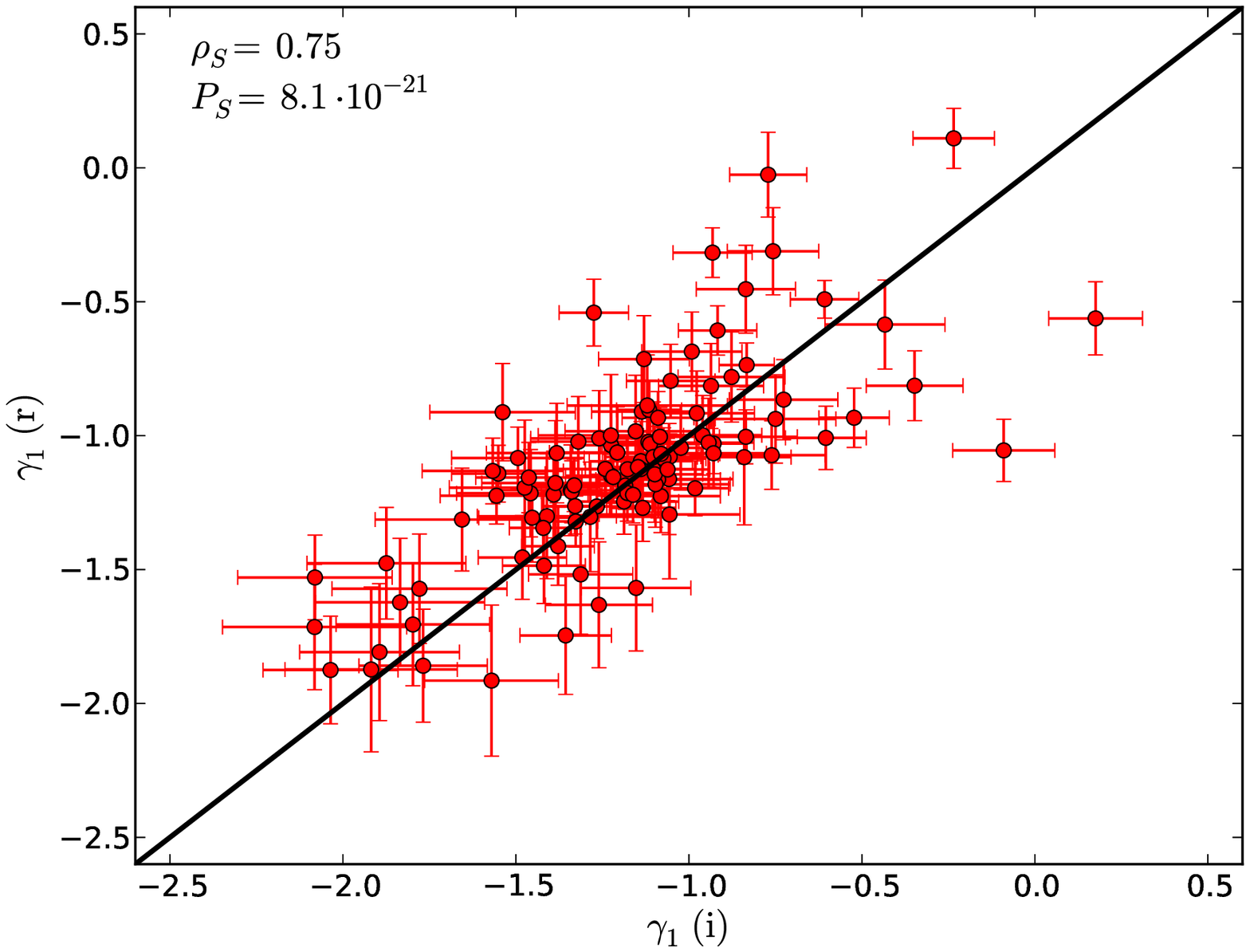}}

\subfloat{%
	\includegraphics[width=.4\textwidth]{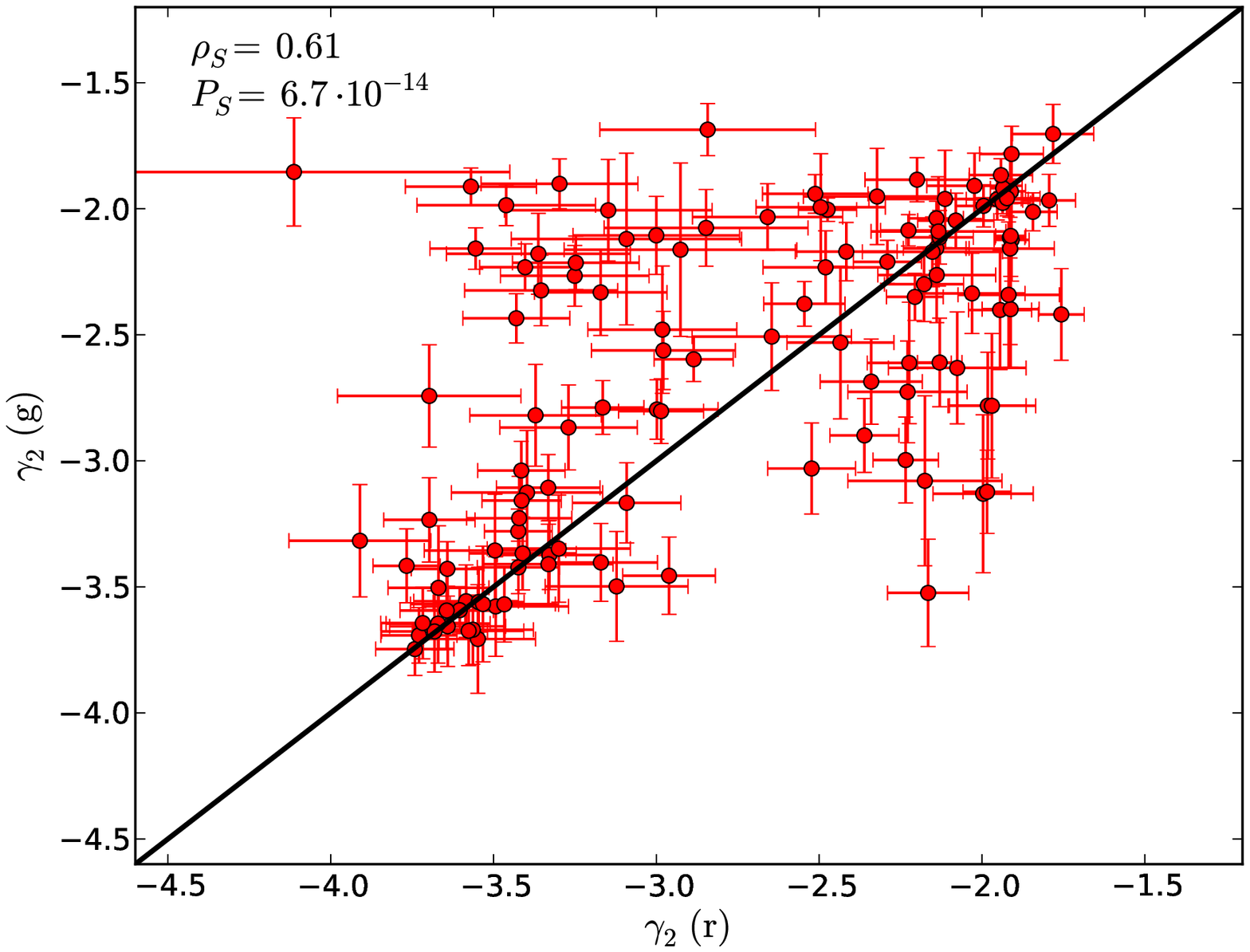}}
\quad
\subfloat{%
	\includegraphics[width=.4\textwidth]{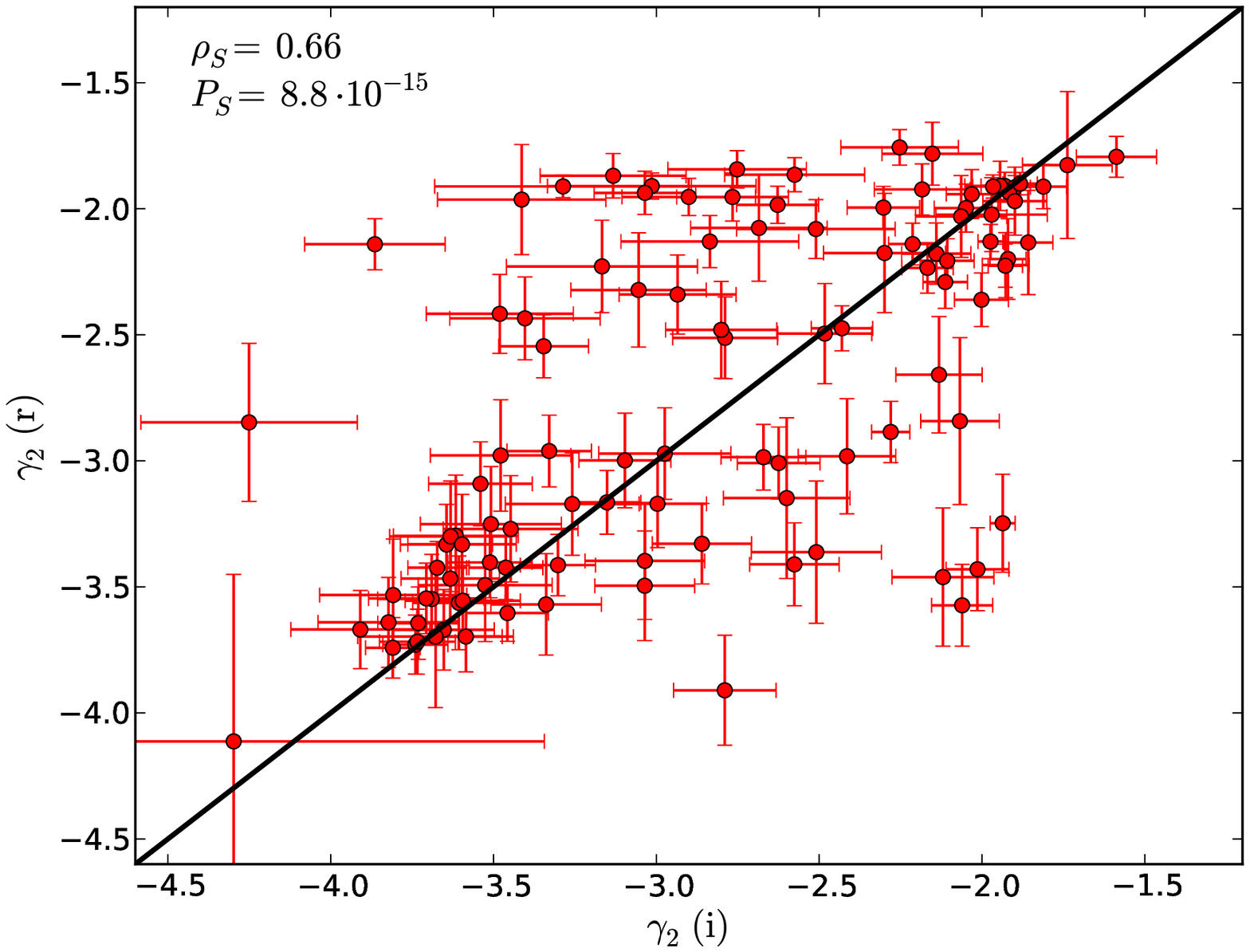}}
\caption{Comparison of the fit parameters of the broken power law PSD (equation \ref{eq:psdfit}) in various PS1 bands. The data of all objects from the PSD sample with fit parameters in both considered bands are shown. The black line corresponds to the one to one relation. The Spearman correlation coefficient and respective p-value are shown in each panel.}
	\label{fig:cmpnubr}
\end{figure*}

In this work we determined PSDs in different optical bands by modelling the $g_{\mathrm{P1}}$, $r_{\mathrm{P1}}$, $i_{\mathrm{P1}}$ and $z_{\mathrm{P1}}$ band flux light curves of our AGN sample with a CARMA(p,q) process. Fitting the derived PSDs with a broken power law after equation \ref{eq:psdfit} we obtained values for the amplitude $A$, the break frequency $\nu_{\mathrm{br}}$, the low frequency slope $\gamma_{1}$ and the high frequency slope $\gamma_{2}$. We find that the optical PSDs for the different bands are generally very similar, reflecting the fact that our AGNs vary approximately simultaneous in these bands. The fitted parameters of the power law PSD are contrasted for the $g_{\mathrm{P1}}$, $r_{\mathrm{P1}}$ and $i_{\mathrm{P1}}$ bands in Fig. \ref{fig:cmpnubr}. Obviously the parameters in the various bands are highly correlated and very close to the one to one relation, except for the slope $\gamma_{2}$. The break frequency and the low frequency slope exhibit no dependency on the radiation wavelength. Considering the amplitude $A$ we observe that the values are systematically shifted upwards the one to one relation when comparing a "bluer" band on the y-axis with a "redder" band on the x-axis. This resembles the findings presented in section \ref{sec:rmswave} using the excess variance as variability estimator and therefore provides an independent confirmation of the result that AGNs are more variable in the "bluer" bands. In contrast, regarding the high frequency slope $\gamma_{2}$ we see that there are significant deviations in the fitted values for the different bands. This may either indicate that the PSD is wavelength dependent at high variability frequencies or that our fitting method is significantly less robust in determining the high frequency slope from the data at these low variability amplitudes. The latter is clearly the case for many of the objects located far from the one to one relation. In fact the large uncertainty on the PSD at low variability amplitudes often does not allow a firm determination of the high frequency slope, causing the different slopes for the various PS1 bands. The aforementioned statements are qualitatively the same when comparing the fitted parameters with the ones obtained in the $z_{\mathrm{P1}}$ band, which is why we do not show the results here.            

\end{appendix}

\end{document}